\documentclass[aps,prd,showpacs,amsfonts,notitlepage,amssymb,amsmath,nofootinbib,superscriptaddress]{revtex4-1}
\usepackage[utf8]{inputenc}
\usepackage[english]{babel}
\usepackage[titletoc,toc,title]{appendix}
\usepackage{amsmath}
\usepackage{amsfonts}
\usepackage[colorlinks=true,linkcolor=blue,citecolor=blue,urlcolor=blue]{hyperref}
\usepackage{amssymb}
\usepackage{graphicx}
\usepackage{enumerate}
\usepackage{csquotes}
\usepackage[caption=false]{subfig}
\usepackage{dsfont}
\usepackage{color}
\usepackage{bbold}
\usepackage{mathtools}
\usepackage{tensor}

\newcommand{\eq}[1]{Eq.~(\ref{#1})}

\newcommand{\abs}[1]{\ensuremath{\left\lvert #1 \right\rvert}}

\newcommand{\dd}{\ensuremath{\mathrm{d}}}
\definecolor{darkcyan}{rgb}{0.,0.5,0.5}
\newcommand{\ee}{e} 
\newcommand{\ii}{i} 

\begin{document}

\title{Controlling and observing nonseparability of phonons created \\ 
in time-dependent 1D atomic Bose condensates}
\date{\today}

\author{Scott Robertson\thanks{scott.robertson@th.u-psud.fr}}
\affiliation{Laboratoire de Physique Th\'eorique, CNRS, Univ. Paris-Sud, Universit\'e Paris-Saclay, 91405 Orsay, France}
\author{Florent Michel\thanks{florent.michel@th.u-psud.fr}} 
\affiliation{Laboratoire de Physique Th\'eorique, CNRS, Univ. Paris-Sud, Universit\'e Paris-Saclay, 91405 Orsay, France}
\affiliation{Center for Particle Theory, Durham University, South Road, Durham, DHA 3LE, UK} 
\author{Renaud Parentani\thanks{renaud.parentani@th.u-psud.fr}}
\affiliation{Laboratoire de Physique Th\'eorique, CNRS, Univ. Paris-Sud, Universit\'e Paris-Saclay, 91405 Orsay, France}

\begin{abstract}
We study the spectrum and entanglement of phonons produced by temporal changes in homogeneous one-dimensional atomic condensates. To characterize the experimentally accessible changes, we first consider the dynamics of the condensate when varying the radial trapping frequency, separately studying two regimes: an adiabatic one and an oscillatory one. Working in momentum space, we then show that {\it in situ} measurements of the density-density correlation function can be used to assess the nonseparability of the phonon state after such changes. We also study time-of-flight (TOF) measurements, paying particular attention to the role played by the adiabaticity of opening the trap on the nonseparability of the final state of atoms. In both cases, we emphasize that commuting measurements can suffice to assess nonseparability. Some recent observations are analyzed, and we make proposals for future experiments. 
\end{abstract}

\maketitle

\section{Introduction}

One of the main predictions of quantum field theory concerns the possibility of producing pairs of particles 
by exciting vacuum fluctuations. This requires coupling the quantum field to a strong external classical field.
The simplest example consists in working with a homogeneous but time-dependent background.
In this case, by solving the linear equation of motion of the quantum field,
one easily verifies that the temporal variations of the background induce a parametric amplification of vacuum fluctuations 
which results in the production of pairs of particles with opposite momenta,
such as occurs in an expanding universe~\cite{Parker-1968,Birrell-Davies,Fulling}. 
A similar phenomenon takes place in static external fields which have sufficiently strong spatial gradients, such as 
the production of pairs of charged particles in a constant electric field~\cite{Schwinger-1951,Greiner} 
or the Hawking effect in the vicinity of a black hole horizon~\cite{Hawking-1975}. 
(See \cite{Primer} for the relationship of these latter phenomena to each other, as well as to pair production by an accelerating mirror.)

However, it turns out that the experimental validation of these predictions is very difficult, and so far
we are not aware of any direct detection using elementary particles. To circumvent the difficulties 
related to the high energies at play in particle physics, it has been proposed to use quasi-particles describing collective excitations of some medium~\cite{Unruh-1981,AnalogueGravity-LivingReview}. In fact, recently there have been several experiments aimed at observing the pair creation of quasi-particles propagating in a background with either temporal~\cite{Wilson-et-al-2011,Jaskula-et-al,Lahteenmaki-et-al} or strong spatial gradients~\cite{Lahav-et-al-2010,Nguyen-et-al,Steinhauer-2016}. 

Two aspects of such experiments deserve close attention. The first concerns the dynamical response of the medium. Indeed, when dealing with quasi-particles, one controls only indirectly the temporal (or spatial) dependence of the macroscopic system which acts as a background field. For instance, when considering the dynamical Casimir effect (DCE) in an atomic Bose gas~\cite{Fedichev-Fischer,Carusotto-DCE} (the main focus of the present work), the condensed portion of the gas acts as the background, whereas one has direct control only over the external potential. Modifying this potential induces a response in the condensate, the dynamics of which are governed by the Gross-Pitaevskii equation and may lead to nontrivial evolutions. 

The second issue concerns the origin of the detected particles. For, if the scattering of vacuum fluctuations gives rise to pairs of correlated outgoing quanta, so does the scattering of a nonzero incident distribution (such as a thermal bath). Hence, in practice, both processes produce correlated pairs. But we wish to be able to show that a part of the signal {\it must} have originated in vacuum fluctuations, as this is the component with no classical counterpart. 
This requires measurements able to distinguish between the spontaneous and stimulated channels (sourced respectively by vacuum and by an incident incoherent distribution)~\cite{Werner,Simon,Horodecki-Review}.
\footnote{In this work, we shall neglect all possible couplings between the 2-mode systems under study, namely phonon pairs with opposite momenta, and other degrees of freedom. These would induce decoherence effects which can destroy the entanglement that would otherwise be present, see Refs.~\cite{Busch-Parentani-2013,Busch-Parentani-Robertson,Busch-Carusotto-Parentani,Swislocki-Deuar,Zin-Pylak} for applications in condensed matter systems, and Refs.~\cite{Campo-Parentani-2005,Campo-Parentani-2008-I,Campo-Parentani-2008-II,Adamek-Busch-Parentani} for applications in early cosmology.} In this paper, two observables shall be used: when working {\it in situ}, we shall use the two-point function of density fluctuations in the atomic gas, whereas the two-body distribution of the atoms' momenta will be used when considering time-of-flight (TOF) experiments after having opened the trap~\cite{Pitaevskii-Stringari-BEC}. We shall determine under which conditions precise measurements of each of these observables is sufficient to assess the nonseparability of the phonon state after DCE.  Surprisingly, 
we shall show that instantaneous measurements involving only commuting variables can be sufficient~\footnote{{\it Added note:} After having completed this work, in~\cite{Robertson-Michel-Parentani-steering} we extended the analysis here presented to a degree of entanglement of bipartite states which is stronger than nonseparability, namely the possibility of steering the outcome of one subsystem by performing measurements on the other.  We also considered inhomogeneous transonic flows in view of the above-mentioned analogy with black holes.}. 

In this work we consider density perturbations in a cigar-shaped ultracold atomic Bose gas. 
That is, we focus on elongated clouds with transverse dimensions much smaller than their longitudinal extensions. More precisely, we work in the 1D mean field regime~\cite{Menotti-Stringari,Tozzo-Dalfovo,Gerbier} where the transverse excitations can be neglected. 
In Section \ref{sec:Condensate} we study the dynamical response of such an elongated cloud when modifying the harmonic trap
frequency $\omega_\perp$, which fixes its transverse extension~\cite{Kagan-Surkov-Shlyapnikov}. 
In Section \ref{sec:Phonons} we study the longitudinal excitations (phonons) excited 
by the resulting time dependence of the atomic cloud, including what it means for pairs of phonon modes to be in a nonseparable state. Section \ref{sec:InSitu} deals with the determination of the phonon state via {\it in situ} measurements, 
whereas in Section \ref{sec:TOF} we examine the measurements of atoms after TOF and how these can be used to infer the phonon state before the expansion of the cloud. Our main results are summarized in Section \ref{sec:Conclusion}, along with their 
implications for future experiments. In Appendix~\ref{app:num_sol_GPE} we compare our analytical approximations with the results of 
numerical resolutions of the Gross-Pitaevskii equation, while in Appendix~\ref{app:Thermal} 
we present some results when the initial phonon state is a thermal bath.  


\section{Dynamical response of elongated condensates 
\label{sec:Condensate}}

In this Section, we examine the properties of a cylindrical condensate and its response to a varying harmonic potential~\cite{Kagan-Surkov-Shlyapnikov}. Since 
the condensate plays the role of the background field in the pair production of phonons, this is a necessary first step to know what variations of the background are experimentally accessible.  

\subsection{Gaussian approximation and stationary configuration}

To characterize the background solutions we use the Bogoliubov approximation and 
write the field operator for a Bose condensed atomic gas in the form
\begin{equation}
\hat{\Phi} = \Phi_{0} \left( \mathbb{1} + \hat{\phi} \right) \,,
\label{eq:Bog_approx}
\end{equation}
where $\Phi_{0}$ is a $c$-number and $\hat{\phi}$ describes relative 
linear perturbations.
The mean field $\Phi_{0}$ obeys 
(in units where $\hbar = 1$):
\begin{equation}
i\partial_{t}\Phi_{0} = -\frac{1}{2m} \nabla^{2}\Phi_{0} + V \Phi_{0} + g \left|\Phi_{0}\right|^{2} \Phi_{0} \,,
\label{eq:GPE}
\end{equation}
where $m$ is the atomic mass, $V$ is an external potential, 
and $g$ is the two-atom coupling constant, related to the $s$-wave scattering length $a_{s}$ via $g = 4\pi a_{s}/m$ (see, e.g., \cite{Dalfovo-et-al-1999,Pitaevskii-Stringari-BEC}).
Throughout this work, we assume $g > 0$. 
The external potential is taken to be harmonic, and assuming rotational symmetry 
it can be decomposed into a radial and a longitudinal part:
\begin{equation}
V\left(\vec{x}\right) = \frac{1}{2}m \omega_{\perp}^{2} r^{2} + \frac{1}{2}m \omega_{\parallel}^{2} z^{2} \,.
\end{equation}
The ratio $\omega_{\parallel}/\omega_{\perp}$ determines the shape of the condensate: when $\omega_{\parallel}/\omega_{\perp} \ll 1$, the condensate is very elongated in the longitudinal direction $z$, 
becoming cigar-shaped.  For simplicity, we shall assume $\omega_{\parallel} = 0$, so that the ground state 
of the condensate is homogeneous in $z$. 
If we further assume that the condensate is stationary and cylindrically symmetric, 
then $\Phi_{0}$ becomes a function of only the radial coordinate $r$.  Even in this case, 
the exact solutions of Eq.~(\ref{eq:GPE}) are 
complicated by the nonlinear term, see Appendix~\ref{app:num_sol_GPE} 
for details.

Here we adopt the procedure of~\cite{Gerbier}: we approximate the atomic number density by a Gaussian: 
\begin{equation}
\rho_0(r) = \left|\Phi_{0}\right|^{2} \approx \frac{n_{1}}{\pi \sigma^{2}} e^{-r^2/\sigma^{2}}\, ,
\label{eq:gaussian_ansatz}
\end{equation}
where $n_1$ is the linear atom density and $\sigma$ the width of the condensate. 
This ansatz becomes exact when $g=0$. We thus expect it to be a good approximation when $n_{1} a_{s}$ is small enough. 
In a stationary configuration, the left-hand side 
of Eq. (\ref{eq:GPE}) is simply $\mu\, \Phi_{0}$, where $\mu$ is the energy per atom (equal to the chemical potential at zero temperature). 
Multiplying Eq. (\ref{eq:GPE}) by $\Phi_{0}^{\star}$ and integrating over $r$, we find that $\mu$ is equal to the effective potential
\begin{equation}
V_{\rm eff}(\sigma) = \frac{1}{2} m \omega_{\perp}^{2} \sigma^{2} + \frac{1 + 4n_{1}a_{s}}{2 m \sigma^{2}} \,. 
\label{eq:Veff}
\end{equation}
An approximation of the ground state of the condensate for a given linear density $n_1$ is obtained by minimizing $V_{\rm eff}$ with respect to $\sigma$, yielding the optimal width $\sigma_0$ and corresponding value of $\mu$~\cite{Gerbier}:  
\begin{subequations}\begin{alignat}{2}
\sigma_{0} &= a_{\perp} \left( 1 + 4 n_{1} a_{s} \right)^{1/4} \,, \\
\mu &= \omega_{\perp} \left( 1 + 4 n_{1} a_{s} \right)^{1/2} \,,
\end{alignat}\label{eq:cond_ground_state}\end{subequations}
where $a_{\perp} \equiv \left(m \omega_{\perp}\right)^{-1/2}$ is the harmonic oscillator length.  
Although these expressions have been derived in the Gaussian approximation of Eq. (\ref{eq:gaussian_ansatz}),
the comparison with numerical results of the full Gross-Pitaevskii equation (\ref{eq:GPE}) in Appendix~\ref{app:num_sol_GPE} 
indicates that they are quite robust (in agreement with \cite{Gerbier}).

\begin{figure}
\includegraphics[width=0.5\columnwidth]{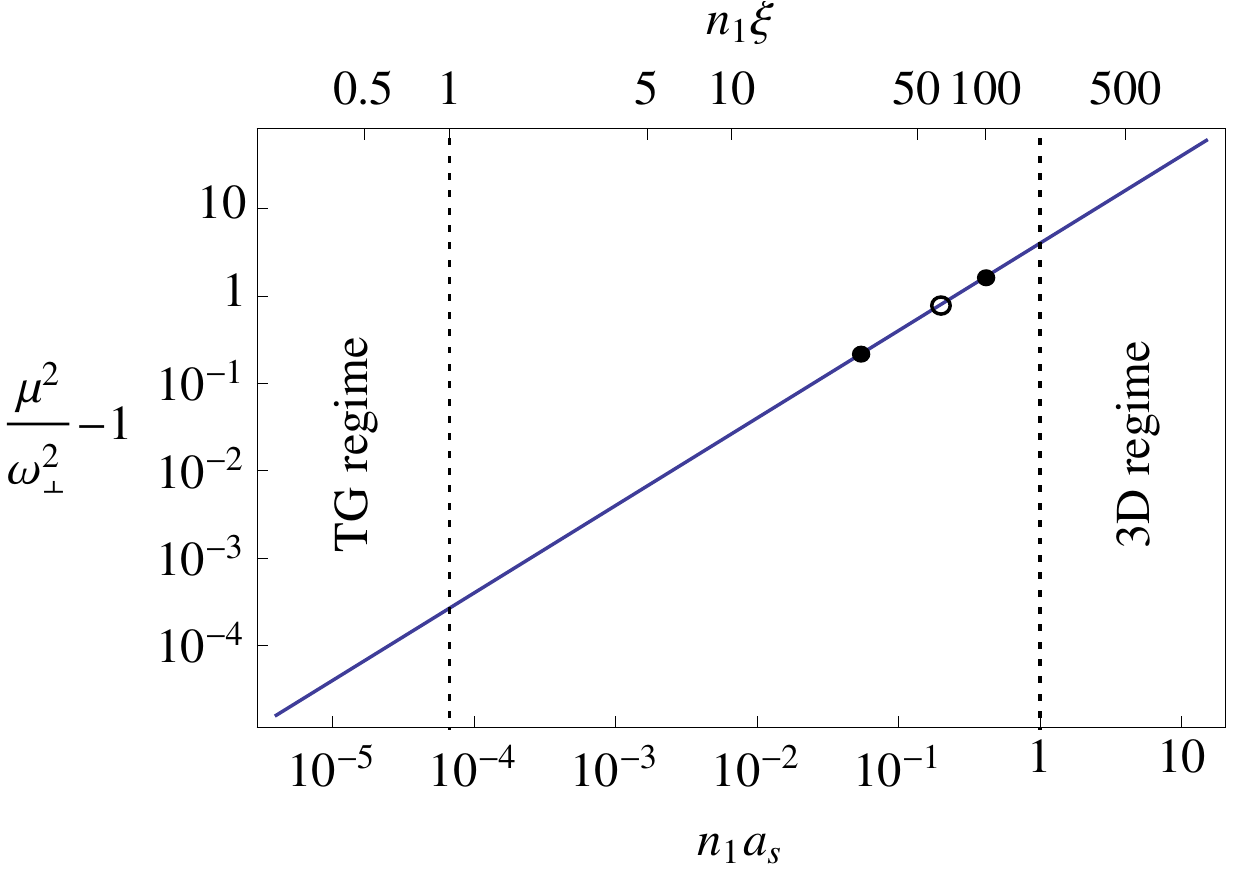}
\caption{The chemical potential as a function of the adimensionalized linear densities, $n_{1} a_{s}$ and $n_{1} \xi$, in the Gaussian approximation, see \eq{eq:gaussian_ansatz}. 
For clarity we have plotted $\mu^2/\omega_{\perp}^{2}-1$, which in this approximation is linear in $n_{1} a_{s}$. 
We have marked the region $n_{1} a_{s} \gtrsim 1$ 
(where the gas exhibits 3D behavior) 
and the region $n_{1} \xi \lesssim 1$ (where the gas exhibits Tonks-Girardeau behavior~\cite{Petrov-Shlyapnikov-Walraven}). 
To relate $n_{1} \xi$ to $n_{1} a_s$, see \eq{eq:1Dxi}, we have used the value of $\left(a_{s}/a_{\perp}\right)^{2} = 3 \times 10^{-5}$ corresponding to the experiment of~\cite{Steinhauer-2016}. The two black dots (roughly) correspond to the subsonic ($n_{1} a_{s} \approx 0.4$) and supersonic ($n_{1} a_{s} \approx 0.05$) sides of the analogue black hole realised in \cite{Steinhauer-2016}, while the circle gives a rough indication of the position of the modulated DCE experiment of \cite{Jaskula-et-al} ($n_{1} a_{s} \approx 0.2$).
\label{fig:chemical_potential}}
\end{figure}

In the Gaussian approximation, 
$\mu/\omega_{\perp}$ is a function of only $n_{1} a_{s}$. By contrast, the relationship between the density $n_{1}$ and the (effective 1-dimensional) healing length $\xi$ also depends on the ratio $a_{s}/a_{\perp}$: 
\begin{equation}
n_{1} a_{s} \, \sqrt{1 + 4 n_{1} a_{s}} = 2 \left(a_{s}/a_{\perp}\right)^2 \left(n_{1} \xi\right)^{2} \,.  
\label{eq:1Dxi}
\end{equation}
This extra dependence stems from the fact that $\xi$ refers to the dynamics of phonons, which shall be studied in the next Section.
In Fig.~\ref{fig:chemical_potential} we represent the chemical potential $\mu$ as a function of 
$n_{1} a_s$, as given by the second of Eqs. (\ref{eq:cond_ground_state}). On the upper horizontal axis is shown $n_{1} \xi$ of Eq. (\ref{eq:1Dxi}), with $\left(a_{s}/a_{\perp}\right)^{2} = 3 \times 10^{-5}$ as is approximately the case in the experiment of~\cite{Steinhauer-2016} and is of the right order of magnitude for the experiments of \cite{Jaskula-et-al}. The validity of the 1D mean field regime requires $n_{1} a_{s} \lesssim 1$ and $n_{1} \xi \gg 1$.  In Fig.~\ref{fig:chemical_potential} these limits are indicated by the dotted vertical lines. We also indicate the values of $n_{1} a_s$ for Refs.~\cite{Steinhauer-2016,Jaskula-et-al}.

\subsection{Time-dependent response
\label{sub:response}}

Let us now suppose that the trapping frequency $\omega_{\perp}$ varies in time. Using again the Gaussian approximation 
of \eq{eq:gaussian_ansatz} with $\sigma$ time dependent, one obtains the equation of motion
\begin{equation}
m \ddot{\sigma} = - \partial_\sigma V_{\rm eff}(\sigma) \,,
\label{eq:sigma_eom}
\end{equation} 
see Appendix~\ref{app:num_sol_GPE} for the derivation. 
The condensate width $\sigma$ thus behaves like the position of a classical point particle of mass $m$ in the potential $V_{\rm eff}$.
Remarkably, although this equation has been derived in the context of the Gaussian approximation, it is an {\it exact} consequence of equation (\ref{eq:GPE}) in a harmonic radial potential, as found in~\cite{Kagan-Surkov-Shlyapnikov}. This is due to the existence of an exact scale invariance parametrized 
by $\sigma(t)$, the time-dependent radial density profile being given by
\begin{equation}
\rho(t,r) = \frac{\rho_{0}\left({r}/{\sigma(t)}\right)}{\sigma^{2}(t)} , 
\end{equation}
where $\rho_{0}$ 
is a solution of the time-independent Gross-Pitaevskii equation and $\sigma$ satisfies Eq.~\ref{eq:sigma_eom}. 

To study the evolution of $\sigma(t)$, it 
is convenient to express $\omega_{\perp}^{2}(t) = \omega_{\perp,0}^{2}\,  \lambda(t)$, where $\omega_{\perp,0}$ is some reference value (commonly its initial value). Eq. (\ref{eq:sigma_eom}) then gives 
\begin{equation}
\frac{1}{\omega_{\perp,0}^{2}}\frac{\ddot{\sigma}}{\sigma_{0}} = - \lambda(t) \frac{\sigma}{\sigma_{0}} + \frac{\sigma_{0}^{3}}{\sigma^{3}} \,, \label{eq:sigma_eom2}
\end{equation}
where $\sigma_{0}$ is the equilibrium value corresponding to $\omega_{\perp,0}$, see \eq{eq:cond_ground_state}. 
This suggests that we 
use the adimensionalized time variable $\omega_{\perp,0} t$ and width $\sigma(t)/\sigma_{0}$. 
When $\lambda$ is constant, \eq{eq:sigma_eom2}  
can be further simplified by noting that $\sigma^{2}$ exactly obeys a harmonic equation:
noting that Eq.~(\ref{eq:sigma_eom}) implies conservation of $E \equiv m \dot{\sigma}^2/2 + V_{\rm eff}(\sigma)$, it can be rewritten in the form 
\begin{equation}
\partial_{t}^{2} \left( \sigma^{2} - \frac{E}{m\, \omega_{\perp}^{2}} \right) = - \left(2 \omega_{\perp} \right)^{2} \left( \sigma^{2} - \frac{E}{m\, \omega_{\perp}^{2}} \right) \,, 
\label{eq:sigmasq_eom}
\end{equation}
so that $\sigma^{2}$ oscillates with frequency $2\omega_{\perp} = 2\, \sqrt{\lambda}\, \omega_{\perp,0}$. This implies that, while the oscillations of $\sigma$ are not exactly sinusoidal, its oscillation frequency is always $2\omega_{\perp}$. (The latter result is intuitive: a single particle in the same potential would oscillate around $r=0$ at frequency $\omega_{\perp}$, while the condensate wave function 
recovers its initial form when all atoms are in the radially opposite position, and thus oscillates at twice this frequency.)  

\begin{figure}
\includegraphics[width=0.45\columnwidth]{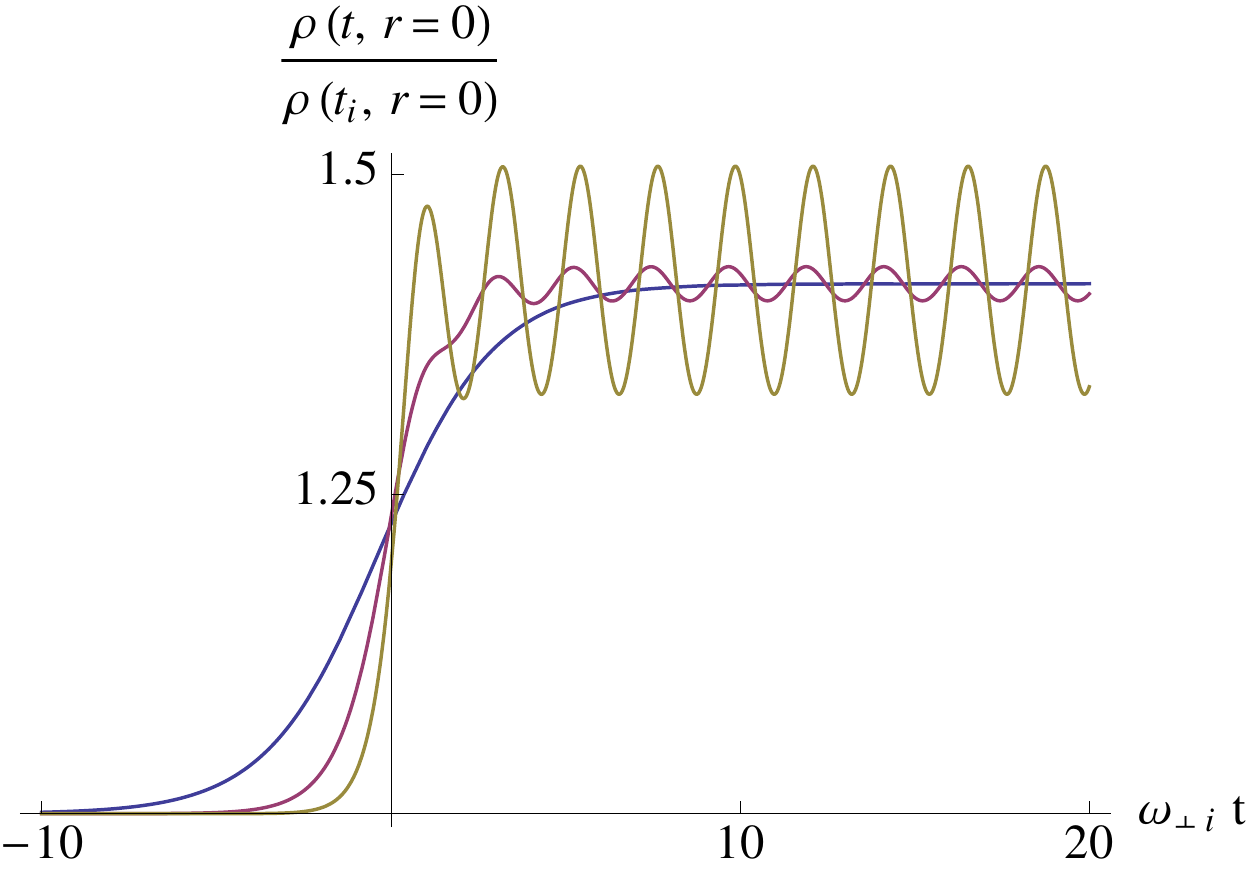} \, \includegraphics[width=0.45\columnwidth]{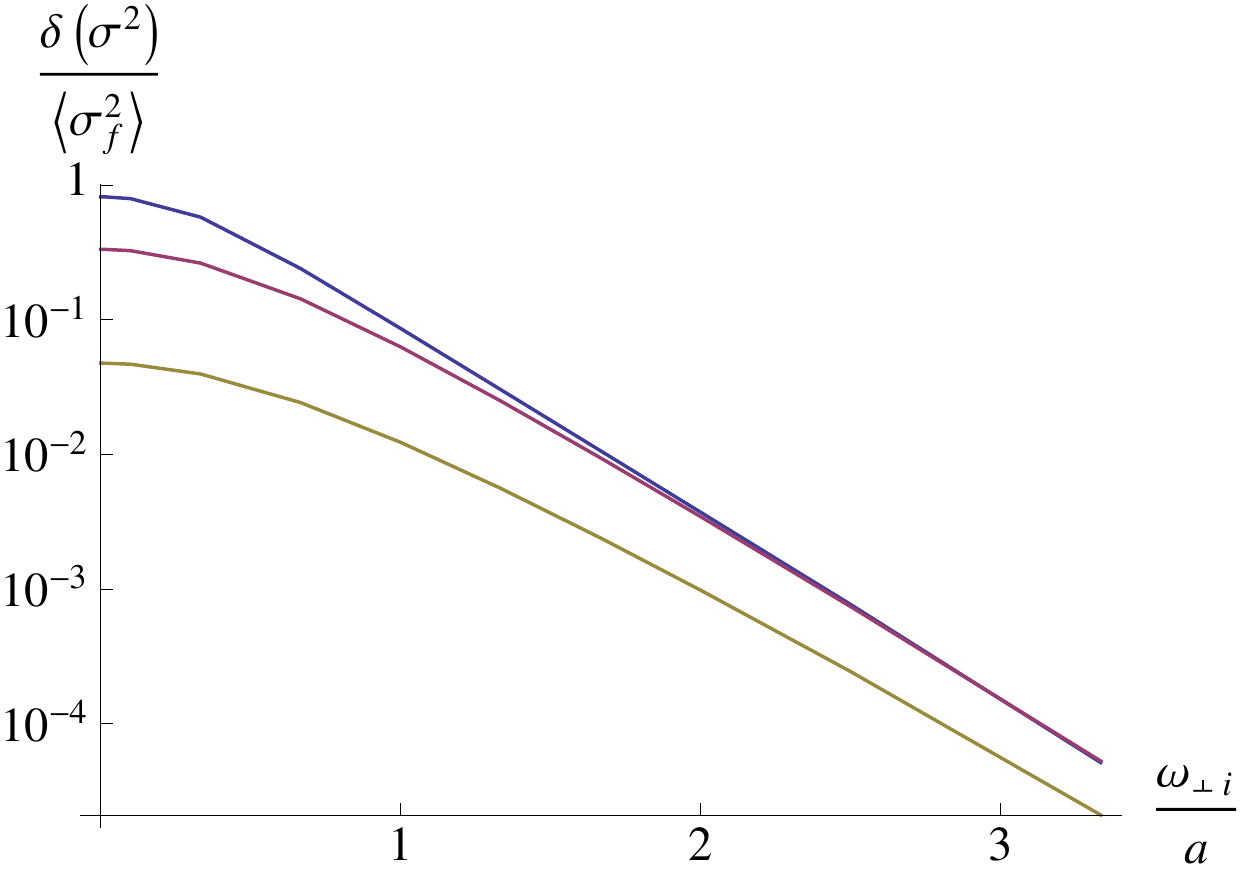}
\caption{The response of the condensate to a temporal change in the radial trap frequency $\omega_{\perp}(t)$.  On the left are plotted the resulting variations in the central density $\rho(t, r= 0)$ when $\omega_{\perp}^{2}$ is varied according to Eq. (\ref{eq:omega_perp_tanh}) with $\omega_{\perp f}^{2}/\omega_{\perp i}^{2} = 2$.  The three curves correspond to $a/\omega_{\perp i} = 0.3$ (blue curve), $0.6$ (purple curve) and $1.0$ (yellow curve).  We note that, while for the slowest variation 
$\rho(t, r= 0)$ varies adiabatically, oscillations appear for larger $a$, and are more pronounced the more rapid is the change in $\omega_{\perp}$.  In the limit of a sudden change, i.e. 
$a/\omega_{\perp i} \rightarrow \infty$, their amplitude reaches a maximum, so that after the change 
$\rho(t, 0)/\rho(t_i, 0)$ oscillates between $1$ and another (computable) value. On the right is shown, on a logarithmic scale, the relative amplitude of oscillations in $\sigma^{2}$ after the change, as a function of the inverse rate $\omega_{\perp i}/a$.  There, the three curves correspond to $\omega_{\perp f}^{2}/\omega_{\perp i}^{2} = 10$ (blue), $2$ (purple) and $1.1$ (yellow).  We note that, for $\omega_{\perp i}/a \gtrsim 2$, these curves become proportional to $\mathrm{exp}\left(-\pi \omega_{\perp i}/a\right)$, whereas for $\omega_{\perp i}/a \rightarrow 0$ they approach values that can be calculated assuming a sudden change in $\omega_{\perp}^{2}$.
\label{fig:tanh_response}}
\end{figure}

Let us use Eq.~\ref{eq:sigma_eom2} to determine (numerically) how the condensate responds to a temporal change in $\omega_{\perp}$.
For concreteness, we assume that $\omega_{\perp}^{2}$ varies according to a hyperbolic tangent\footnote{The hyperbolic tangent has the advantage that $\omega_{\perp}$ is asymptotically constant, and the solutions are thus fairly straightforward to analyze. 
One could also consider the case where $\omega_{\perp}$ is modulated in time, as in the second experiment of~\cite{Jaskula-et-al}.  We have looked briefly at this case, and our numerical simulations indicate that $\sigma(t)$ behaves in a complicated way. This case requires further study.} in time: 
\begin{equation}
\frac{\omega_{\perp}^{2}(t)}{\omega_{\perp i}^{2}} = \frac{1}{2}\left(\frac{\omega_{\perp f}^{2}}{\omega_{\perp i}^{2}} + 1\right) + \frac{1}{2}\left(\frac{\omega_{\perp f}^{2}}{\omega_{\perp i}^{2}} - 1\right) \, \mathrm{tanh}\left(a t\right) \,,
\label{eq:omega_perp_tanh}
\end{equation}
where $\omega_{\perp i}$ and $\omega_{\perp f}$ are the initial and final values of $\omega_{\perp}(t)$, respectively.
Here we work with $\omega_{\perp,0}= \omega_{\perp i}$ so that $\lambda(t) \to 1$ for $t \to -\infty$.
In the left panel of Figure \ref{fig:tanh_response} are shown examples in which $\omega_{\perp f}^{2}/\omega_{\perp i}^{2}$ is fixed at $2$, as in the first experiment of \cite{Jaskula-et-al}.  The three variations of $\omega_{\perp}$ differ only in the rate $a$, which describes the rapidity of the change. The plot itself shows the corresponding time dependence of the central density $\rho(t, r = 0)$. As can be seen, the rate $a$ is crucial in determining the final state of the condensate.  For, when $a/\omega_{\perp i}$ is small enough (i.e., $a/\omega_{\perp i} \ll 1$), the change is adiabatic: the condensate remains in its ground state throughout the change, with its final width and chemical potential simply corresponding to Eqs. (\ref{eq:cond_ground_state}) with the new values of $\omega_{\perp}$ and $a_{\perp}$.  However, as $a/\omega_{\perp i}$ is increased, the condensate cannot keep up with the change in $\omega_{\perp}$ and ends up with an energy which is larger than the final ground state energy. Its central density thus oscillates around the value of the final ground state. As could have been expected, the amplitude of these oscillations increases with $a/\omega_{\perp i}$, saturating at a value corresponding to a sudden change, where one extreme of the oscillations in $\sigma$ is equal to its initial value. 

In the right panel of Fig.~\ref{fig:tanh_response}, we plot the relative amplitude in the final oscillations of $\sigma^{2}$ as a function of the inverse rate $\omega_{\perp i}/a$ for various overall changes $\omega_{\perp f}^{2}/\omega_{\perp i}^{2}$.
$\sigma^{2}$ is used because, as implied by Eq.~(\ref{eq:sigmasq_eom}), when $\omega_{\perp}$ is constant $\sigma^{2}$ undergoes oscillations that are exactly sinusoidal. This illustrates the validity domains of two regimes discussed in~\cite{Kagan-Surkov-Shlyapnikov}: the sudden limit where the relative amplitude $\delta \sigma^2/\langle \sigma_f^2 \rangle$ becomes independent of $a$ for $a/\omega_{\perp i} \rightarrow \infty$; and the adiabatic regime where this relative amplitude is proportional to $\mathrm{exp}\left(-\pi \omega_{\perp i}/a\right)$ in the limit $a/\omega_{\perp i} \rightarrow 0$.

\subsection{Opening the trap adiabatically
\label{sub:opening}}

Let us end this section by considering the case where the final value of $\omega_{\perp}$ is zero. This corresponds to the opening of the trap, as performed in TOF
measurements.  We can immediately appreciate the qualitative difference of this case by considering the effective potential (\ref{eq:Veff}), which for $t \to \infty$ ends up 
with only the $1/\sigma^{2}$ repulsive term.  The condensate thus no longer oscillates, but expands indefinitely, and at large enough $\sigma$ when the repulsive interaction between the atoms becomes negligible, they will simply expand freely and $\sigma$ will increase linearly in time as if it were the position of a free particle.

\begin{figure}
\includegraphics[width=0.45\columnwidth]{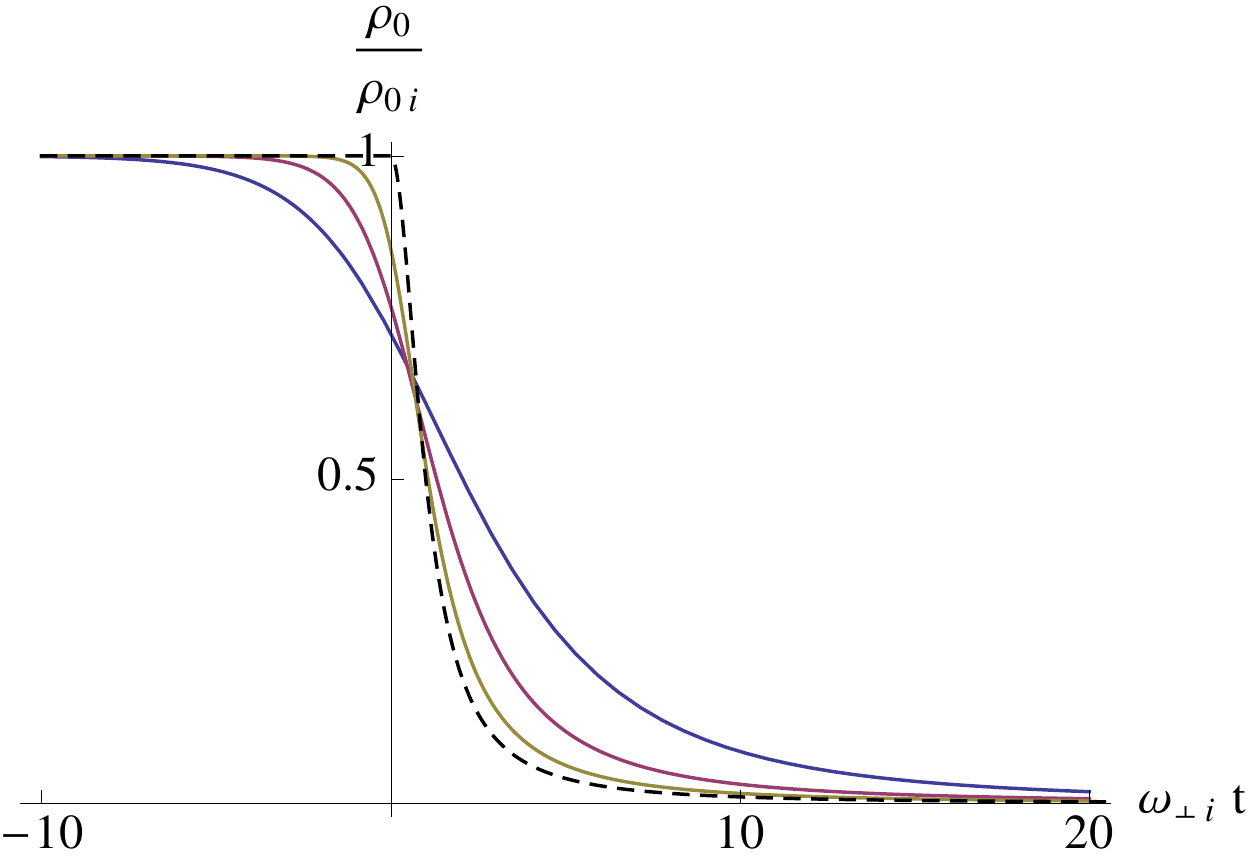}
\caption{The response of the condensate to a temporal change in $\omega_{\perp}$ in which the trap is switched off, i.e. the final value of $\omega_{\perp}$ is zero. 
As in Fig. \ref{fig:tanh_response}, the variation in $\omega_{\perp}^{2}$ follows \eqref{eq:omega_perp_tanh}.
Plotted are the resulting variations in the central density $\rho_{0}$.  The black dashed curve corresponds to the limiting case of a sudden switch-off, i.e., $a/\omega_{\perp i} \rightarrow \infty$.  
The three solid curves correspond to three different values of the rate $a$: $a/\omega_{\perp i} = 0.3$ (blue), $0.6$ 
(purple) and $1.2$ (yellow).
\label{fig:tanh_Tozzo_response}}
\end{figure}

To make contact with the former analysis, we continue to assume a hyperbolic tangent profile of the form (\ref{eq:omega_perp_tanh}), though 
the vanishing of $\omega_{\perp f}$ means that $a/\omega_{\perp i}$ is now the only independent parameter.
From Fig.~\ref{fig:tanh_Tozzo_response} it is clear that  $a/\omega_{\perp i}$ 
significantly affects the rate of expansion of the condensate.  In particular, a sudden opening of the trap (which is the standard case, see e.g.~\cite{Tozzo-Dalfovo}) 
results in the most rapid expansion. Importantly, it is possible to make the expansion arbitrarily slow by lowering $a$. 
As we shall see in Section \ref{sec:TOF}, in certain cases controlling 
the adiabaticity of the expansion  
is a necessity if one wishes to be able to assess the nonseparability of the phonon state that existed before opening the trap. 
Let us briefly explain why. 

Although the condensate 
response shown in Fig.~\ref{fig:tanh_Tozzo_response} is independent of the value of $n_1 a_s $ when using the adimensional units $\rho(t)/\rho_i$ and $\omega_{\perp i} t$, 
the {\it phonon} response depends strongly on $n_1 a_s$ in the same units. 
This is because, while $\omega_{\perp}$ is the natural frequency describing the response of the condensate to a sudden change, 
the natural frequency for describing the phonon response is $m c^{2}$, where $c$ is the low-frequency speed of longitudinal phonons (as we shall see in Section \ref{sec:Phonons}). Since the one-dimensional healing length $\xi  \equiv 1/mc$, \eq{eq:1Dxi} implies that the two time scales are related by 
\begin{equation}
\frac{\omega_{\perp}}{m c^{2}} = \frac{\sqrt{1 + 4 n_{1} a_{s}}}{2 n_{1} a_{s}} \,. 
\label{eq:timescaleratio}
\end{equation}
The larger is this factor, the faster the condensate expansion will appear from the phonon point of view, and (as we shall see in Section \ref{sec:TOF}) this rapid expansion can have a significant effect on the phonon state, including its degree of nonseparability. 
Importantly, in the 1D regime $n_{1} a_{s} \lesssim 1$ in which we work, 
this factor can be large. As a result, when considering phonon excitations for a given value of $k \xi$, a sudden opening of effectively 1D condensates leads to larger non-adiabatic effects than for thick ones in the 3D regime with $n_1 a_s > 1$. 
It is thus imperative to reduce the expansion rate through a controlled slow opening of the trap, so that the expansion is adiabatic from the phonon point of view and the phonon state can be preserved. (Again, we shall study these issues more fully in Section \ref{sec:TOF}.)  
We finally notice that the ratio $a_{s}/a_{\perp}$ plays here no role (at least within the Gaussian approximation).

\section{Pair creation of phonons and their nonseparability 
\label{sec:Phonons}}

In this Section, we recall two fundamental aspects of linear perturbations (phonons) propagating on top of the condensate: 
the response of phonons to a time-varying background (as provided by the condensate), and what it means for phonon pairs to be in a nonseparable state. Although it contains no original results, this Section provides key notions and equations needed in the sequel. 

\subsection{Equations of motion
\label{subsec:Equations_of_motion}}

Let us return to the Bogoliubov approximation of Eq.~(\ref{eq:Bog_approx}), for $z$-independent but time-dependent mean fields $\Phi_{0}$.  To simplify the analysis, we assume that the relative perturbation $\hat{\phi}$ is uniform in the transverse directions, i.e. that the absolute perturbation $\delta\hat{\Phi}(t,r,z) = \Phi_{0}(t,r)\,  \hat{\phi}(t,z)$ has the same transverse profile as $\Phi_{0}$. The validity of this ansatz is discussed in Appendix~\ref{app:num_sol_GPE}. Using this factorization, $\hat{\phi}$ obeys the 1-dimensional Bogoliubov-de Gennes equation:
\begin{equation}
i\,\partial_{t}\hat{\phi} = - \frac{1}{2 m} \partial_{z}^{2}\hat{\phi} + g_{1} n_{1} \left( \hat{\phi} + \hat{\phi}^{\dagger} \right) \,,
\label{eq:BdGeqn}
\end{equation}
where $n_{1}$ is the linear atom density in the longitudinal direction, and $g_{1} \equiv g/2\pi\sigma^{2}$ is the effective one-dimensional two-atom coupling constant ($\sigma$ being the cylindrical width of the condensate in the Gaussian approximation, see Eq. (\ref{eq:gaussian_ansatz})). Note that, while the linear density $n_{1}$ is a constant, $g_{1}$ varies as the inverse square of the condensate width, and is thus time-dependent.

The homogeneity of the background entails the decomposition of the system into two-mode sectors $(k,-k)$ 
of opposite wave vectors. It is thus useful to perform the spatial Fourier transform: for a gas of extension $L$ in the $z$ direction, we define $\hat{\phi}_{k}$ such that 
\begin{align} \label{eq:inv_Fourier} 
\hat{\phi}(z) = \sum_{k \in 2 \pi \mathbb{Z} / L} \hat{\phi}_{k} e^{i k z} / \sqrt{N}, 
\end{align}
where $N$ is the total number of atoms in the gas and serves here to normalize the plane wave eigenfunctions in such a way that the equal time commutator is 
\begin{equation}
\left[ \hat{\phi}_{k}(t) \,, \, \hat{\phi}_{k^{\prime}}^{\dagger}(t) \right] = \delta_{k,k^{\prime}} \,.
\end{equation}
Then the operator $\hat{\phi}_{k}$ ($\hat{\phi}_{k}^{\dagger}$) destroys (creates) an atom carrying momentum $k$.

To get the time dependence of these operators, we plug \eq{eq:inv_Fourier} 
in \eq{eq:BdGeqn} and obtain 
\begin{equation}
i \, \partial_{t} \left[ \begin{array}{c} \hat{\phi}_{k} \\ \hat{\phi}_{-k}^{\dagger} \end{array} \right] 
= \left[ \begin{array}{cc} \Omega_{k} & m c^{2} \\ -m c^{2} & -\Omega_{k} \end{array} \right] \left[ \begin{array}{c} \hat{\phi}_{k} \\ \hat{\phi}_{-k}^{\dagger} \end{array} \right] \,,
\label{eq:phi-k_eqn}
\end{equation}
where $m c^{2} \equiv g_{1} n_{1}$ and $\Omega_{k} \equiv k^{2}/2m + g_{1} n_{1}$. To identify the consequences of the time dependence of $g_{1}(t)$, it is appropriate to perform the Bogoliubov transformation:
\begin{equation}
\left[ \begin{array}{c} \hat{\phi}_{k} \\ \hat{\phi}_{-k}^{\dagger} \end{array} \right]
\equiv \left[ \begin{array}{cc} u_{k} & v_{k} \\ v_{k} & u_{k} \end{array} \right] \left[ \begin{array}{c} \hat{\varphi}_{k} \\ \hat{\varphi}_{-k}^{\dagger} \end{array} \right]\, , 
\label{eq:varphi_phi}
\end{equation}
where
\begin{equation}
\begin{array}{c} u_{k} \\ v_{k} \end{array} = \frac{ \sqrt{\Omega_{k} - m c^{2}} \pm \sqrt{\Omega_{k} + m c^{2}}}{2\sqrt{\omega_{k}}}\, ,
\label{eq:u-v_defn}
\end{equation}
and
\begin{eqnarray}
\omega_{k}^{2} & = & c^{2} k^{2} + \left(\frac{k^{2}}{2m}\right)^{2} \,.  
\label{eq:om_defn}
\end{eqnarray}
Irrespective of the time dependence of $c^2(t) = g_1(t) n_1/m$, one verifies that at all times  $u_{k}^{2}(t)-v_{k}^{2}(t)=1$ 
as required by the Bose commutation relations. Then Eq.~(\ref{eq:phi-k_eqn}) gives 
\begin{equation}
i\,\partial_{t} \left[ \begin{array}{c} \hat{\varphi}_{k} \\ \hat{\varphi}_{-k}^{\dagger} \end{array} \right]
= \left[ \begin{array}{cc} \omega_{k} & -i \frac{\partial_t \omega_k}{2 \omega_k} \\ 
-i \frac{\partial_t \omega_k}{2 \omega_k} 
& -\omega_{k} \end{array} \right] \left[ \begin{array}{c} \hat{\varphi}_{k} \\ \hat{\varphi}_{-k}^{\dagger} \end{array} \right] \,.
\label{eq:varphi_eom}
\end{equation}
Whenever the system is stationary, the operators $\hat{\varphi}_{k}$ and $\hat{\varphi}_{-k}^{\dagger}$ decouple, and oscillate at 
constant frequencies $\pm \omega_{k}$:
\begin{alignat}{2}
\hat{\varphi}_{k}(t) = \hat{b}_{k} \, e^{-i\omega_{k} t} \,, & \qquad \hat{\varphi}_{-k}^{\dagger}(t) = \hat{b}^{\dagger}_{-k} \, e^{i\omega_{k} t} \,.
\label{eq:varphi_t-dep}
\end{alignat}
The operators $\hat{b}_{k}$ and $\hat{b}_{k}^{\dagger}$ 
are, respectively, annihilation and creation operators for collective excitations (phonons) of momentum $k$ relative to the condensate. 
Note that, in the limit of small $k$, $\omega_{k} \approx c k$, i.e. 
$c$ is just the speed of low-momentum phonons. 
Moreover, recalling that the (one-dimensional) 
healing length is $\xi \equiv 1/mc$, Eq.~(\ref{eq:om_defn}) can be written in dimensionless form: 
\begin{equation}
\left( \frac{\omega_{k}}{mc^{2}} \right)^{2} = \left( k \xi \right)^{2} + \frac{1}{4} \left( k \xi \right)^{4} \,.
\label{eq:om_defn_adim}
\end{equation}
Here we see the origin of our claim at the end of Section \ref{sec:Condensate} that, from the phonon point of view, the natural frequency is $m c^{2}$. 

When $c^2 $ varies, the system (\ref{eq:varphi_eom}) becomes nontrivial 
and encodes a mode mixing between the phonon modes of momenta $k$ and $-k$, which entails amplification. To characterize this mode mixing, it is convenient to introduce the coefficients $\alpha_{k}(t)$ and $\beta_{k}(t)$ \cite{Busch-Parentani-Robertson}:
\begin{eqnarray}
\hat{\varphi}_{k}(t) & = & \left(\alpha_{k}(t) \hat{b}_{k}^{\mathrm{in}} + \beta_{k}^{\star}(t) \hat{b}_{-k}^{\mathrm{in}\dagger}\right) \mathrm{exp}\left(-i\int^{t} \omega_{k}(t^{\prime}) \, \dd t^{\prime}\right) \,, \nonumber \\  
\hat{\varphi}_{-k}^{\dagger}(t) & = & \left( \alpha_{k}^{\star}(t) \hat{b}_{-k}^{\mathrm{in}\dagger} + \beta_{k}(t) \hat{b}_{k}^{\mathrm{in}}\right) \mathrm{exp}\left(i\int^{t} \omega_{k}(t^{\prime}) \, \dd t^{\prime}\right) \,, 
\end{eqnarray}
where $\hat{b}_{k}^{\mathrm{in}}$ and $\hat{b}_{-k}^{\mathrm{in}\dagger}$ are defined such that, as $t \to -\infty$,
\begin{alignat}{2}
\hat{\varphi}_{k}(t) \sim \hat{b}_{k}^{\mathrm{in}} \, e^{-i \omega_{k} t} \,, & \qquad \hat{\varphi}_{-k}^{\dagger}(t) \sim \hat{b}_{-k}^{\mathrm{in}\dagger} \, e^{i \omega_{k} t} \,,
\end{alignat}
or equivalently, $\alpha_{k} \to 1$ and $\beta_{k} \to 0$ as $t \to -\infty$. Then the evolution of the operators $\hat{\varphi}_{k}, \hat{\varphi}_{-k}^{\dagger}$ is completely determined by the equations of motion for $\alpha_{k}$ and $\beta_{k}$, which are
\begin{eqnarray}
\partial_{t}\alpha_{k} & = &  - \frac{\partial_t \omega_k}{2 \omega_k} 
\mathrm{exp}\left(2i\int^{t}\omega_{k}(t^{\prime}) \, \dd t^{\prime}\right) \beta_{k} \,, \nonumber \\ 
\partial_{t}\beta_{k} & = & - \frac{\partial_t \omega_k}{2 \omega_k}  
\mathrm{exp}\left(-2i\int^{t}\omega_{k}(t^{\prime}) \, \dd t^{\prime}\right) \alpha_{k} \,. 
\label{eq:alpha_beta_eom}
\end{eqnarray}
Once we know how $c$ varies with time, we can use Eqs.~(\ref{eq:om_defn}) and (\ref{eq:alpha_beta_eom}) 
to get the resulting evolution 
of $\alpha_{k}$ and $\beta_{k}$, 
and hence of $\hat{\varphi}_{k}$ and $\hat{\varphi}_{-k}^{\dagger}$. 

If we assume that $c^2$ reaches another constant at late time, the initial and final mode operators 
are related by the $U(1,1)$ linear transformation 
\begin{equation}
\left[ \begin{array}{c} \hat{b}_{k}^{\rm out} \\ \hat{b}_{-k}^{\rm out \dagger}  \end{array} \right]
= \left[ \begin{array}{cc} \alpha_{k} & \beta_{k}^{\star}  \\ 
\beta_{k} & \alpha_{k}^{\star} \end{array} \right] 
\left[ \begin{array}{c} \hat{b}_{k}^{\rm in} \\ \hat{b}_{-k}^{\rm in \dagger} \end{array} \right] \,.
\label{eq:DCE}
\end{equation}
When the state is vacuum at early time, at late time the mean number of phonons with wave number $k$ is given by $n_{k} = n_{-k} = \left| \beta_{k} \right|^{2}$.  In  Sec.~\ref{sub:InSitu_DCE_Tanh} we shall present how this expression is modified when some phonons are present in the initial state, see Eq.~(\ref{eq:n_mod}).  Because phonons are produced in pairs, the transformation of Eq.~(\ref{eq:DCE}) introduces correlations between the (1-mode) phonon states with wave number $k$ and those with wave number $-k$.  
In fact, given the right conditions, it is possible that the 2-mode state $(k,-k)$ becomes nonseparable.
Here, it is appropriate to recall this notion, for we shall find in later sections that it can be inferred from suitable measurements.

\subsection{Nonseparability of bipartite systems} 

Roughly speaking, a multipartite quantum state is said to be nonseparable whenever the correlations between its constituent parts are too strong to be described by classical statistics (for more details, see~\cite{Werner,Simon}). 
Restricting our attention to a two-mode state (e.g., particles of opposite momentum produced by DCE), 
each single-mode subsystem has its corresponding quantum amplitude operators $\hat{b}_{j}$ and $\hat{b}_{j}^{\dagger}$ ($j=k,-k$), 
satisfying the standard bosonic commutation relations $\left[ \hat{b}_{j} \, , \, \hat{b}_{j^{\prime}}^{\dagger} \right] = \delta_{j,j^{\prime}}$. The density matrix $\hat \rho_{k,-k}$ of a two-mode system is said to be {\it separable} when it can be written as~\cite{Simon} 
\begin{equation}
\hat \rho_{k,-k} = \sum_a P_a \, \hat \rho_{k}^a \otimes \hat \rho_{-k}^a\ , 
\label{eq:nonsep1}
\end{equation}
where $\hat \rho_j^a$ are the density matrices of the single-mode subsystems, and where the $P_a$ are real numbers such that $0 \leq P_{a} \leq 1$. Then $\sum_{a} P_{a} = 1$, and the two-mode state has 
the properties of a probability distribution.  Such states can be considered as classically correlated in that they can be obtained by putting separately the subsystems $k$ and $-k$ in the (quantum) states described by $\hat \rho_{k}^a$ and $\hat \rho_{-k}^a$, where the state $a$ is determined using a random number generator distributed according to the probability distribution $P_a$. Conversely, {\it nonseparable} states possess correlations which cannot be accounted for by classical means. In fact, the nonseparability of the state is a necessary condition for obtaining a violation of 
Bell inequalities~\cite{Werner,Campo-Parentani-2006} based on operators acting separately on the subsystems $k$ and $-k$. 

Assuming homogeneity of the system, the only nonvanishing expectation values at quadratic order in the operators $\hat{b}_{\pm k}$ are 
\begin{alignat}{2}
n_{j} = \Big\langle \hat{b}_{j}^{\dagger} \hat{b}_{j} \Big\rangle \,, & \qquad c_k 
 = \Big\langle \hat{b}_{k} \hat{b}_{-k} \Big\rangle \,.
\label{eq:n_c_defn}
\end{alignat}
The first gives the mean occupation number of quanta in a single mode, which in our case means the number of quasi-particles of type $j$. 
The second determines the correlations between the two modes, and is generally a complex number. Importantly, these 
$c$-numbers contain enough information for assessing the nonseparability of $\hat \rho_{k,-k}$. Indeed, the condition
\begin{equation}
\left| c_k \right|^{2} - n_{k} n_{-k} > 0
\label{eq:nonsep}
\end{equation}
is sufficient for the state $\hat \rho_{k,-k}$ to be nonseparable~\cite{Campo-Parentani-2005,Adamek-Busch-Parentani,deNova-Sols-Zapata}.  In addition, when the state $\hat \rho_{k,-k}$ is Gaussian, condition (\ref{eq:nonsep}) is also necessary for nonseparability.  To appreciate the fact that nonseparable states are very peculiar, and therefore difficult to realize and to identify, one should also recall that $\left| c \right|^{2}$ is bounded from above. As a result, nonseparable states live in a relatively small domain where one has
\begin{equation}
{\rm min}\left\{ \frac{1}{n_{k}}, \, \frac{1}{n_{-k}} \right\} \geq \frac{\left| c_k \right|^{2}}{n_{k} n_{-k}} - 1 > 0 \,.  
\label{eq:nonsep2}
\end{equation}
In the limit of large occupation numbers, $n \gg 1$, one sees that $\left| c_k \right|^{2}/n_{k} n_{-k}$ can be larger than $1$ only by a term in $1/n$. Furthermore, if it is known due to reasons of symmetry that $n_{k} = n_{-k}$, 
then it is convenient to introduce the parameter $\Delta_k  \equiv n_k - \left| c_k \right|$, for which the two conditions of~(\ref{eq:nonsep2}) are 
equivalent to 
\begin{equation}
-\frac{1}{2} < n_{k} - \sqrt{n_{k}\left(n_{k}+1\right)} \leq \Delta_k < 0 \,.  
\label{Delt}
\end{equation}

The knowledge of $\Delta_{k}$ seems to involve the measurement of noncommuting operators, for neither of the number operators $n_{\pm k} = \hat{b}_{\pm k}^{\dagger} \hat{b}_{\pm k}$ commutes with the correlation operator $\hat{c}_{k} = \hat{b}_{k} \hat{b}_{-k}$, nor even the Hermitian part of $\hat{c}_{k}$ with its anti-Hermitian part.  However, while measuring the $c$-numbers $n_{k} = n_{-k}$ and $c_{k}$, see Eqs.~(\ref{eq:n_c_defn}), requires the performance of noncommuting measurements, we shall see that the nonseparability condition~(\ref{Delt}) can be tested via measurements in which noncommutativity plays no role.


\section{{\it In situ} measurements of the phonon state 
\label{sec:InSitu}}

In this Section we consider measurements made directly on the Bose gas while it is still in the trap.  The key observable is the instantaneous atom density, particularly fluctuations about its mean.  We shall see how density correlations are related to the phonon state, and how they are affected by phonon pair creation due to temporal variations of the condensate.

\subsection{Analyzing the density-density correlation function}

\subsubsection{Generalities} 

As an operator, the total (1-dimensional) number density of atoms in the gas is given by
\begin{eqnarray}
\hat{n}_{1}(t,z) & = & \int 2\pi r \, \dd r \, \hat{\Phi}^{\dagger}(t,z,r) \hat{\Phi}(t,z,r) \nonumber \\ 
& \approx & n_{1} \left( \mathds{1} + \hat{\phi}^{\dagger}(t,z) + \hat{\phi}(t,z) \right) \,.
\end{eqnarray}
To obtain the second line we have used the decomposition (\ref{eq:Bog_approx}) and the $r$-independence of $\hat{\phi}$, and neglected the nonlinear contribution from the fluctuations. Using the constancy of the background density $n_{1}$ in homogeneous systems~\footnote{The reader is directed to \cite{Carusotto-BH,Busch-Parentani-2013,deNova-Sols-Zapata-CS,deNova-Sols-Zapata,Boiron-et-al,Steinhauer-2015,Steinhauer-2016} for related considerations when working with stationary inhomogeneous systems, in particular with a black hole configuration. We hope to return to this case in a future work.}, the relative density fluctuation is
\begin{eqnarray}
\frac{\delta \hat{n}_{1}(t,z)}{n_{1}} & \equiv & \frac{\hat{n}_{1}(t,z) - n_{1}}{n_{1}} \nonumber \\ 
& = & \hat{\phi}(t,z) + \hat{\phi}^{\dagger}(t,z) \,.
\label{eq:drho}
\end{eqnarray}
Since $\hat{n}_{1}(t,z)$ commutes with $\hat{n}_{1}(t,z^{\prime})$ for all $z,z^{\prime}$, 
within the framework of quantum mechanics 
it is possible to precisely measure $\hat{n}_{1}(t,z)$ 
for all $z$ at any given time.
This can be done by taking an {\it in situ} image of the Bose gas (as in \cite{Schley-et-al}).
For each image, this serves as a single measurement of $\hat{n}_{1}(t,z)$. Such a measurement is destructive, but on performing the experiment many times and collecting an ensemble of measurements of $\hat{n}_{1}(t,z)$, 
one has access to both the expectation value $n_{1} \equiv \left\langle\hat{n}_{1}\right\rangle$ 
and the variance of $\delta\hat{n}_{1}(t,z) $. 

We define the Fourier transform of $\hat{\phi}$ by inverting Eq.~\eqref{eq:inv_Fourier}, so that:
\begin{align} \label{eq:direct_Fourier}
\hat{\phi}_k(t) = \frac{\sqrt{N}}{L} \int_0^L e^{-i k z} \hat{\phi}(t,z) \, \mathrm{d}z,
\end{align} 
for $k \in 2 \pi \mathbb{Z} / L$. 
The quantities $\delta \hat{n}_{1,k}$ and $\hat{\varphi}_k$ are defined similarly, with $\hat{\phi}$ replaced by $\delta \hat{n}_1$ 
and $\hat{\varphi}$. 
From Eqs. (\ref{eq:drho}) and (\ref{eq:varphi_phi}), we obtain
\begin{equation}
\frac{\delta \hat{n}_{1,k}(t)}{n_{1}} = \hat{\phi}_{k}(t) + \hat{\phi}^{\dagger}_{-k}(t) = \left( u_{k} + v_{k} \right) \left( \hat{\varphi}_{k}(t) + \hat{\varphi}_{-k}^{\dagger}(t) \right) \,.
\label{dnk} 
\end{equation}
Note that taking the hermitian conjugate of this operator is 
equivalent to changing the sign of $k$, a result of the fact that the relative density fluctuation of Eq. (\ref{eq:drho}) is itself a hermitian operator and measurements of it are thus real quantities.  It is straightforward to show that this operator commutes with its hermitian conjugate, and the following correlation function is thus well-defined:
\begin{eqnarray}
G_{2,k}(t) & \equiv & \frac{\left\langle \left| \delta\hat{n}_{1,k}(t) \right|^{2} \right\rangle}{n_{1}^{2}} \nonumber \\ 
& = & \left( u_{k}+v_{k} \right)^{2} \left( 2 n_{k} + 1 + 2 \, \mathrm{Re}\left[ c_{k} \, e^{-2i\omega_{k} t} \right] \right) \,, \label{eq:G2k_n_c} 
\label{eq:drho-drho-corr}
\end{eqnarray}
where in the last line we have substituted the expressions in Eqs. (\ref{eq:varphi_t-dep}), along with 
the definitions of $n_k$ and $c_k$ in Eqs. (\ref{eq:n_c_defn}) and the fact that isotropy~\footnote{Homogeneity in the mean also implies $\langle \hat{b}_{\pm k} \rangle = 0$. If this were not true, the expectation value of $\langle \hat{b}_{\pm k} \rangle$ would break homogeneity. This more general case can, however, be treated similarly by considering only the connected part of the correlation function, i.e. by separating the coherent part from the fluctuations, see Appendix C of~\cite{Macher-Parentani-BEC}. When dealing with the connected part of the correlation function, the criterion based on the sign of $\Delta_k$ 
is equivalent to the generalized criterion used in \cite{Carusotto-BH}, see also Table I of~\cite{deNova-Sols-Zapata}. \label{fn:connected}}
implies $n_{k} = n_{-k}$.
The last expression, {\it stricto sensu}, applies 
only when the background has reached a 
stationary state, 
so that phonon excitations can be analyzed in terms of 
stationary modes $e^{\pm i\omega_{k} t}$. (A generalized version of this equation shall be studied in the next subsection). 
The mean occupation number $n_{k}$ determines the time-averaged mean of $G_{2,k}(t)$, while the magnitude and phase of the correlation $c_{k}$ respectively determine the amplitude and phase of the oscillations of $G_{2,k}(t)$ around its mean value.
Note also the additional constant term `$+1$'. We shall see below that this term is a measurable quantity which encodes the vacuum fluctuations of the phonon field. 

For homogeneous systems with a stationary background, one is thus able, 
irrespective of the complexity of the actual state $\hat \rho_{k,-k}$, 
to extract both $n_{k}$ and $c_{k}$ from $G_{2,k}(t)$ of Eq.~(\ref{eq:G2k_n_c})
through repeated measurements at a series of different times. 
Since $G_{2,k}(t)$ encodes three real numbers, 
at least three such times are necessary to be able to extract $n_{k}$ and $c_{k}$. 
Three times may be sufficient if the measurements are accurate and the times well-chosen (e.g. they are not separated by an integral number of oscillation periods). 
It should also be pointed out that measurements of $G_{2,k}$ 
which allow the extraction of 
$n_{k}$ and $c_{k}$ 
do not commute with each other, in conformity with the above mentioned fact that $\hat b_k \hat b_{-k}$ does not commute with $\hat b^\dagger_k \hat b_{k}$. 
Indeed, such measurements must be performed at different times, and for a stationary background it is straightforward to show that 
\begin{equation}
\left[ \frac{\delta\hat{n}_{1,k}(t)}{n_{1}} \,, \frac{\delta\hat{n}_{1,-k}(t^{\prime})}{n_{1}} \right] = -2i \left(u_{k}+v_{k}\right)^{2} \, \mathrm{sin}\left(\omega_{k} (t-t^{\prime})\right) \,.
\label{eq:dn1_comm}
\end{equation}

\subsubsection{Measuring thermal and vacuum fluctuations} 

\begin{figure}
\includegraphics[width=0.45\columnwidth]{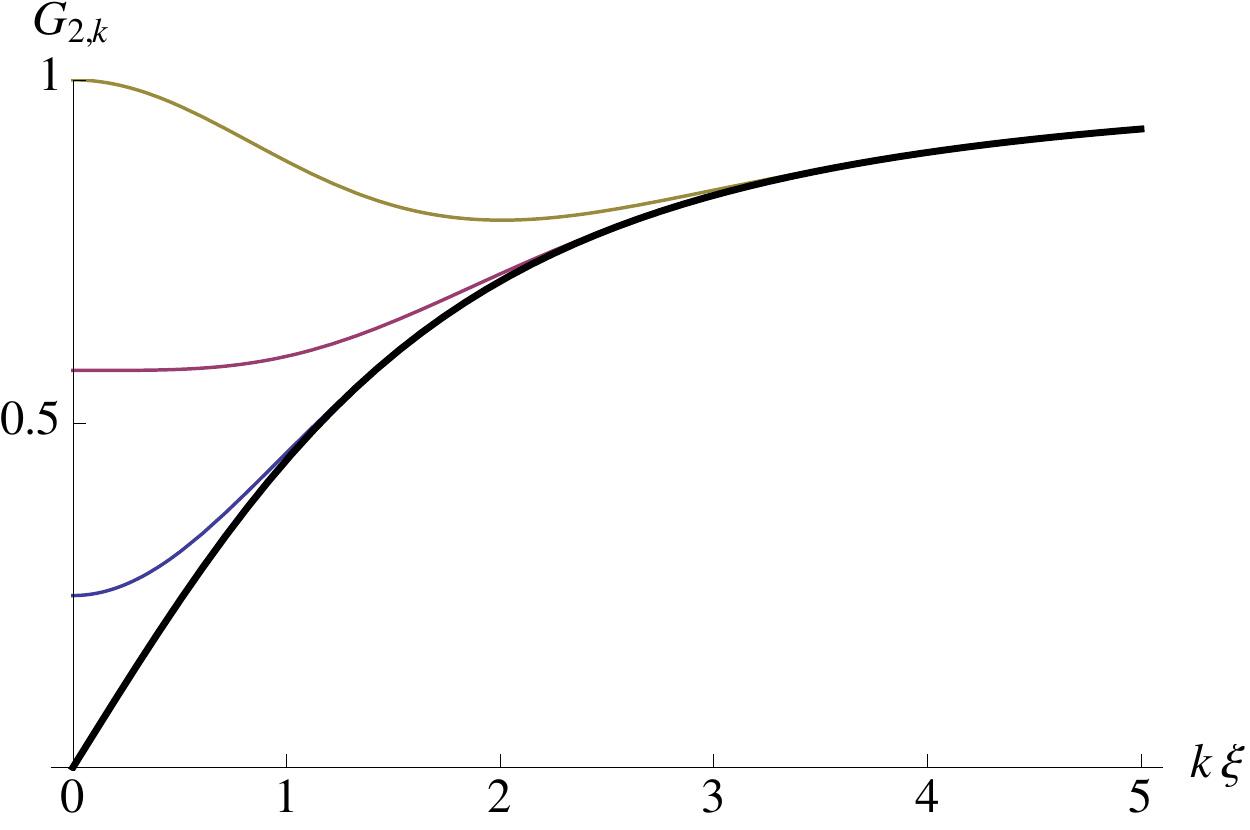}
\caption{The density-density correlation function $G_{2,k}$ in a thermal BEC, before any stimulation of phonons.
This function is time-independent and, when considered as a function of
$k \xi$, depends only on $T/mc^{2}$. As explained in the text, it does not depend on the thickness of the condensate. 
Here, the various curves correspond to four different values of the temperature: $T/mc^{2} = 0$ (thick black), $0.25$ (blue), $1/\sqrt{3}$ (purple) and $1$ (yellow).  The purple curve marks the transition between having a minimum at $k=0$ for $T/mc^{2} < 1/\sqrt{3}$ and a local maximum for $T/mc^{2} > 1/\sqrt{3}$. We note that all curves approach $T/mc^{2}$ in the limit $k \xi \to 0$, and they approach $\left(u_{k} + v_{k}\right)^{2}$ when $\omega_{k}/T \gg 1$.
\label{fig:InSitu_NoDCE}}
\end{figure}

To illustrate these concepts, let us consider the simplest example: a stationary system in a thermal state, with temperature $T$. 
This is characterized by no correlations ($c_{k} = 0$) and by the expectation number
\begin{equation}
2 n_{k} + 1 = \mathrm{coth}\left(\frac{\omega_{k}}{2T}\right) \,.
\end{equation}
Then we have simply
\begin{equation}
G_{2,k} = \left(u_{k} + v_{k}\right)^{2} \mathrm{coth}\left(\frac{\omega_{k}}{2T}\right) \,.
\label{eq:G2k_hom_stat1}
\end{equation}
Substituting the expressions (\ref{eq:u-v_defn}) and (\ref{eq:om_defn}) and rearranging slightly, this is
\begin{equation}
G_{2,k} = \frac{k\xi/2}{\sqrt{1+\left(k\xi/2\right)^{2}}} \, \mathrm{coth}\left( \frac{k\xi \, \sqrt{1+\left(k\xi/2\right)^{2}}}{2T/mc^{2}} \right) \,.
\label{eq:G2k_hom_stat2}
\end{equation}
This is a one-parameter family of functions of the adimensionalized momentum $k\xi = k/mc$,
where the parameter is the adimensionalized temperature $T/mc^{2}$. Note that $n_{1} a_{s}$ is included in the definitions of $\xi$ and $c^{2}$, so that, 
to the extent that the Gaussian approximation is valid, 
$G_{2,k}$ is independent of the thickness of the condensate when expressed in terms of $k\xi$ and $T/mc^{2}$. Examples of this function for various temperatures are shown in Figure \ref{fig:InSitu_NoDCE}. 

It is of interest to take the low- and high-momentum limits of Eq. (\ref{eq:G2k_hom_stat2}).  For $\omega_{k}/T \ll 1$, we have $\mathrm{coth}\left(\omega_{k}/2T\right) \approx 2T/\omega_{k} + \omega_{k}/6T$, so that Eq. (\ref{eq:G2k_hom_stat2}) becomes
\begin{equation}
G_{2,k} \rightarrow \frac{T}{mc^{2}} + \frac{mc^{2}}{4T} \left( \frac{1}{3} - \left(\frac{T}{mc^{2}}\right)^{2} \right) \left(k\xi\right)^{2} \,.
\end{equation}
In particular, in the limit $k \to 0$, $G_{2,k}$ simply approaches the adimensionalized temperature $T/mc^{2}$.  We can thus determine the temperature of the gas by examining the low-momentum density fluctuations. 
$k = 0$ is a local extremum whose nature is determined by the sign of the coefficient of the $k^{2}$ term.

On the other hand, for $\omega_{k}/T \gg 1$, we have $\mathrm{coth}\left(\omega_{k}/2T\right) \approx 1$, so that Eq. (\ref{eq:G2k_hom_stat2}) becomes
\begin{equation}
G_{2,k} \approx 
\left(u_{k}+v_{k}\right)^{2} = \frac{k\xi/2}{\sqrt{1+\left(k\xi/2\right)^{2}}} \,.
\end{equation}
In this zero temperature limit, $G_{2,k}$ characterizes the mean amplitude of vacuum fluctuations.
Note that this measurable quantity originates in the constant term in the last line of Eq. (\ref{eq:G2k_n_c}), and thus owes its existence to the non-commutativity of the quantum amplitudes $\hat{b}_{k}$ and 
$\hat{b}_{k}^{\dagger}$. 
Note also that taking the limits 
$T \to 0$ then 
$k \to 0$, $G_{2,k}$ goes to $T/mc^{2} = 0$ at $k=0$ while the derivative $\partial_k G_{2,k}$ remains finite.

\subsubsection{Assessing nonseparability} 

Using measurements of $G_{2,k}(t)$, we are also able to assess whether or not the state $(k,-k)$ is nonseparable using condition (\ref{Delt}). 
Moreover, it is not necessary to extract both $n_{k}$ and $c_{k}$ in order to assess nonseparability; from Eq. (\ref{eq:G2k_n_c}), it is clear that it is sufficient to find that, for some value of $t$, $G_{2,k}(t)$ satisfies the inequality
\begin{equation}
G_{2,k}(t) < \left(u_{k} + v_{k}\right)^{2} \,.
\label{nonseptr}
\end{equation}
This is one of our key observations: whatever 
the state $\hat \rho_{k,-k}$, 
nonseparability can be ascertained by examining the behavior of a well-defined observable quantity, 
without any need to perform measurements of non-commuting variables. (Preliminary versions of this assertion were used in~\cite{Busch-Parentani-2013,Busch-Parentani-Robertson,Busch-Carusotto-Parentani} but without emphasis on the absence of measurements of non-commuting variables.)~\footnote{It has been noticed in~\cite{Finke-Jain-Weinfurtner} that the measurements proposed in~\cite{Busch-Parentani-2013,Busch-Parentani-Robertson,Steinhauer-2015} are commuting. The authors then raise the question of whether these measurements could be accounted for by an alternative model based on commuting dynamical variables. 
In answer to this question,
it should be noticed that the contribution 
of vacuum fluctuations $G_{2,k} = \left(u_{k}+v_{k}\right)^{2}$ 
cannot be accounted for by their model. 
In fact, 
precise measurements of $G_{2,k}$ (for sufficiently high $k$ so that $n_k \ll 1$) 
in a homogeneous incoherent ($c_k = 0$) state prior to DCE should be able to rule it out. 
As a result, 
we conclude that 
this model cannot be used to address the question of the nonseparability of the state,
and therefore does not invalidate our claim that 
 commuting measurements can be sufficient 
to assess nonseparability.
}

To conclude this discussion, it is perhaps interesting to consider the inequality~(\ref{nonseptr}) 
from a slightly different point of view. One might ask in abstract terms why repeated measurements of $\delta\hat{n}_{1,k}(t)$ of \eq{dnk} 
are sufficient to assess nonseparability. The reason is that measuring the density fluctuation $\delta\hat{n}_{1,k}(t) \propto \hat b_k e^{- i\omega_k t} + \hat b_{-k}^\dagger e^{i\omega_k t} $ amounts in effect to using an interferometer,  
in that what is recorded in these measurements is a superposition of two channels (here the destruction of a phonon of momentum $k$, and the creation of a phonon of momentum $-k$). When changing $t$ it amounts to varying the relative phase between them. Then, if $t$ is such that (\ref{nonseptr}) 
is satisfied, the interferences are so destructive that they reveal the existence of the nonseparability of the state. The analogy with the measurements of Ref.~\cite{Lopes-et-al} is striking.

\subsection{DCE due to a one-time change in $c^2$
\label{sub:InSitu_DCE_Tanh}}

Now suppose, having started in the incoherent thermal state just described, that there is a change in the system such that $c^{2}$ varies from one constant value to another, and that this final value of $c^{2}$ is constant for $t \to \infty$. 
According to Fig. \ref{fig:tanh_response}, such a variation of $c^{2}$ will occur iff the rate of change is small compared with $\omega_{\perp}$ throughout the evolution. This is quite feasible in the 1D regime, where $\omega_{\perp}/mc^{2}$ is typically large; there thus exist variations which are slow from the point of view of the condensate, but fast from the point of view of the phonons. 

\begin{figure}
\includegraphics[width=0.45\columnwidth]{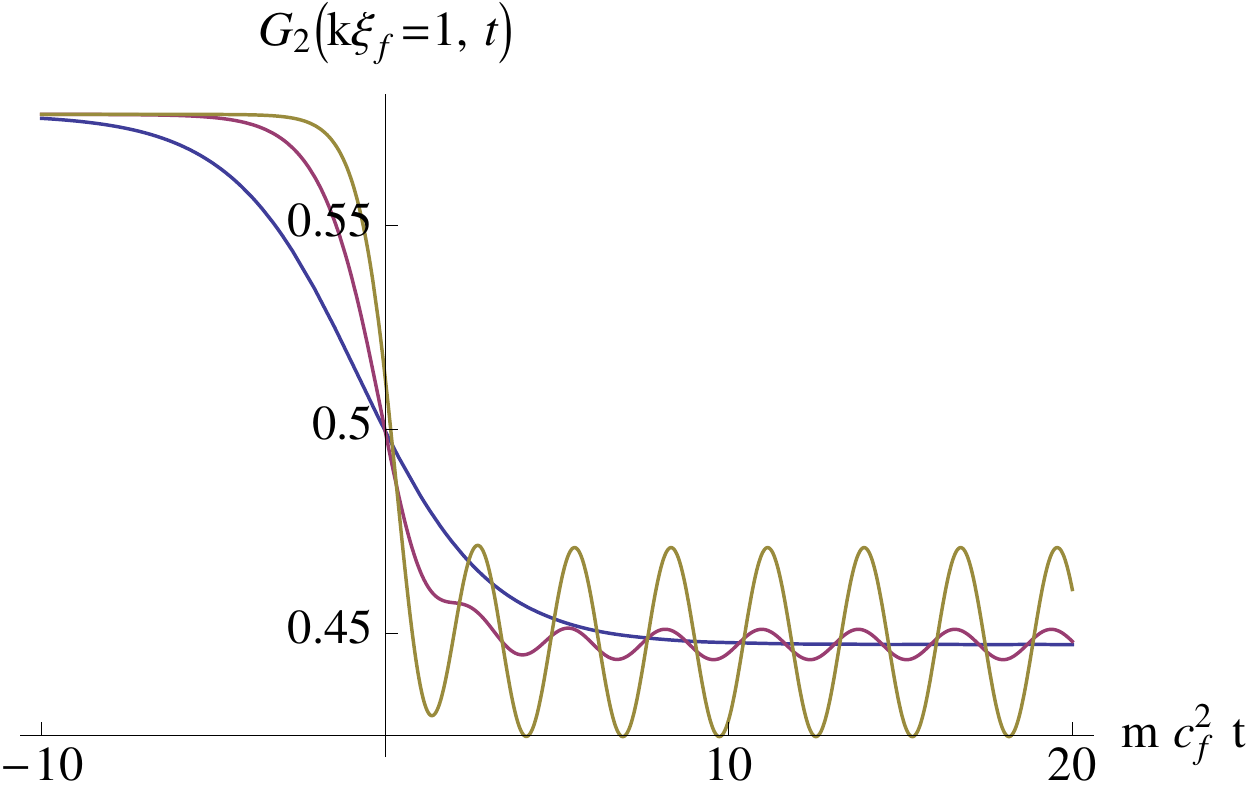} \, \includegraphics[width=0.45\columnwidth]{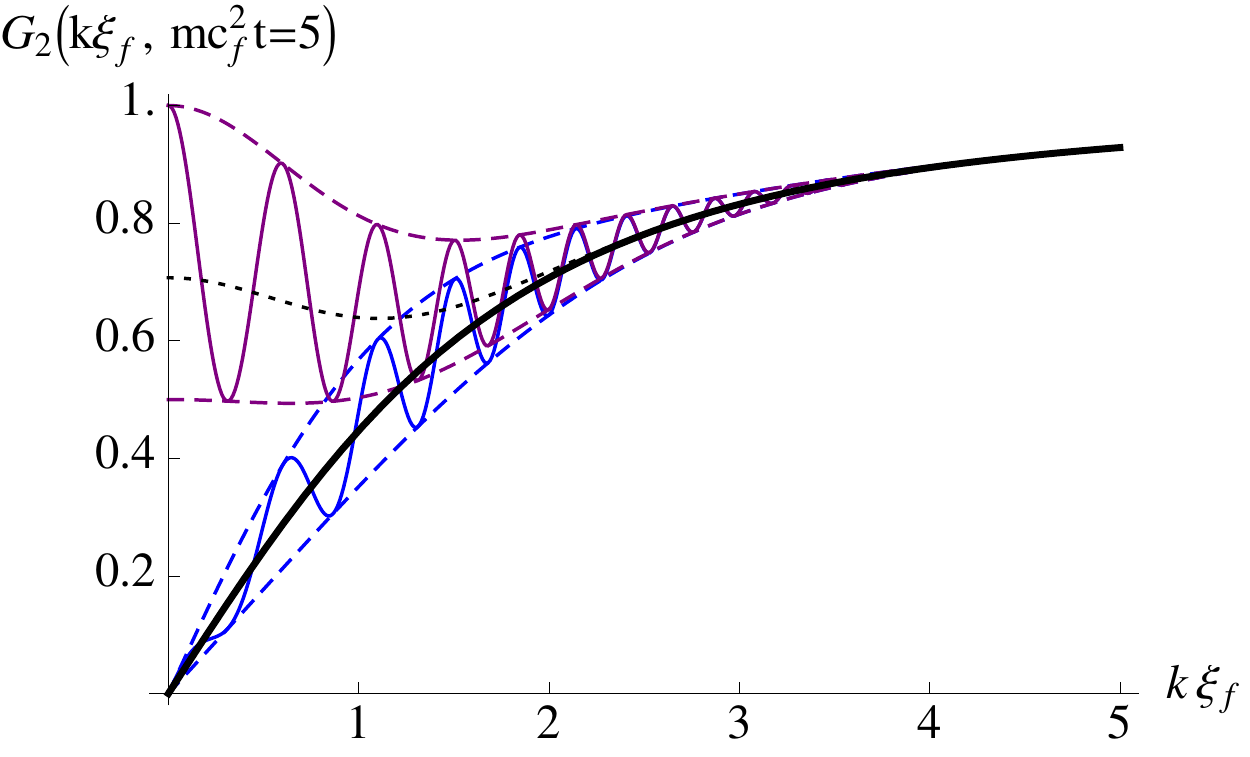}
\caption{The response of the {\it in situ} 
correlation function of Eq.~(\ref{eq:G2_alpha_beta}) to a smooth variation in $c^{2}$ described by 
\eqref{cdep}. On the left are shown three curves showing the evolution of $G_{2}$ with time, for a fixed wave vector $k \, \xi_{f} = 1$, for a relative change $c_{i}^{2}/c_{f}^{2} = 1/2$, and at zero temperature. 
The three curves differ only in the rate of change,
namely $a/\omega_{k i} = 0.3$ (blue), $0.6$ (purple) and $1.0$ (yellow). 
We see that, just as with the response of the condensate width in Fig. \ref{fig:tanh_response}, the final $G_{2}$ can either settle down on a single value or oscillate around that value, depending here on the ratio $a/\omega_{k i} $.
On the right, we have fixed $m c_{f}^{2} t = 5$ and allowed the wave vector $k$ to vary.  The thick black curve 
gives the zero-temperature adiabatic solution, which is equal to the nonseparability threshold, see condition~(\ref{nonseptr}). 
The solid blue and purple curves correspond to the rate $a = \omega_{k}(k \xi_{f} = 3)$, with $T_{i}/m c_{i}^{2} = 0$ (blue) and $T_{i}/m c_{i}^{2} = 1$ (purple). 
The dotted curve is the adiabatic solution at $T_{i}/m c_{i}^{2} =1$. 
To facilitate the reading, we have used dashed curves to show the maximum and minimum values reached as $G_{2}$ oscillates in time.  
For the purple curve, the high initial temperature has destroyed the nonseparability for long wavelengths, but it is preserved for $k \xi_{f} \gtrsim 1.5$.  In both cases, the nonseparability is much less visible for $k \xi_{f} > 3$ as the evolution there becomes adiabatic, so the number of excited phonon pairs drops rapidly to zero.
\label{fig:InSitu_G2t_Tanh}}
\end{figure}

The phonon state is probed 
by the operators $\hat{\varphi}_{\pm k}(t)$, whose equation of motion is 
Eq. (\ref{eq:varphi_eom}). 
This leads to nontrivial 
values of the Bogoliubov coefficients $\alpha_{k}(t)$ and $\beta_{k}(t)$, and
the correlation function of 
the first line of 
Eq. (\ref{eq:G2k_n_c}) 
becomes
\begin{equation}
G_{2,k}(t) = \left(u_{k}(t) + v_{k}(t)\right)^{2} \left( \left| \alpha_{k}(t) \right|^{2} + \left| \beta_{k}(t) \right|^{2} + 2 \, \mathrm{Re}\left\{ \alpha_{k}(t) \beta_{k}^{\star}(t) e^{-2i\int^{t} \omega_{k}(t^{\prime}) dt^{\prime}} \right\} \right) \left( 2 n_{k}^{\rm in} + 1\right) \,.
\label{eq:G2_alpha_beta}
\end{equation}
To parameterize the adiabatic evolution of the background, we consider a hyperbolic tangent dependence of $c^{2}$:
\begin{equation}
\frac{c^{2}(t)}{c^{2}_{f}} = \frac{1}{2} \left( 1 + \frac{c^{2}_{i}}{c^{2}_{f}} \right) + \frac{1}{2} \left( 1 - \frac{c^{2}_{i}}{c^{2}_{f}} \right) \mathrm{tanh}\left(a t\right) \,.
\label{cdep}\end{equation}
Note that we normalize $c^{2}$ with respect to its {\it final} value.  Extending this to all quantities has the advantage that the final profile $\left(u_{k}+v_{k}\right)^{2}$ becomes a universal function of $k \xi_{f}$ (where $\xi_{f}$ is the final value of the healing length), allowing a straightforward assessment of the nonseparability of the final state.  In Figure \ref{fig:InSitu_G2t_Tanh} are shown some examples of the variation of $G_{2,k}(t)$ in time at fixed $k$, and with $k$ at fixed time.  When considering the variation in time, we note the similarity with the response of the condensate density in Fig. \ref{fig:tanh_response}:
\begin{itemize}
\item when $a/\omega_k \ll 1$, the phonons of wave vector $k$ are not excited above their initial occupation number;
\item when $a/\omega_k \gtrsim 1$, 
correlated excitations are produced in this 
2-mode sector. 
\end{itemize}
Indeed, comparing Eq. (\ref{eq:G2_alpha_beta}) to Eq. (\ref{eq:G2k_n_c}), we can write 
the final values of $n_{k}$ and $c_{k}$ in terms of $\alpha_{k}$ and $\beta_{k}$ (see also \cite{Busch-Parentani-Robertson}):
\begin{subequations}\begin{alignat}{2}
2 n_{k} + 1 &= \left( \left| \alpha_{k} \right|^{2} + \left| \beta_{k} \right|^{2} \right) \left( 2 n_{k}^{\rm in} + 1 \right) \,, \label{eq:n_mod} \\
c_{k} & = \alpha_{k} \beta_{k}^{\star} \left( 2 n_{k}^{\rm in} + 1 \right) \,.
\end{alignat}\label{eq:n_and_c}\end{subequations}
The adiabatic variation (from the point of view of the phonons) corresponds to the case where $\alpha_{k}$ and $\beta_{k}$ remain $1$ and $0$, respectively, throughout the evolution; thus the phonon content remains as it was before the change, and $G_{2,k}$ varies in time only because $(u_{k}+v_{k})^{2}$ does so.  A non-zero $\beta_{k}$ encodes the level of non-adiabaticity, inducing correlated phonons in the condensate; furthermore, this phonon production is enhanced by the presence of phonons to start with, since both quantities in Eqs. (\ref{eq:n_and_c}) are proportional to $2 n_{k}^{\rm in} + 1$.
As these correlated phonons pass through each other and go in and out of phase, their contribution to the density profile varies, producing the oscillations seen in both panels of Fig. \ref{fig:InSitu_G2t_Tanh}.~\footnote{The oscillations on the right panel, evaluated at fixed time as a function of $k$, are analogous to the Sakharov oscillations~\cite{Mukhanov-PhysRep,Campo-Parentani-2004,Hung-Gurarie-Chin} seen in the anisotropies of the cosmic microwave background. 
The latter are observed at a given time (i.e., on the last scattering surface) and also appear as a function of the wave number.  
}


\begin{figure}
\includegraphics[width=0.45\columnwidth]{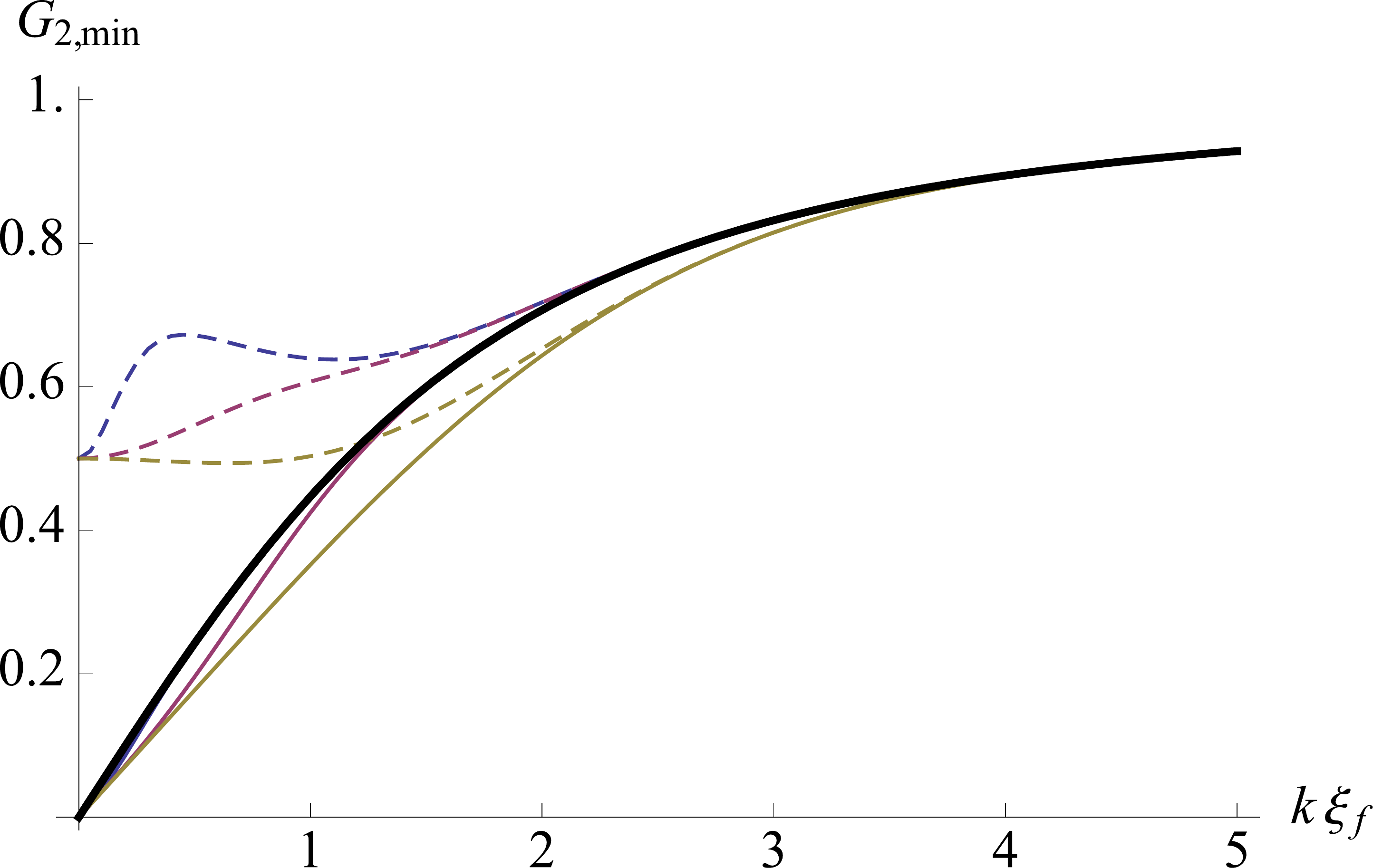} \, \includegraphics[width=0.45\columnwidth]{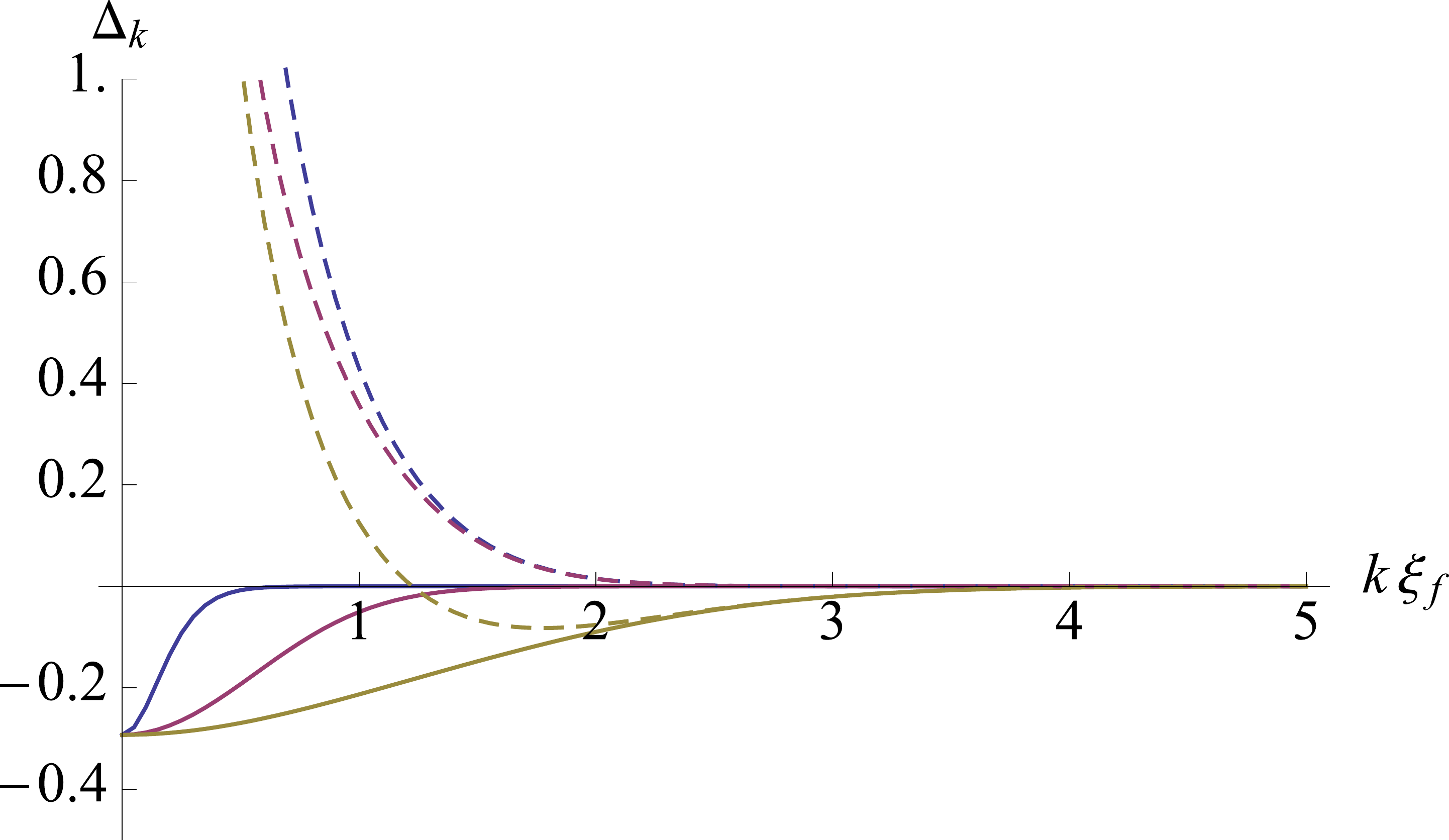}
\caption{Nonseparability of phonon pairs due to a smooth change in $c^{2}$.  In the left panel, we plot the minimum value of $G_{2}$ after the change, i.e. the lower envelope plotted on the right of Fig. \ref{fig:InSitu_G2t_Tanh}. The thick black curve shows again the final profile of $\left(u_{k} + v_{k}\right)^{2}$. 
The change in $c^{2}$ is of the same sort considered in Fig. \ref{fig:InSitu_G2t_Tanh}, except that here we use three different values of the rate $a$, which is set equal to the initial phonon frequency of three different wave vectors: $k \xi_{f} = 0.3$ (blue), $1$ (purple) and $3$ (yellow).  The solid and dashed curves correspond to initial temperatures of $T_{i}/m c_{i}^{2} = 0$ and $1$, respectively.  
On the right panel are shown the corresponding plots of $\Delta_{k} \equiv n_{k} - \left| c_{k} \right|$, which (the state being Gaussian) 
is less than zero iff the two-mode phonon state $(k,-k)$ is nonseparable, see condition~(\ref{Delt}). 
From both plots, we see that a higher rate of change tends to increase the degree of nonseparability, while a higher initial temperature decreases and, as for the blue and purple curves, can even destroy it. 
\label{fig:InSitu_Tanh_nonsep}}
\end{figure}

In the right panel of Fig. \ref{fig:InSitu_G2t_Tanh}, one clearly sees that the lower envelopes of $G_{2, k}$ are below $\left(u_{k}+v_{k}\right)^{2}$ for a certain range of $k$ which depends on the initial temperature. When this is the case, condition~(\ref{nonseptr}) 
implies that the phonon bipartite states are nonseparable. 
In Figure \ref{fig:InSitu_Tanh_nonsep} we look more directly at this nonseparability. 
To this end, in the left panel we show again the minimum value reached by $G_{2,k}$.
It is clear from this plot that a higher rate of change of $c^{2}$ tends to increase the degree of nonseparability, while a higher initial temperature tends to reduce it.  On the right panel is shown the corresponding behavior of $\Delta_{k} \equiv n_{k} - \left| c_{k} \right|$. 

\subsection{DCE due to a modulation of $c^{2}$
\label{sub:DCE_modulated}}

Having considered a transient evolution where the final value of $c^{2}$ is steady for $t \to \infty$, as in the adiabatic variation of the condensate, let us now turn our attention to the outcome of a {\it non}-adiabatic variation of the condensate -- in particular, a situation in which $c^{2}$ oscillates in time, like the final state of the purple and yellow curves in Fig. \ref{fig:tanh_response}.  This case is very similar to the one studied in \cite{Busch-Parentani-Robertson}; here, we simply review the main results and apply them to the first experiment described in~\cite{Jaskula-et-al}.

An oscillation of $c^{2}$ leads to an oscillation of $\omega_{k}$ for any given $k$.  The frequency of that oscillation is the same for all $k$, and sets up a resonance in that mode whose time-averaged frequency $\bar{\omega}_{k}$ is equal to half the oscillation frequency\footnote{This can be viewed as a consequence of energy and momentum conservation: conserving momentum requires phonons to be produced in pairs $\left(k,-k\right)$, while energy conservation leads to a preference for the sum of the phonon frequencies, $2 \bar{\omega}_{k}$, to be equal to the oscillation frequency.}.  The particle content of the resonant mode grows exponentially in time.  Furthermore, the resonance has a finite width in $k$-space that depends on the relative amplitude, $\delta \omega_{k} / \bar{\omega}_{k}$, of the frequency oscillations (which, for dispersive systems as here, will depend both on the relative amplitude of the oscillations in $c^{2}$ and on the wave vector $k$).  More precisely, we have
\begin{equation}
\left.\frac{ \delta k}{k}\right|_{\rm res} = \left| \frac{v_{p}}{v_{g}} \right| \left.\frac{ \delta \omega_{k}}{\omega_{k}} \right|_{\rm res} = \frac{1}{2} \left| \frac{v_{p}}{v_{g}} \right| \left.\frac{\delta \omega_{k}}{\omega_{k}}\right|_{\rm time} \,,
\end{equation}
where  $v_{p}/v_{g}$ gives the ratio of the phase and group velocities. 
The subscript `res' indicates the width of the resonance in Fourier space, and the subscript `time' indicates the 
width of the variation of $\omega_{k}$ in time.  Frequencies lying outside this narrow range 
are also excited, but their particle content oscillates in time, with those further from the resonance oscillating faster.  An important point to note is that, since the modulation of $c^{2}$ is caused by the natural oscillation of the condensate, the oscillation frequency is simply twice the final value of $\omega_{\perp}$, as explained in Sec.~\ref{sub:response}. 
As a result, the resonant phonon mode is 
that for which $\bar{\omega}_{k} = \omega_{\perp f}$.  
The amplitude of the oscillation will depend on how quickly $\omega_{\perp}$ is varied, as we have already seen in Fig. \ref{fig:tanh_response}.

\begin{figure}
\includegraphics[width=0.45\columnwidth]{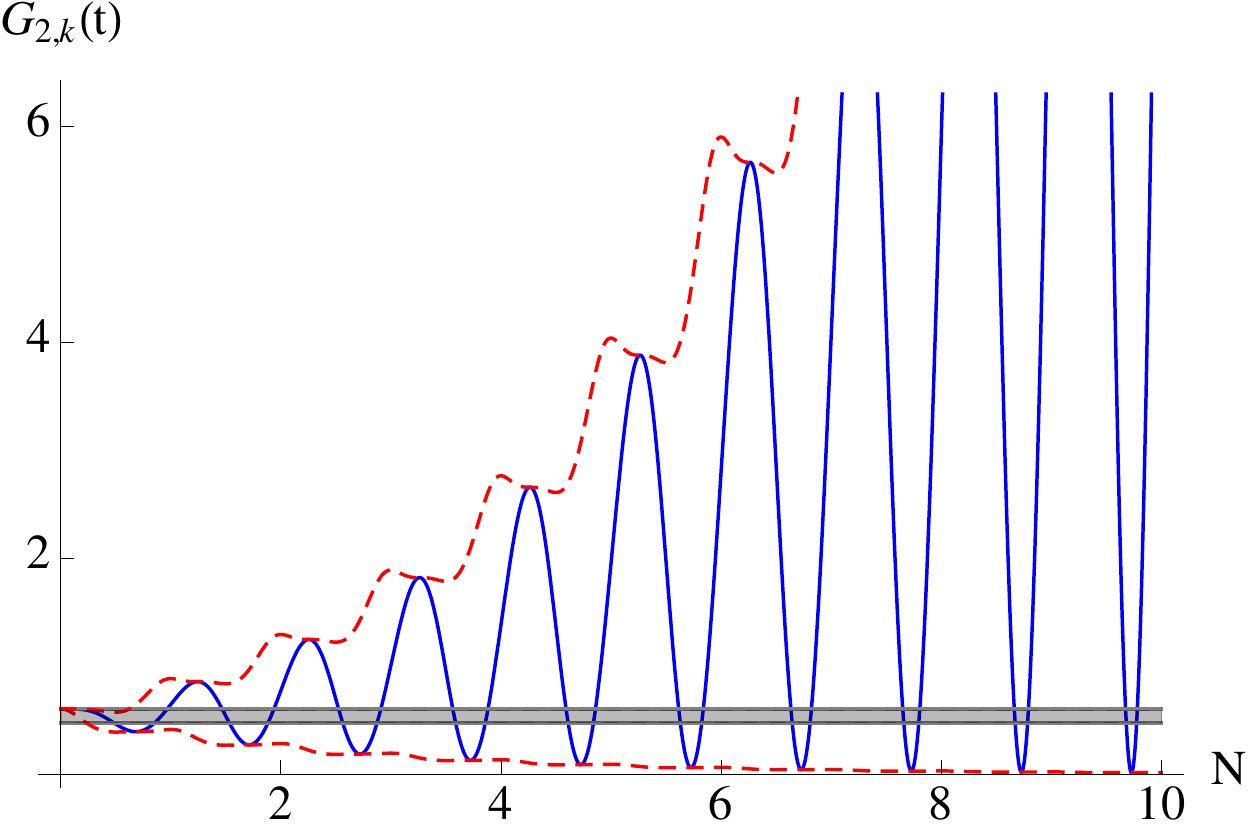} \, \includegraphics[width=0.45\columnwidth]{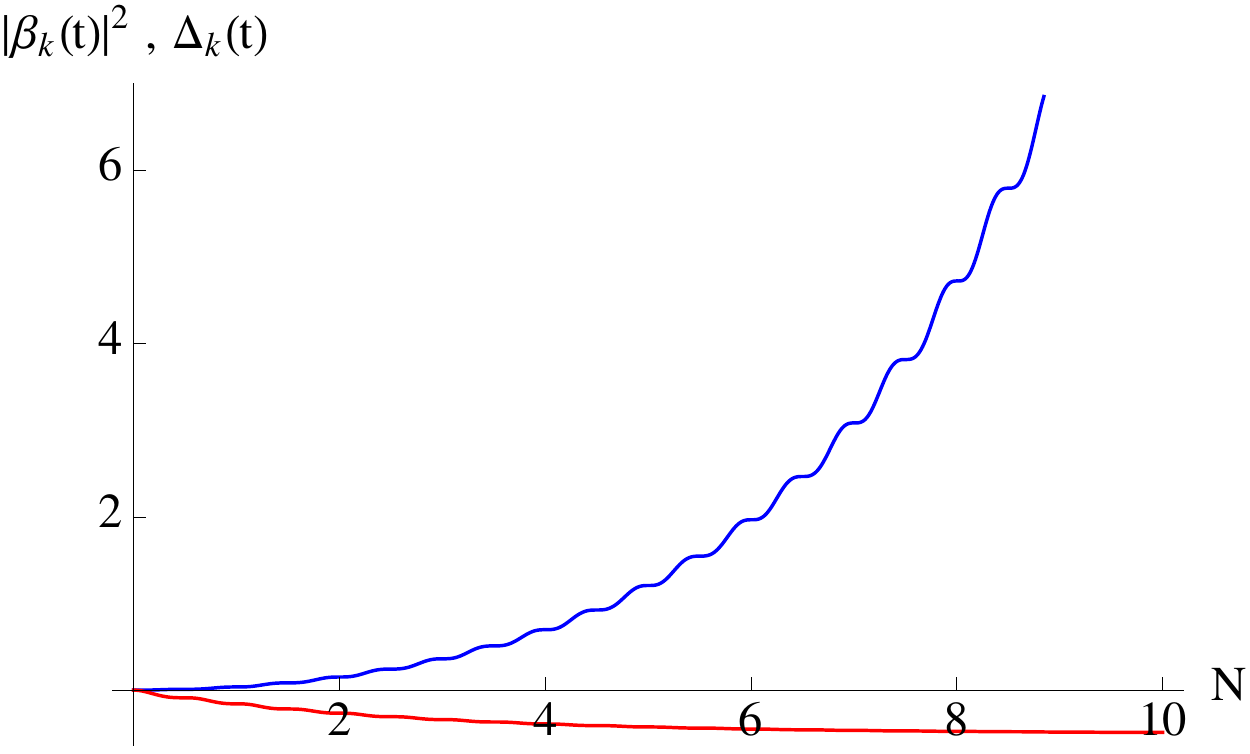}
\caption{{\it In situ} observables for the resonant mode in an oscillating condensate as functions of the number $N$ of oscillations. 
Here, we take experimental parameters similar to those used in \cite{Jaskula-et-al}: $\omega_{\perp f}/\omega_{\perp i} = \sqrt{2}$, $\omega_{\perp \, i}/mc^{2} = 1.5$, $\delta \left(c^{2}\right) / c^{2} = 0.5$, but we work with a vanishing initial temperature. 
On the left, in blue, is shown $G_{2,k}(t)$, while its upper and lower envelopes are shown in dashed red.  The thick gray 
line shows $\left(u_{k}+v_{k}\right)^{2}$, which is smeared due to the fact that it oscillates in response to the oscillations in $c^{2}$.  On the right are shown $\left|\beta_{k}(t)\right|^{2}$ (in blue) and, for $T=0$, $\Delta_{k}(t)$ (in red).  Note, in the left plot, that the fraction of an oscillation for which $G_{2,k}(t) < \left(u_{k}+v_{k}\right)^{2}$ gets smaller with increasing $n_{k}$, since this causes the amplitude to increase but does not affect the minimum, which saturates at zero.  When measuring $G_{2,k}(t)$, then, the nonseparability of the state becomes less visible with time.
\label{fig:InSitu_exp}}
\end{figure}

As an example, in Figure \ref{fig:InSitu_exp} are shown the evolutions 
of various {\it in situ} observables for the resonant mode, for a sudden change in which $\omega_{\perp f}/\omega_{\perp i} = \sqrt{2}$ (corresponding to the first experiment of \cite{Jaskula-et-al}).  On the left is shown $G_{2,k}(t)$, along with its upper and lower envelopes and the nonseparability threshold $\left(u_{k}+v_{k}\right)^{2}$. Note that the latter quantity is smeared because of the fact that $\left(u_{k}+v_{k}\right)^{2}$ oscillates in time in response to the oscillations in $c^{2}$. On the right are shown $\left|\beta_{k}(t)\right|^{2}$ and, for zero temperature, $\Delta_{k}(t)$ of \eq{Delt}. It is quite clear that the amplitude of the oscillations in $G_{2,k}(t)$ increase exponentially in time, as expected for a resonant mode, 
while the minimum of $G_{2,k}(t)$ saturates at zero (since $G_{2,k}(t) \geq 0$). These two observations lead to a third: while the degree of nonseparability increases in time in the sense that $\Delta_{k}$ becomes more negative, 
the visibility of the nonseparability in $G_{2,k}(t)$ actually {\it decreases} 
as $n_{k}$ becomes large, for 
the fraction of an oscillation for which $G_{2,k}(t) < \left(u_{k}+v_{k}\right)^{2}$ becomes smaller and smaller. 
For the sake of obtaining a visibly nonseparable state, then, it is best to stop the growth when $\left|\beta_{k}\right|^{2} \lesssim 1$. Using the values of the experiment of \cite{Jaskula-et-al}, this means that the optimal number of oscillations should satisfy $N\lesssim 5 $.
\begin{figure}
\includegraphics[width=0.45\columnwidth]{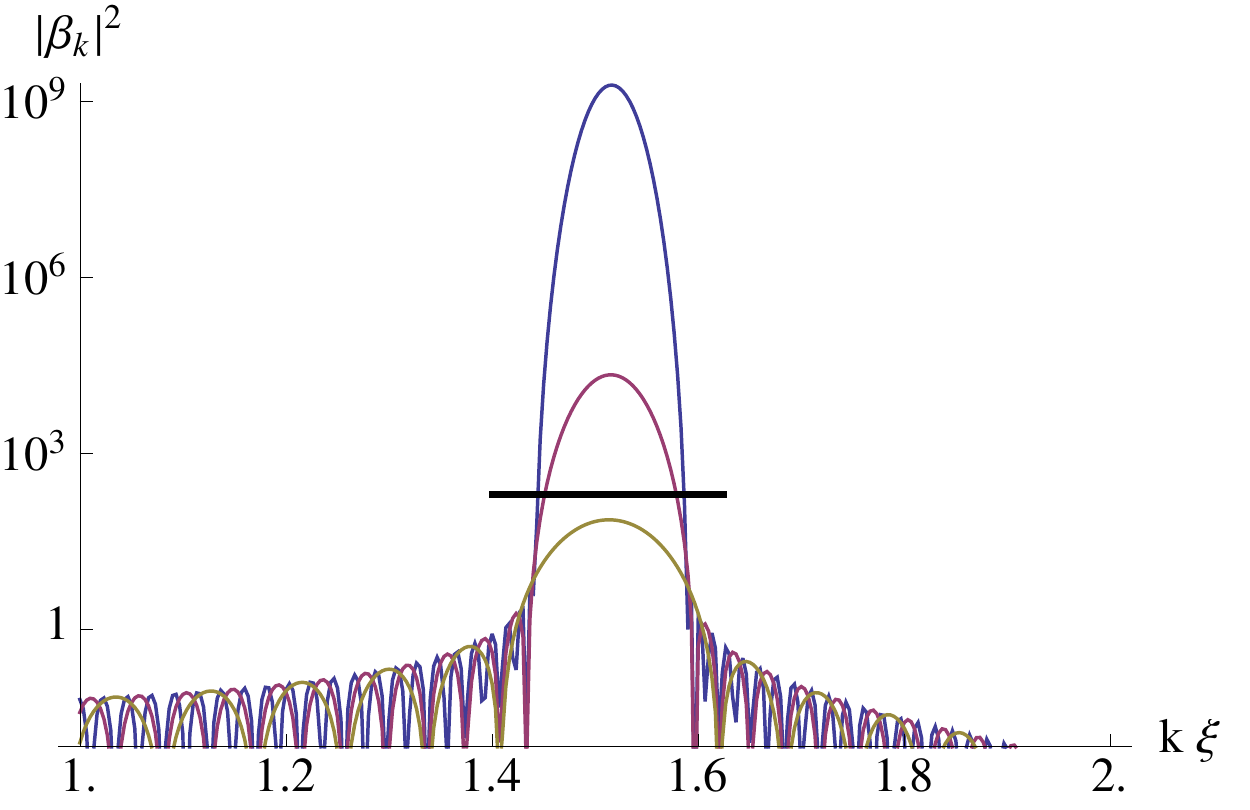} \, \includegraphics[width=0.45\columnwidth]{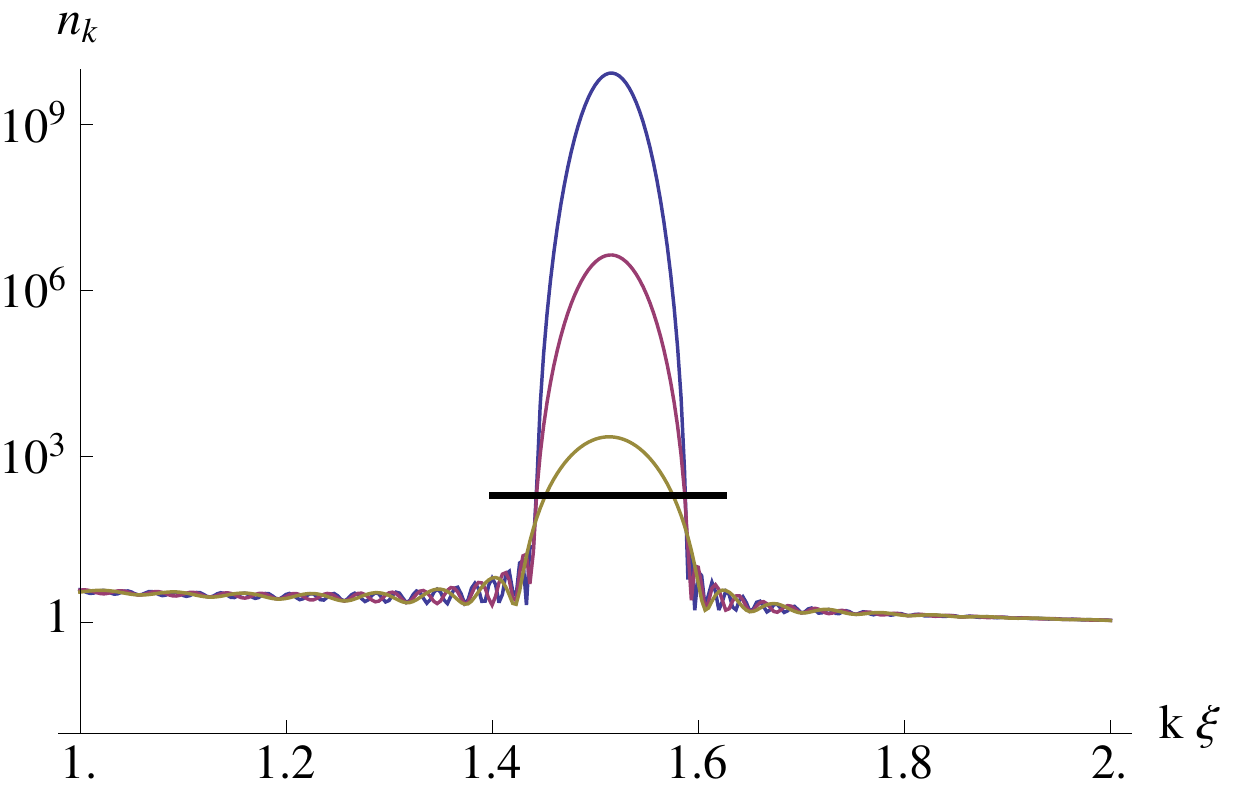}
\caption{Mean occupation number near the resonant mode due to natural oscillations of the condensate, as a function of $k \xi$ where $\xi$ is here 
the upper limit of the healing length during the modulation. 
We use the same parameters as in Fig. \ref{fig:InSitu_exp}. 
The left plot shows $\left|\beta_{k}\right|^{2}$, or equivalently, $n_{k}$ at zero temperature; the right plot shows the corresponding $n_{k}$ at temperature $T/\omega_{\perp} = 2.8$, as in \cite{Jaskula-et-al}.  The blue curves correspond to the number of oscillations $N = 60$, in accordance with \cite{Jaskula-et-al}, while the red and yellow curves correspond to $N = 30$ and $N = 15$, respectively.  The horizontal black line 
corresponds to a theoretical estimate for the maximum possible $\left|\beta_{k}\right|^{2}$ before the resonant mode begins to interact nonlinearly with other modes. 
\label{fig:InSitu_Jaskula}}
\end{figure}
  
In the actual experiment of \cite{Jaskula-et-al}, however, the trapping frequency is held at its 
final value for an equivalent of around $60$ oscillations of the condensate, after which the trap is opened and the condensate is allowed to expand freely.  The measured distribution of atom velocities is very broad 
(see Fig. 1(c) of \cite{Jaskula-et-al}), in contrast with the expected peak at $\bar{\omega}_{k} = \omega_{\perp f}$ 
and, even more strikingly, failing to show signs of the condensate itself at $k=0$.  To investigate this, we have performed numerical simulations using roughly the same parameters.  The results for $\left|\beta_{k}\right|^{2}$ are shown in Figure \ref{fig:InSitu_Jaskula}, as well as the corresponding value of $n_{k}$ for the initial temperature $T/\omega_{\perp} = 2.8$.  The blue curve is that expected for the experiment of \cite{Jaskula-et-al}.  Note that $\left|\beta_{k}\right|^{2}$ reaches $10^{9}$ at the maximum of the resonance.  Also plotted, in red and yellow, are the expected values of $\left|\beta_{k}\right|^{2}$ after $30$ and $15$ oscillations (i.e. for $1/2$ and $1/4$ of the time after the sudden change in $\omega_{\perp}$).  The black horizontal segment gives 
an estimate for the maximum allowed value of $\left|\beta_{k}\right|^{2}$ before nonlinear effects come into play, and above which the phonons in the resonant mode can no longer be treated as a perturbation on top of the condensate.  This is derived from the condition that the integral over the peak in $k$-space, which gives the total number of produced phonons per unit length, should be significantly smaller than $n_{1}$, the number of atoms per unit length; in particular, the black line occurs where $\left|\beta_k\right|^{2} \left.\delta k\right|_{\rm res} \approx n_{1}/10$, or
\begin{equation}
\left|\beta\right|^{2} \approx \frac{1}{\left.\delta\omega_{k}/\omega_{k}\right|_{\rm time}}  \cdot  \frac{1}{a_{s}/a_{\perp}} \cdot \frac{n_{1}a_{s}}{2\left(1+4n_{1}a_{s}\right)^{1/4}} \,.  
\end{equation}
We conjecture that, in the experiment of \cite{Jaskula-et-al}, there was an initial build-up of phonons in the resonant mode, but that the occupation number grew so large that the resonant mode began to interact nonlinearly with the condensate\footnote{This can be considered as an analogue of preheating in the early universe \cite{Kofman-Linde-Starobinsky}, where the energy content of a single mode grows so large that nonlinear effects cause it to leak into many other modes, see Refs.~\cite{Busch-Parentani-Robertson,Swislocki-Deuar,Zin-Pylak} for studies of these dissipative effects in condensed matter systems.}.
It would thus be very interesting to observe the broadening of the distribution of the atom velocities when increasing the number of oscillations $N$. We conclude that, in order to observe the effects of modulated DCE and to hope to see nonseparability, one should either consider the behavior at early time or opt for a smaller 
change in $\omega_{\perp}$.


\section{Measuring the state of atoms after TOF
\label{sec:TOF}}

In this Section we consider the switching off of the trap, the subsequent expansion of the cloud, 
and the atom counting measurements able to infer the initial velocities of the atoms from their time of flight.  We begin by noting that the expansion itself, as a temporal variation of the condensate, will tend to produce correlated pairs.  We then examine to what extent this secondary effect can pollute the results of any prior pair production process.

\subsection{DCE due solely to the cloud expansion}

After having opened the trap and let the cloud expand, the observable quantities that could be measured are given by the
distribution of the number of atoms carrying a momentum $k$. As in the experiments of~\cite{Jaskula-et-al}, 
we integrate over the perpendicular directions so as to count only atoms with longitudinal momentum $k$. The two observables we shall use are 
\begin{eqnarray}
n_k^{a} &=& \langle \hat{n}_{k}^{a} \rangle \,, \nonumber \\  
C_k^{a} &=& \langle \hat{n}_{k}^{a} \hat{n}_{-k}^{a} \rangle \,, 
\label{eq:nandcsq}
\end{eqnarray}
where $\hat{n}_{k}^{a} = \hat{a}_{k}^{\dagger} \hat{a}_{k}$ 
is the atom number operator 
and $\hat{a}_k$ ($\hat{a}_{k}^{\dagger}$) destroys (creates) an atom carrying momentum $k$. 
The relationship with the $c$-number quantities of Eq.~(\ref{eq:nandcsq}) and those entering Eqs. (\ref{eq:n_c_defn}) and (\ref{eq:nonsep}) is easily made by considering Gaussian isotropic ($n_k^{a} = n_{-k}^{a}$) states.
For these, one finds that $|c_k^{a}| = \sqrt{C_k^{a} - \left(n_k^{a}\right)^2} $,
so that
\begin{eqnarray}
\Delta_k^{a} = n_k^{a} - \sqrt{C_k^{a} - \left(n_k^{a}\right)^2} = \frac{2 \left(n_k^{a}\right)^2 - C_k^{a} }{n_k^{a} + \sqrt{C_k^{a} - \left(n_k^{a}\right)^2}} \, . 
\label{Deltasq} 
\end{eqnarray}
We see that once more it is the relative strength of the correlation term (here $C_k^{a}$) with respect to a function of the occupation number (here $2 \left(n_k^{a}\right)^2$) which fixes the sign of $\Delta_k^{a}$. Being positive definite, the denominator plays no role in determining this sign. We also notice that the operators $\hat C_k^{a} = \hat{n}_{k}^{a} \hat{n}_{-k}^{a}$ and $\hat{n}_{\pm k}^{a}$ commute, and yet, 
just as for the density perturbations measured {\it in situ} (see Section \ref{sec:InSitu}), 
their precise measurement allows us to assess the nonseparability of Gaussian two-mode atomic states $(k,-k)$.   

To illustrate this, we begin our analysis 
by considering the case where both the condensate and the phonons within it are initially in their ground state. (In Appendix~\ref{app:Thermal} we extend this analysis to a thermal bath of phonons.) 
From the point of view of the phonons, as explained in Sec. \ref{sub:opening}, 
the expansion will be seen as adiabatic for 
frequencies much higher than the expansion rate, and sudden by phonons with frequencies much lower than the expansion rate.  We thus expect the latter to be excited by the expansion itself, resulting in a nontrivial 
state of atoms at the corresponding momenta.

\begin{figure}
\includegraphics[width=0.45\columnwidth]{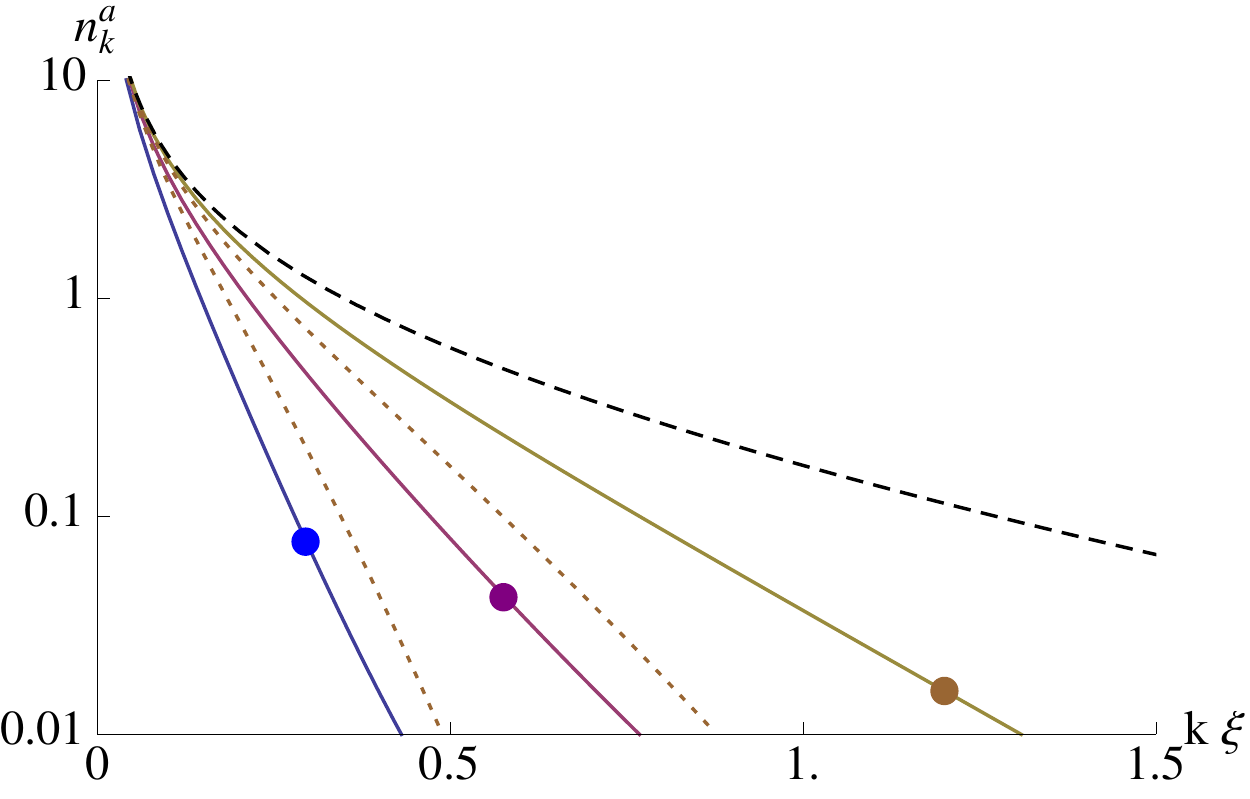} \, \includegraphics[width=0.45\columnwidth]{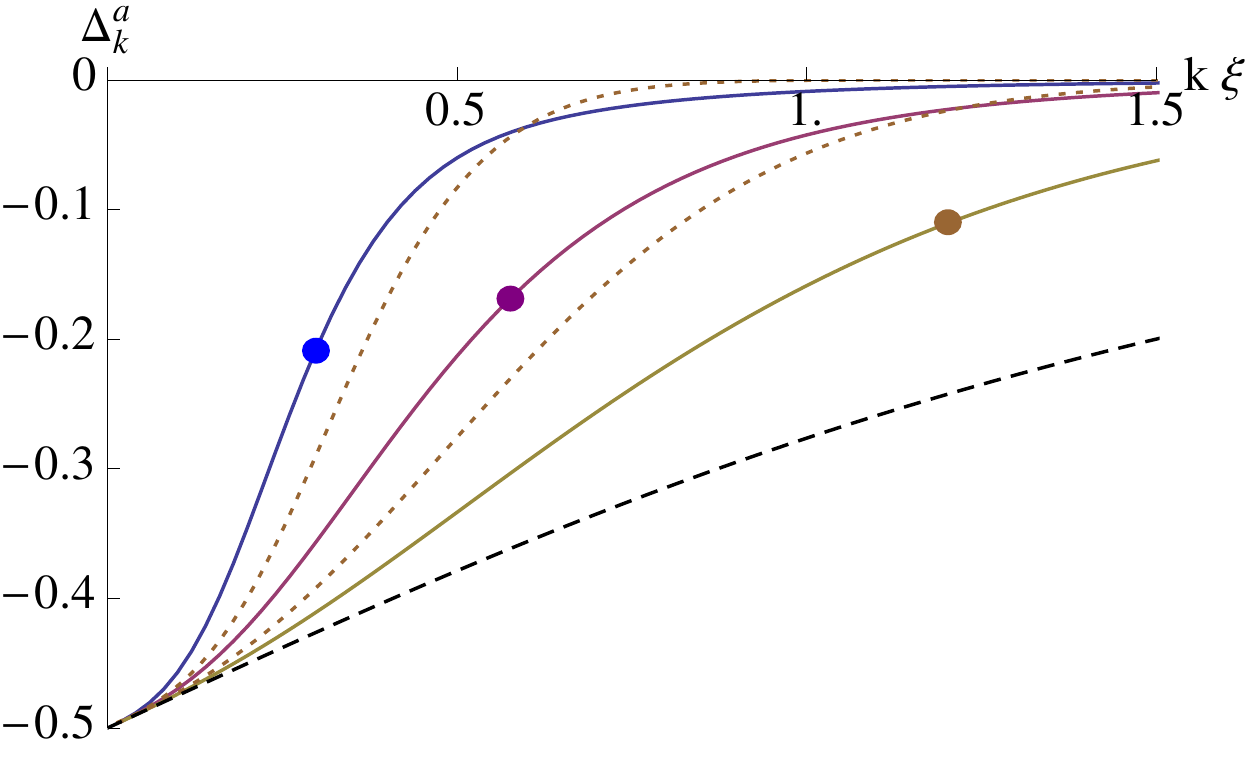}
\caption{Opening the trap, at zero temperature.  The condensate and phonons are initially in their ground state, and the condensate expands in response to a reduction in $\omega_{\perp}$, which vanishes asymptotically.  
The final atomic state is shown as a function of $k \xi$, where $\xi$ is the healing length in the condensate before the opening of the trap.  
On the left is plotted the final occupation number of atoms, while on the right is plotted the final nonseparability parameter $\Delta_{k}^{a}$ 
of Eq.~(\ref{Deltasq}). 
The black dashed curve on the left plot shows $\left|v_{k}\right|^{2}$, the initial 
number of atoms with wavenumber $k$, while on the right it corresponds to $\left|v_{k}\right|^{2}-\left|u_{k}v_{k}\right|$
which is the value of $\Delta_k^{a}$ 
characterizing the atoms in the depletion. 
The solid curves correspond to a sudden change in $\omega_{\perp}$, so that the cloud expands completely freely.  The various colors correspond to different values of $n_{1} a_{s}$: $10$ (blue), $3$ (purple), and $0.7$ (yellow).  (These are equivalent to initial values of $\omega_{\perp}/mc^{2} = 0.3$, $0.6$ and $1.4$, respectively.)  
The large dots on these curves show where the initial phonon frequency $\omega_{k\, i}$ is equal to $\omega_{\perp}$, and mark the boundary between phonons which see the expansion as sudden and are highly excited ($\omega_{k\, i} \ll \omega_{\perp}$) and phonons which see the expansion as adiabatic and are hardly excited at all ($\omega_{k\, i} \gg \omega_{\perp}$).
The dotted curves, by contrast, correspond to a gradual opening of the trap, with the condensate having initially the same thickness as the solid yellow curve. For these, $\omega_{\perp}$ varies like a hyperbolic tangent (as was also used 
in Fig. \ref{fig:tanh_Tozzo_response}), with 
the rate $a/\omega_{\perp i}$ taking 
the values $0.85$ (right curve) and $0.3$ (left curve). 
\label{fig:TOFalone}}
\end{figure}

In Figure \ref{fig:TOFalone} are shown examples of the final state of atoms after an expansion of the condensate in response to an opening of the trap.
We consider two cases: 
\begin{itemize}
\item a sudden opening of the trap, so that the condensate expands freely with a rate proportional to the initial value of $\omega_{\perp}$; we consider several such values, finding that the final occupation number of atoms grows with $\omega_{\perp}/mc^{2}$;
\item a gradual opening of the trap,  with the initial value of $\omega_{\perp}$ fixed; 
as already seen in Fig. \ref{fig:tanh_Tozzo_response}, this slows the rate of expansion, and we find that the resulting spectrum is qualitatively similar to the case of a sudden opening with a lower initial value of $\omega_{\perp}$.
\end{itemize}
Interestingly, we find that no matter how the trap is opened, the final occupation number is always less than $\left| v_{k} \right|^{2}$, the mean number of atoms initially present in the depletion at zero temperature~\cite{Pitaevskii-Stringari-BEC} 
(which was observed in~\cite{Vogels-et-al}).  
More precisely, it approaches $\left| v_{k} \right|^{2}$ in the limit of a sudden expansion, where 
the initial phonon frequency $\omega_{k \, i}$ is much smaller than the expansion rate. 

The main lesson one draws from Fig. \ref{fig:TOFalone} is the following. If one wants to have an adiabatic opening for a certain range of phonon 
excitations (those with $k\xi < 1$), one could in principle 
work with fat cigar-shaped condensates with $n_1 a_s \gg 1$
and allow $\omega_{\perp}$ to drop suddenly to zero. 
However, for such 
condensates, one cannot neglect the transverse excitations. Hence the only way to have negligible transverse excitations {\it and} an adiabatic opening is by controlling the time dependence of $\omega_{\perp}$, for instance as described in \eq{eq:omega_perp_tanh} when putting  $\omega_{\perp f} = 0$.  Then, if $a \ll \omega_{\perp i}$, the rate of expansion will be controlled by $a$ while $\omega_{\perp i}$ drops out, see Fig.~\ref{fig:tanh_Tozzo_response}. The relevant parameter for phonon excitation will then be $a/mc_{i}^{2}$,
which can be as small as experiment will allow.

\subsection{Combined effects of DCE and expansion}

Since the expansion of the condensate itself induces phonon pair production, we must take its contribution to the final atomic state into account when using TOF measurements to investigate a previous DCE.  For example, let us consider the results presented in Fig.~\ref{fig:InSitu_G2t_Tanh}, in which a slow variation of the condensate is used to excite phonons.  
We now allow the cloud to expand so that individual atoms can be measured, and so we need to subject the condensate to another change corresponding to this expansion.

\begin{figure}
\includegraphics[width=0.45\columnwidth]{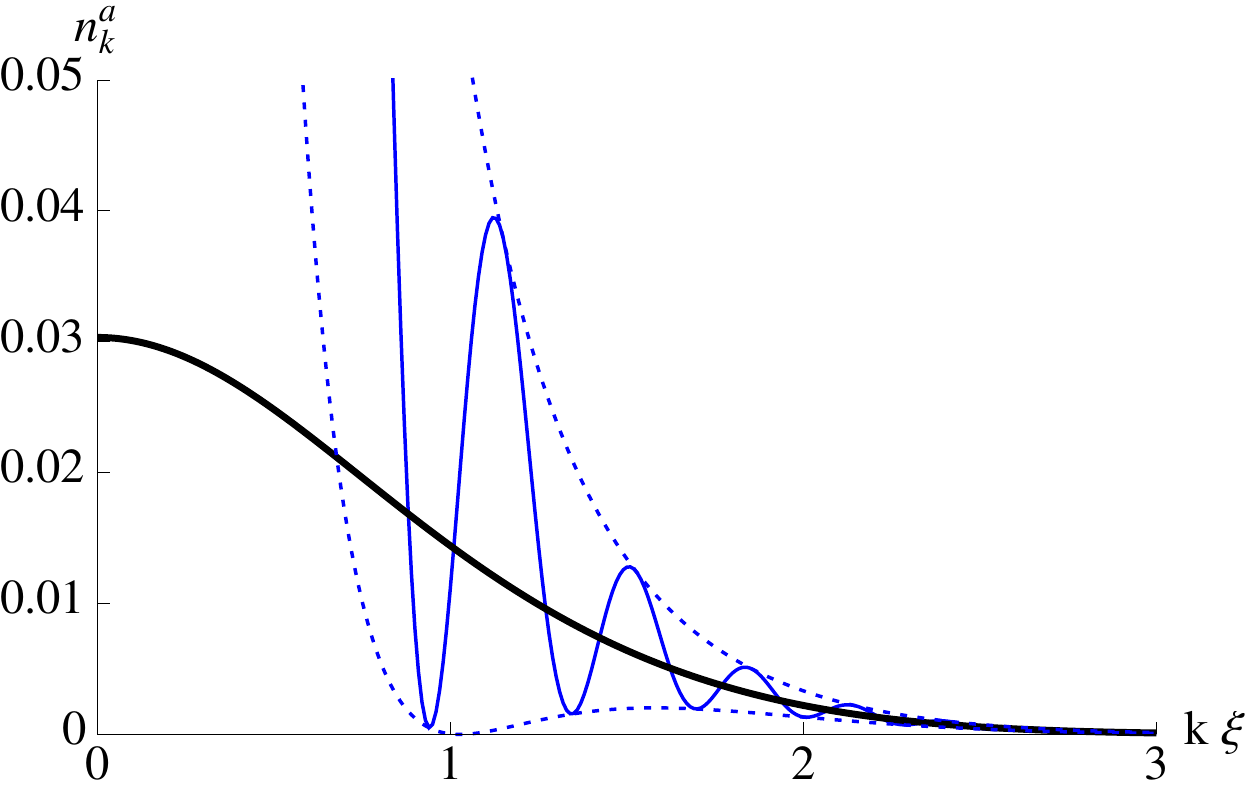} \, \includegraphics[width=0.45\columnwidth]{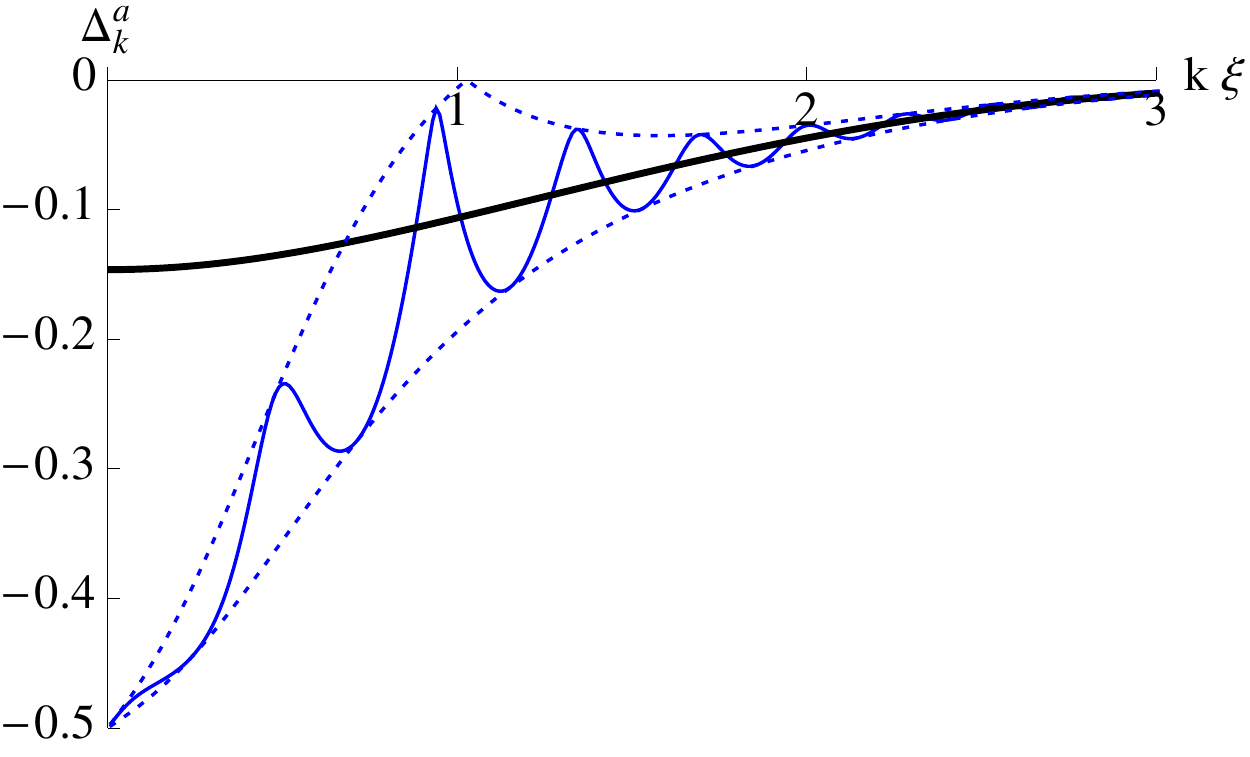} \\
\includegraphics[width=0.45\columnwidth]{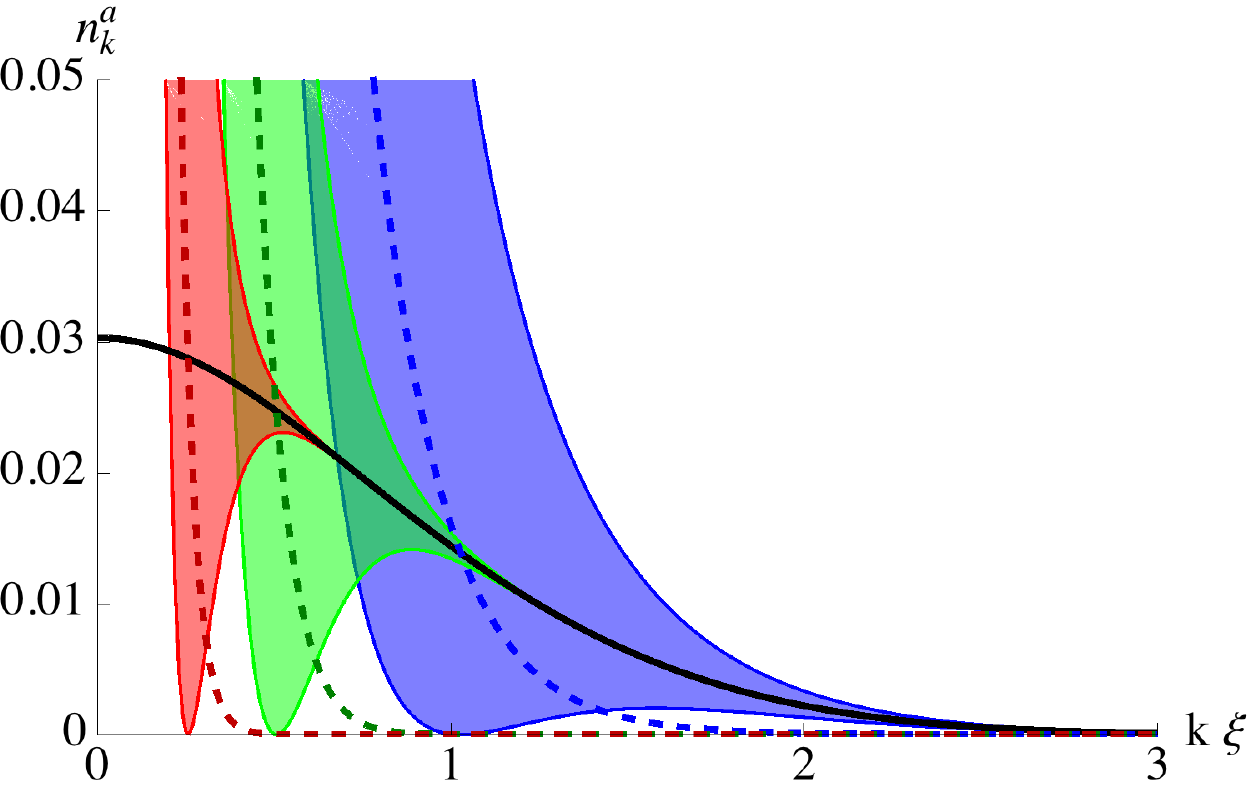} \, \includegraphics[width=0.45\columnwidth]{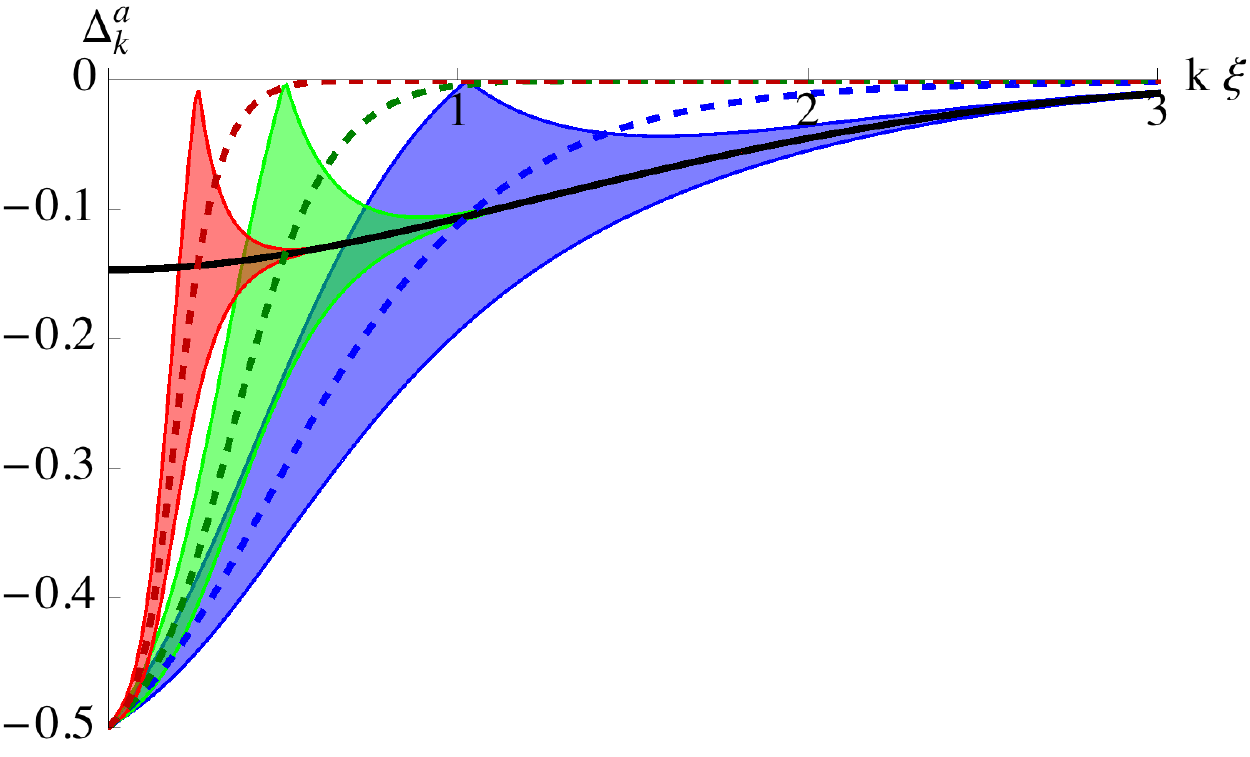}
\caption{Opening the trap after a smooth change in $c^{2}$, at zero temperature.  
We begin with the situation shown in Figure \ref{fig:InSitu_G2t_Tanh}, followed by a variation of $\omega_{\perp}$ that goes to zero asymptotically. 
The final atomic state is shown as a function of $k \xi$, where $\xi$ is the healing length in the condensate after the initial DCE process but before the opening of the trap.  
In the upper row, $\omega_{\perp}$ is switched off suddenly from an initial value of $\omega_{\perp}/mc^{2} = 1$, after a time delay of $m c^{2} \Delta t = 5$ (the same moment at which the {\it in situ} state in Fig. \ref{fig:InSitu_G2t_Tanh} is drawn).  The thick black curves plot the phonon state after the initial DCE, while the solid blue curves show the state of the atoms after the expansion and the dotted curves plot the upper and lower envelopes of the oscillations.
In the lower row, only the envelopes are shown, and we now include thick dashed curves to show the state that would result from the cloud expansion alone.  The blue curves correspond to the same sudden switch-off of $\omega_{\perp}$ as used in the upper row, while the green and red curves correspond instead to a gradual decrease of $\omega_{\perp}$ of the kind used 
in Figs. \ref{fig:tanh_Tozzo_response} and \ref{fig:TOFalone}. 
The rates used for the switch-off are $a/\omega_{\perp} = 0.6$ (green) and $0.2$ (red).  It is clear that slowing the rate of expansion 
greatly reduces its effect, and in this way we can bring the final state of the atoms closer to the state of the phonons that existed before the expansion.
\label{fig:TOF_InSitu_Opening}}
\end{figure}

In Figure \ref{fig:TOF_InSitu_Opening} are shown examples of the atom state after the expansion of the cloud.  In the upper row, $\omega_{\perp}$ is switched off suddenly from an initial value of $\omega_{\perp}/mc^{2} = 1$, $n_1 a_s = 1.2$. 
Because the spectrum of the original DCE and the spectrum due to the expansion alone overlap significantly, 
the final atom state is generally different from the phonon state that existed before the expansion.  They are closer at high $k$, where the spectrum due to the expansion becomes negligible, but for low $k$ the expansion is dominant.
In particular, at low $k$, the final value of the nonseparability parameter $\Delta_k^{a}$ 
is essentially determined by the expansion, and hence no conclusion can be drawn concerning the nonseparability of the phonon state prior to opening the trap.
These findings are reinforced when working with a temperature such that the phonon state after DCE turns out to be separable, 
while the final state of atoms of momentum $\pm k$ is nonseparable, see Appendix~\ref{app:Thermal}. 

We note from the upper row of Fig. \ref{fig:TOF_InSitu_Opening} that the interference between the two processes produces oscillations in the final results which are not present in the results of either process taken on its own.  These vary periodically as a function of the time difference between the original variation of the condensate and its final expansion, but the upper and lower envelopes of the results are independent of this time delay. 
Effectively, then, the results become ``smeared'', and it is convenient to plot the envelopes rather than the oscillations themselves.  Further examples of this are shown in the lower row of the same figure.  There, we plot the envelopes instead of the oscillations within them, and we also show the state that would result due to the expansion alone, in the absence of the original DCE.  It is clear that the larger of the two spectra dominates.

In the lower row of Fig. \ref{fig:TOF_InSitu_Opening}, we also include results corresponding to a gradual opening of the trap, where $\omega_{\perp}$ has the same initial value but varies in the manner of a hyperbolic tangent, as also used 
in Figs. \ref{fig:tanh_Tozzo_response} and \ref{fig:TOFalone}. 
The key lesson is that, if the expansion alone does not excite phonons at a particular wave vector $k$, then the state of the atoms of momentum $k$ after the expansion will be exactly the same as the state of the phonons of momentum $k$ before the expansion.  As a result, by controlling the opening rate, one can squeeze the effects of the expansion into a narrow window at low $k$, thus getting closer agreement between the final atomic and initial phononic states at higher $k$.

\subsection{Combined effects of modulated DCE and opening the trap}

We might also consider the effects of the cloud expansion on the results obtained from the modulated condensate in Sec. \ref{sub:DCE_modulated}.  In this case, however, the results are very similar whether the cloud expansion is taken into account or not, and we are now in a position to understand why.  Since the natural oscillation frequency of the condensate is $2\omega_{\perp}$, the resonant phonon mode occurs at $\omega_{k} \sim \omega_{\perp}$, 
or 
$k \xi \sqrt{1+\left(k \xi\right)^{2}/4} \sim \omega_{\perp}/mc^{2}$.  
But this is a function of $n_{1} a_{s}$, and in the 1D regime, we have (see Eq.~(\ref{eq:timescaleratio})) $\omega_{\perp}/mc^{2} = \sqrt{1 + 4 n_{1} a_{s}}/2n_{1}a_{s} \gtrsim 1$.  This means that the resonance occurs at $k\xi \gtrsim 1$, and the maximum pair production due to the expansion in this regime is $\left|v\left(k\xi = 1\right)\right|^{2} = 0.17$.  Typically, it will be much less than this, and will thus have very little effect on the resonance, where $\left|\beta_{k}\right|^{2} \gtrsim 1$ when the number of oscillations $N$ (in the example of Fig. \ref{fig:InSitu_exp}) is larger than $\sim 5$.  We recover here a situation closer to that of Ref.~\cite{deNova-Sols-Zapata}, where the authors consider the nonseparability after a sudden opening of the trap in a transonically flowing condensate which engenders a resonant analogue Hawking effect.


\section{Conclusion
\label{sec:Conclusion}}

In this paper, we have studied the 
production of phonon pairs in effectively one-dimensional homogeneous Bose-Einstein condensates.  We paid particular attention to realistic scenarios and have identified experimentally accessible observations that can be used to determine the nonseparability of the final state.

We first considered the dynamics of the condensate itself, in order to determine what time-varying backgrounds are available in experiments 
where one typically has control over the harmonic trapping frequency $\omega_{\perp}$.  We restricted our attention to the 1D regime $n_{1} a_{s} \lesssim 1$, where a Gaussian approximation of the transverse profile of the condensate gives accurate results. 
There are two possibilities of interest: a smooth variation of the condensate from one equilibrium state to another, which occurs when the trap frequency is varied adiabatically; and a steady oscillatory regime of the condensate around its equilibrium state, which occurs in response to a fast variation of the trap frequency. 

We then considered the dynamics of the phonon field in response to these time-varying backgrounds, first considering density measurements made {\it in situ} and the two-point density correlation function.  After noting the behavior of this function in the absence of correlated phonon pairs, and that it can be used to characterize both the temperature and the vacuum fluctuations, 
we examined its response to the two variations of the condensate mentioned above. In the first case, 
for a given wave vector, this response is very similar to the response of the condensate to a variation of $\omega_{\perp}$: if the variation in $\omega_{k}$ is slow, it varies smoothly and always remains in the ground state; but if the variation in $\omega_{k}$ is fast, it is excited and oscillates around the ground state.  In particular, when the change is sufficiently abrupt or the temperature sufficiently low, 
the two-point density correlation function oscillates such that it periodically dips below its vacuum value, a behavior that indicates a subfluctuant mode and implies that the 2-mode phonon state $(k,-k)$ is nonseparable. 

We also examined the response of the phonons to a sinusoidal modulation of $\omega_{k}$, which is itself caused by the oscillation 
of the condensate. We focused on the resonance domain where the occupation of the modes grows exponentially with the duration of the modulation. 
We were thus able to conclude that the very broad spectrum observed in 
\cite{Jaskula-et-al} was likely due to a very large occupation of the resonant mode, 
which had the time to decay into a broad spectrum in a process akin to the preheating phase in the inflationary scenario. 
We also indicated that a much smaller duration of the oscillatory phase (10 times smaller) should have resulted in a much clearer signal
displaying nonseparability in a small range of frequencies centered around the resonant one. 
It would be particularly interesting to perform observations of the broadening of the spectrum associated with increasing the 
duration of the modulation so as to probe the first effects of nonlinearities. 

Finally, we considered the opening of the harmonic trap and the complete expansion of the cloud into individual atoms, which are examined using TOF measurements.  We 
first noted that this expansion itself generates a time-varying background, and began by looking at the pair production associated with the expansion on its own. As a result, depending on the thickness of the cloud, and its temperature, the final state of atoms carrying momentum $\pm k$ can end up entangled even without any DCE prior to the opening of the trap.
We then combined this with a DCE performed previously, to see its effect on the final atomic state.  We found that, when the two are comparable, they interfere quite strongly, and the results are ``smeared'' by oscillations that depend on the lapse of time between the original DCE and the expansion.  The effects of the final expansion can, however, be tamed by controlling the rate of opening  the trap. Then, whenever the pair production due to this slower expansion is much smaller than the pair production due to the original DCE, the final state of the atoms is the same as the initial state of the phonons. In particular, one can then assess nonseparability by considering the relative importance of the strength of the atomic correlations between $k$ and $-k$ with respect to a certain function of the mean occupation number of atoms with these momenta. 

We are currently studying the more elaborate case of opening a trap in a condensate with a 
transonic flow, which corresponds to an analogue black hole. 


\section*{
Acknowledgments.} 
We are grateful to Jeff Steinhauer, Silke Weinfurtner, Chris Westbrook, Denis Boiron, 
Ivar Zapata, Gora Shlyapnikov, Ted Jacobson and Antonin Coutant 
for interesting discussions and suggestions.
This work was supported by the French National Research Agency by the Grant ANR-15-CE30-0017-04 associated with the project HARALAB.  

\begin{appendices}


\section{Validity of the Gaussian approximation
\label{app:num_sol_GPE}}

This appendix is devoted to the comparison between results drawn from the Gaussian approximation of Eq.~(\ref{eq:gaussian_ansatz}) and by solving numerically the 3-dimensional Gross-Pitaevskii equation (GPE) with cylindrical symmetry. We first focus on the stationary background solution, briefly recalling the results of~\cite{Gerbier} relevant for our purpose and showing how they compare to numerical solutions. 
We then turn to the dispersion relation. Besides variations of its shape due to 3-dimensional effects, it shows additional branches which become relevant for high linear densities $n_1$, while their energies, expressed in units of the healing length and sound velocity, become very large for thin condensates. Finally, we briefly comment on the dynamics when the strength of the trapping potential is varied in time, recalling the findings of~\cite{Kagan-Surkov-Shlyapnikov} which justify the use of Eq.~\eqref{eq:sigma_eom} in the main text.

\subsection{Stationary background solution}

In~\cite{Gerbier} is presented a generic procedure to estimate the relationship between the chemical potential $\mu$ and linear density $n_1$ of a cylindrically symmetric solution of the GPE, based on minimization of the value of $\mu$ on a set of trial Gaussian condensate wave functions. 
The stationary GPE reads
\begin{align}
\mu \, \Phi_0 = - \frac{1}{2 \, m \, r} \, \partial_r \left( r \, \partial_r \Phi_0 \right) + V(r) \, \Phi_0 + g \, \abs{\Phi_0}^2 \, \Phi_0.
\end{align}
Multiplying it by $r \, \Phi_0^\star$ 
and integrating over $r$ with the boundary conditions $r \, \Phi_0^\star (r) \, \partial_r\Phi_0(r) \mathop{\to}_{r \to 0,\infty} 0$ 
gives 
\begin{align}
\mu = \frac{\int_0^\infty \left( \frac{1}{2m} \abs{\partial_r \Phi_0}^2 + V(r) \, \abs{\Phi_0}^2 + g \, \abs{\Phi_0}^4 \right) r \, \dd r}
{\int_0^\infty \abs{\Phi_0}^2 \, r \, \dd r}.
\end{align}
Using the ansatz of 
Eq.~\eqref{eq:gaussian_ansatz} 
and assuming $V$ is harmonic with frequency $\omega_\perp$, the integrals can be performed explicitly; one obtains
\begin{align} \label{eq:app_mu_G}
\mu = \frac{1}{2} \, m \, \omega_\perp^2 \, \sigma^2
	+ \frac{1 + 4 \, n_1 \, a_s}{2 \, m \, \sigma^2} .
\end{align}
One recovers the effective potential of Eq.~\eqref{eq:sigma_eom}. 
When looking for the ground-state solution, $\mu$ should be minimized with respect to $\sigma$ at fixed $n_1$.~\footnote{This may also be justified using Hamilton's equations on the integrated action when seeing $\mu$ as a Lagrange multiplier.} 
This gives
\begin{align} \label{eq:app_sigma_G}
\sigma_G^4 = \frac{1 + 4 \, n_1 \, a_s}{m^2 \, \omega_\perp^2} \, ,
\end{align}
and
\begin{align}
\mu_G = m \, \omega_\perp^2 \, \sigma_G^2,
\end{align}
where the index $G$ indicates that these quantities are evaluated using Gaussian trial wave functions. 

\begin{figure}
\includegraphics[width = 0.49\linewidth]{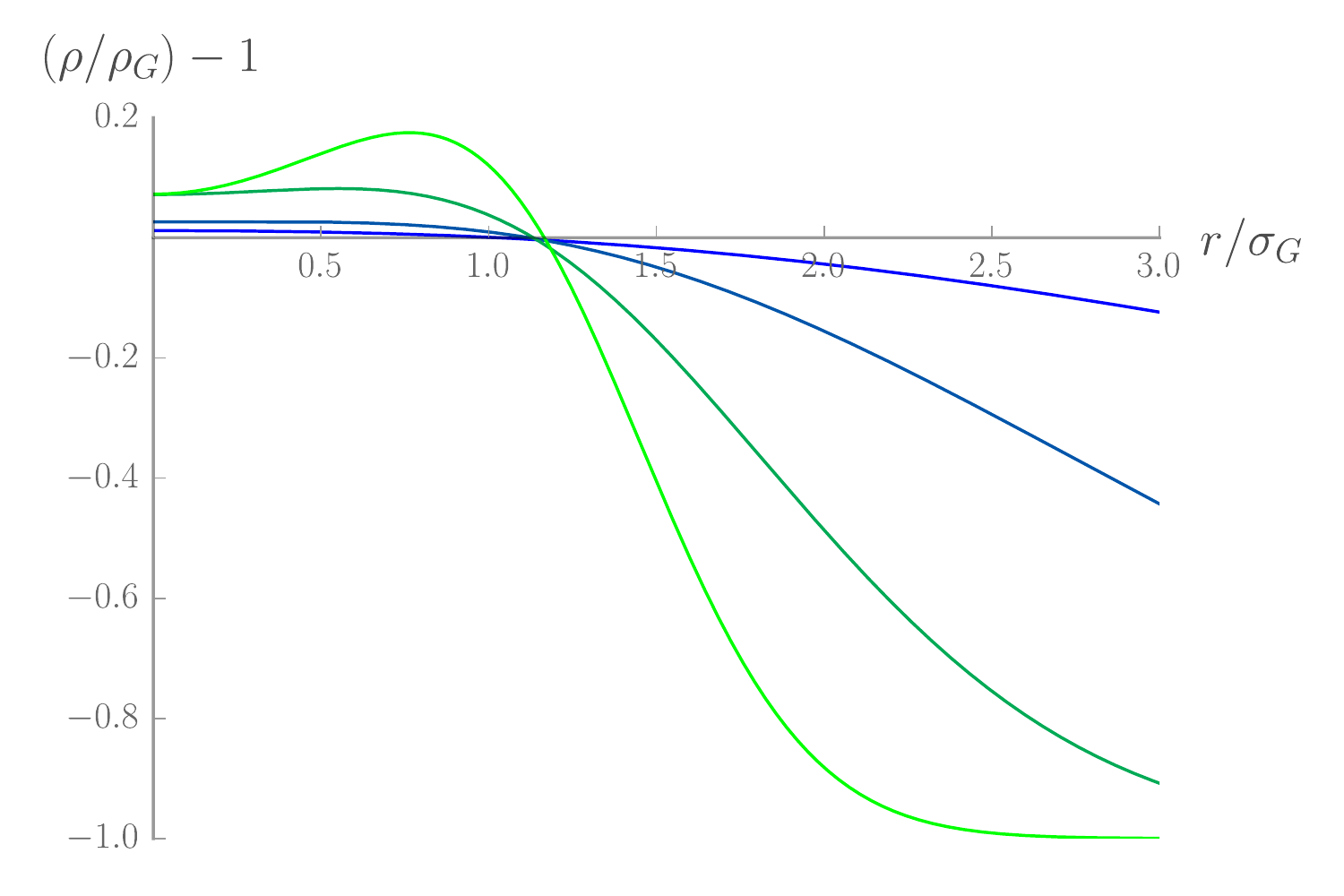}
\caption{Relative difference between the numerical density profile and the Gaussian approximation.
The chemical potential $\mu$ takes the values $1.01 \omega_\perp$, $1.1 \omega_\perp$, $1.42 \omega_\perp$, and $2.475 \omega_\perp$ from blue to green, corresponding to $n_1 \, a_{s} \approx 0.01$, $0.05$, $0.25$, and $1.25$, respectively. 
}\label{fig:app_stat_profiles}
\end{figure}

To verify the validity of this approximation, we solved numerically the stationary GPE with cylindrical symmetry. We used a relaxation method~\cite{Numerical_Recipes} based on the Gross-Pitaevskii action. Starting from an initial Gaussian wave function, a better approximation of a configuration extremizing the discretized action $S^{(d)}$ is obtained by solving the linear equation $S^{(d)}_2 \, \delta \Phi_0 = - \alpha \, S^{(d)}_1$, where $S^{(d)}_2$ denotes the matrix of second derivatives of $S^{(d)}$ with respect to $\Phi_0$ and $S^{(d)}_1$ its first derivative. In this expression, $\delta \Phi_0$ is the difference between the approximate and trial solutions, and $\alpha$ is a coefficient which can be tuned to improve the convergence speed or stability. In practice, setting $\alpha$ to $1$ offers a good trade-off in most relevant cases. The process is then repeated $N$ times, with $N$ chosen large enough for the results to show no visible deviation when doubling its value. 

Results for the density profile in the radial direction are shown in Fig.~\ref{fig:app_stat_profiles}. 
We show the relative difference between the local density $\rho$ computed numerically and its value $\rho_G$ found using the Gaussian approximation of~\cite{Gerbier}, for several values of the quantity $n_1 \, a_s$ which controls the transition between the 1D and 3D regimes. 
For thin condensates, i.e., $n_1 \, a_s < 0.05$, relative deviations become important only for large values of $r / \sigma_G$ where the density is small. 
For instance, they become larger (in absolute value) than $0.1$ only for $r > 1.52 \,  \sigma_G$, where $\rho_G / \rho_G(0) < 0.1$. 
More important deviations occur for thicker condensates, the solution for a given value of $n_1$ becoming more extended than its Gaussian approximation because of stronger repulsive interactions between atoms, but falling off more rapidly for $r/\sigma_{G} \to \infty$. 

\begin{figure}
\includegraphics[width = 0.49\linewidth]{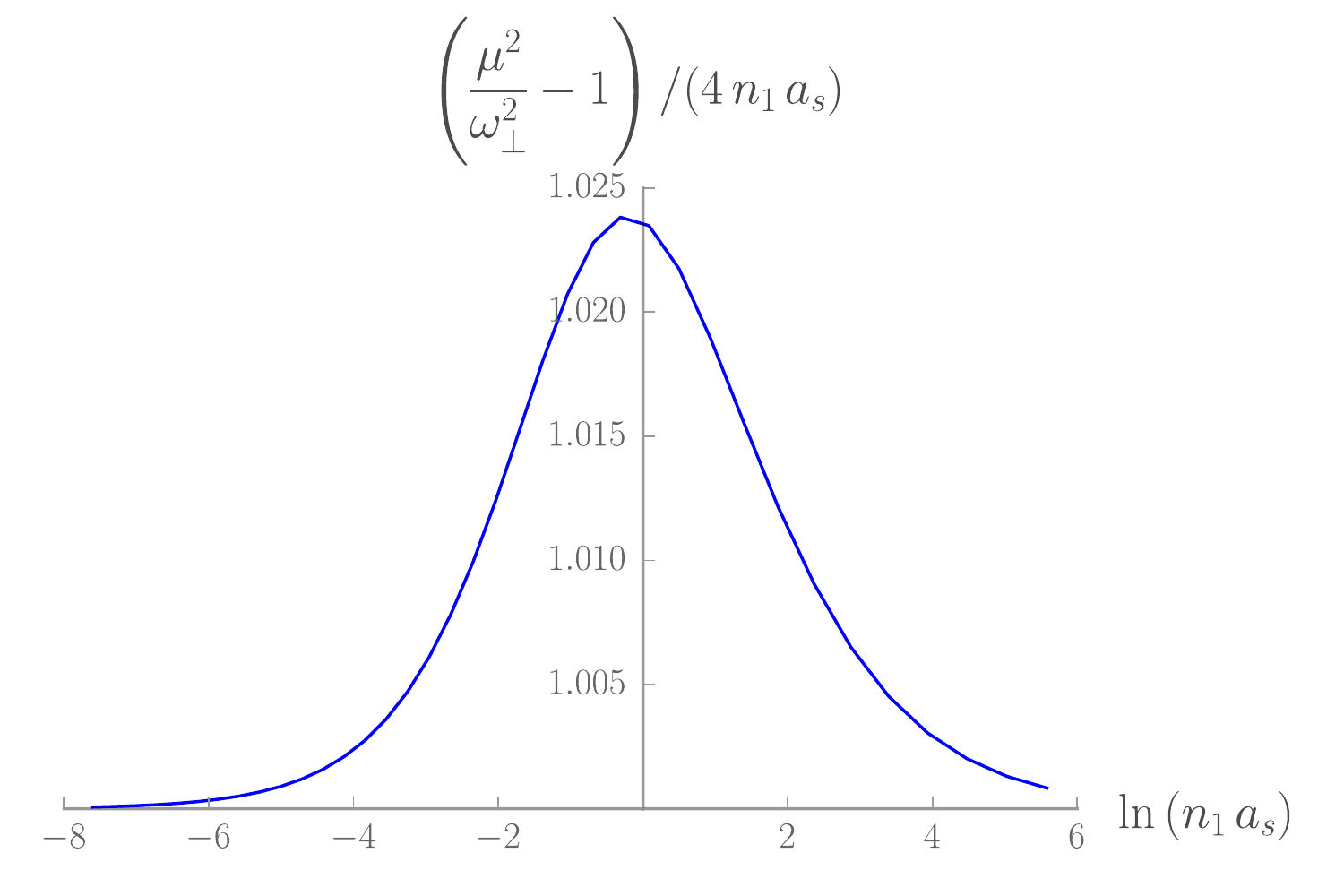}
\caption{Plot of the quantity $\left( \mu^2 / \omega_\perp^2 - 1 \right) / (4 \, n_{1} \, a_s)$, 
which is identically equal to $1$ in the Gaussian approximation. 
The distance from unity is linear in $n_1 \, a_s$ when the latter becomes much smaller than $1$.}\label{fig:app_mu}
\end{figure}

In Fig.~\ref{fig:app_mu} we compare the values of $\mu$ obtained with the two approaches. 
To make the comparison clearer, we show the quantity 
\begin{align}
\delta \equiv
\frac{1}{4 \, n_1 \, a_s} \, \left( \frac{\mu^2}{\omega_\perp^2} - 1 \right),
\end{align}
which is identically equal to $1$ when using Eqs.~\eqref{eq:app_mu_G} and~\eqref{eq:app_sigma_G}. 
One can \textit{a priori} expect the Gaussian approximation to become accurate for $n_1 \, a_s \to 0$, where the solution of the GPE is actually a Gaussian, and $n_1 \, a_s \to \infty$, where a Gaussian ansatz with a large extension gives the Thomas-Fermi result. 
This is confirmed by Fig.~\ref{fig:app_mu}, which indicates that $\delta$ 
goes to $1$ in these two limits. 
Moreover, the maximum deviations seem smaller than $0.025$, indicating that the Gaussian approximation is quite accurate, as far as the chemical potential is concerned, for all values of $n_1 \, a_s$.

\subsection{Dispersion relation}

In this subsection we compare the dispersion relations obtained using Eq.~(\ref{eq:BdGeqn}) 
and the 3-dimensional Bogoliubov-de Gennes equation in a time-independent background. 
The first one can be obtained by looking for solutions of Eq.~(\ref{eq:phi-k_eqn}) 
proportional to $\ee^{- \ii \omega \, t}$. 
This gives an algebraic equation which possesses nontrivial solutions if and only if 
\begin{align}\label{eq:app:DR}
\omega^2 = c^2 \, k^2 + \frac{k^4}{4 \, m^2},
\end{align} 
where $c$ is the velocity of long-wavelength waves. 
The latter is given (independently of the Gaussian approximation) by~\cite{Menotti-Stringari} 
\begin{align}
c^2 = n_1 \, \frac{\dd \mu}{\dd n_1}. 
\end{align}
Since the Gaussian approximation gives a very good estimate of the chemical potential, see Fig.~\ref{fig:app_mu}, one can expect that it also describes well the small-$k$ phonons.
Moreover, for large values of $k$ the dispersion relation must become equivalent to the atomic one $\omega \sim k^2 / (2 \, m)$, which coincides with the large-$k$ limit of Eq.~\eqref{eq:app:DR}. 
This equation should thus be accurate in the two limits $k \xi \to 0$ and $\left| k \xi \right| \to \infty$. 

\begin{figure}
\includegraphics[width = 0.49\linewidth]{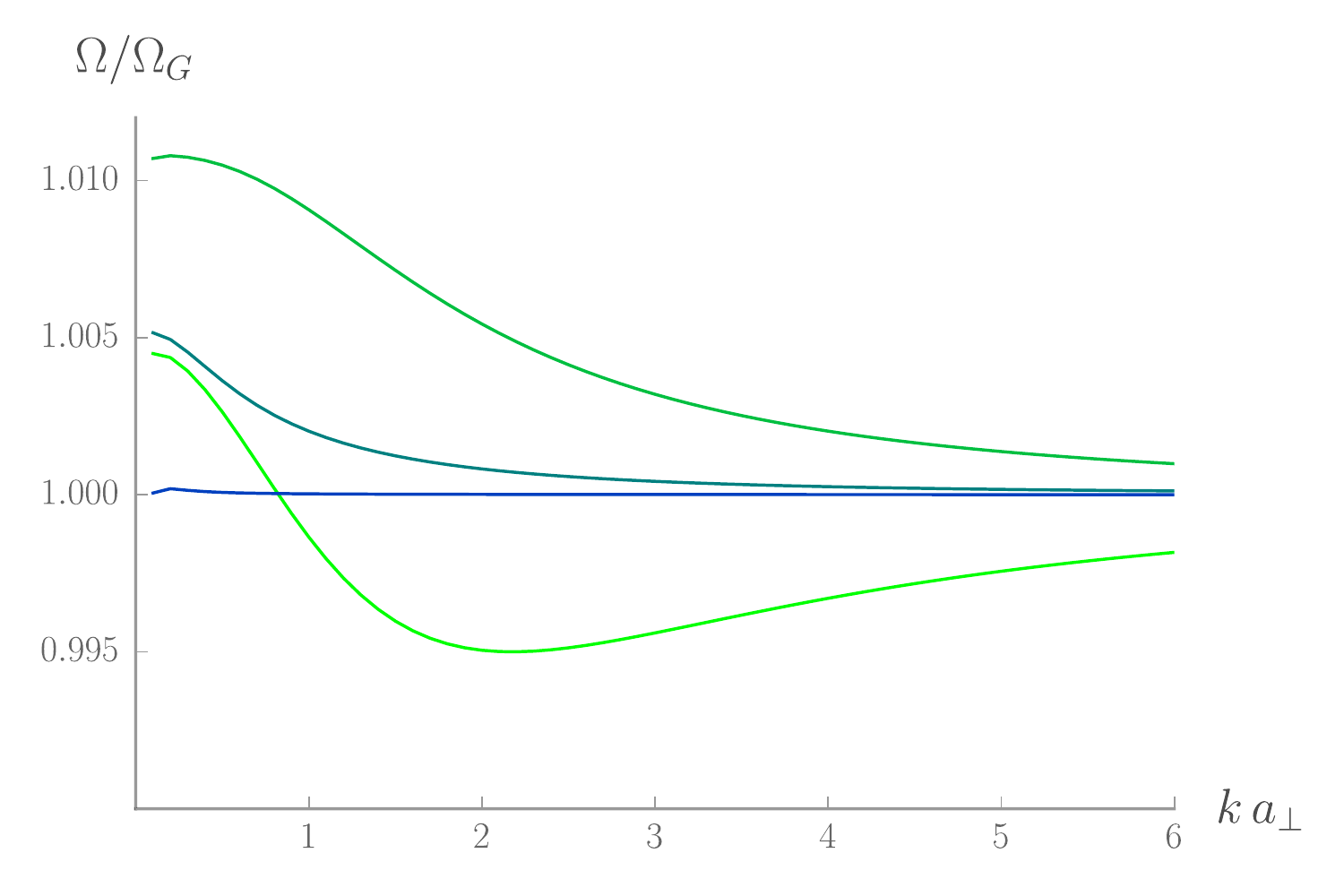} \caption{Plots of the ratio of the frequency $\omega$ numerically computed over the value $\omega_G$ obtained using the approach of Section~\ref{subsec:Equations_of_motion}. 
The background solutions are the same as 
in Fig~\ref{fig:app_stat_profiles}. 
}\label{fig:app_ratio_Omega}
\end{figure}

To verify its validity, 
we solved numerically the 3-dimensional Bogoliubov-de Gennes equation for modes with no angular momentum. Because of the exponential fall-off in $r^2$ of the background solution as $r \to \infty$, it is more practical to work with absolute perturbations rather than the relative perturbations used in the main text (see Eq.~(\ref{eq:Bog_approx})). 
Writing $\Phi = \Phi_0 + \delta \Phi$ and looking for solutions with fixed frequency and angular momentum, of the form
\begin{align} 
\delta \Phi: (t,r,z) \mapsto \ee^{- \ii \mu \, t} \left( U(r) \ee^{\ii \left( k \, z - \omega_t \right)} + W(r)^* \ee^{-\ii \left( k \, z - \omega_t \right)} \right) + \mathcal{O} \left( \delta \Phi^2 \right),
\end{align}
one obtains the following system of equations on $U$ and $W$:
\begin{align} \label{eq:app:BdG_abs}
\frac{1}{2 \, m} \, \partial_r \left( r \, \partial_r  
\begin{pmatrix}
U \\ W
\end{pmatrix}
\right) = 
\begin{pmatrix}
\frac{k^2}{2 \, m} - \omega + V - \mu + 2 \, g \, \abs{\Phi_0}^2 & g \, \Phi_0^2  \\
g \, \Phi_0^{*2} & \frac{k^2}{2 \, m} + \omega + V - \mu + 2 \, g \, \abs{\Phi_0}^2
\end{pmatrix}
\begin{pmatrix}
U \\ W
\end{pmatrix}. 
\end{align}
In general, there are two linearly independent solutions regular at $r=0$, but no nontrivial linear combination of them has asymptotically bounded functions $U$ and $W$. Such a solution exists if and only if $\omega$ is a solution of the dispersion relation. 
Eq.~\eqref{eq:app:BdG_abs} was integrated numerically at fixed $k$ for different value of $\omega$, using an algorithm akin to a shooting method to find these solutions. To make comparison with results from the Gaussian ansatz, we first looked for values of $\omega$ close to those given by Eq.~\eqref{eq:app:DR}. The relative differences are shown in Fig.~\ref{fig:app_ratio_Omega}. 
As expected, they are always relatively small, with a maximum only slightly above $0.01$ for the parameters we considered. 

\begin{figure}
\includegraphics[width = 0.49\linewidth]{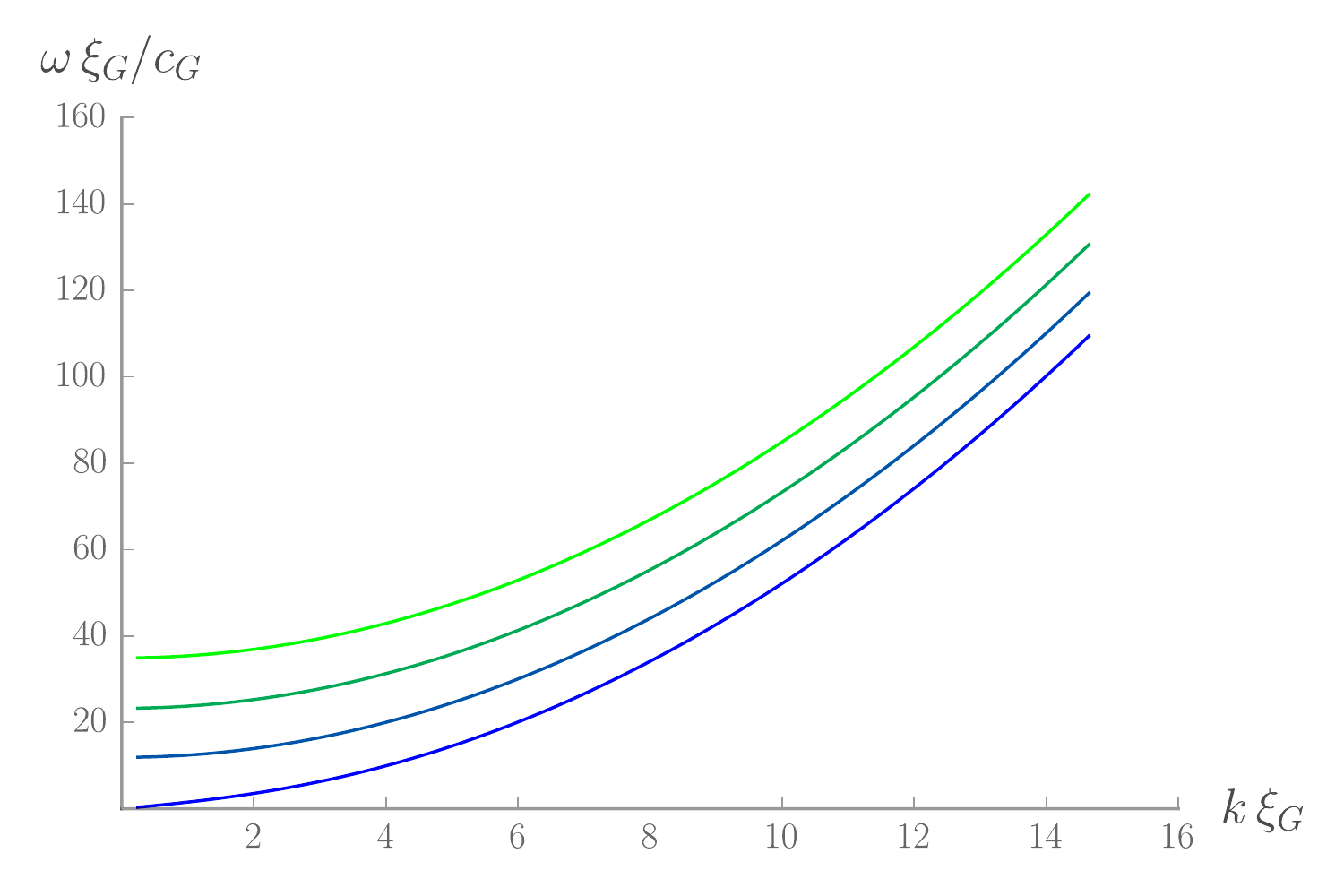}
\includegraphics[width = 0.49\linewidth]{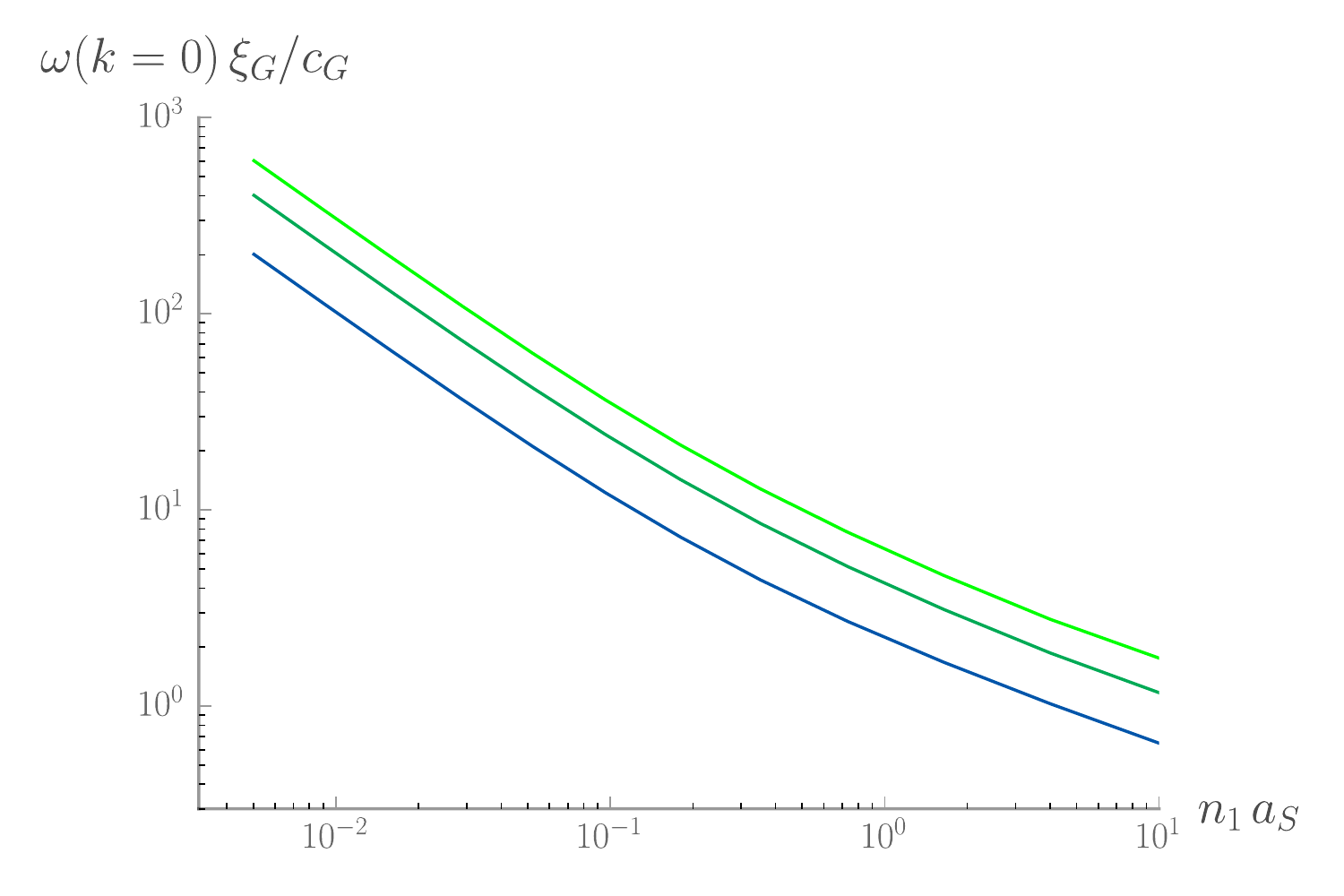}
\caption{Left: Plot of $\omega$ versus $k$ for the first four 
branches of the dispersion relation on the cylindrically-symmetric, node-less stationary solution with $\mu = 1.42 \, \omega_\perp$, corresponding to $n_1 \, a_{s} \approx 0.25$. 
The wave vector $k$ and angular frequency $\Omega$ are adimensionalized using the healing length $\xi_G$ and effective 1D sound speed $c_G$ derived using the Gaussian approximation. 
Right: Angular frequency at $k= 0$ for the second to fourth branches as a function of $n_1 \, a_s$, in logarithmic scale.}\label{fig:app_DR_4_branches}
\end{figure}

When looking for solutions further away from those of Eq.~\eqref{eq:app:DR}, 
we found additional branches with larger frequencies for the same wave vector, see the left panel of Fig.~\ref{fig:app_DR_4_branches}. 
In the thin condensate limit $n_1 \, a_s \to 0$, 
the full dispersion relation can be computed analytically, giving
\begin{align}
\omega = 2 \, n \, \omega_\perp
+ \frac{k^2}{2 \, m}, \, n \in \mathbb{N}. 
\end{align}
Interestingly, we found that the difference in $\omega$ between two consecutive branches remains nearly constant when increasing $n_1 \, a_s$ in the range $0 \leq n_{1} \, a_{s} \leq 10$. 
However, it varies significantly when expressed in units of the healing frequency, see the right panel of Fig.~\ref{fig:app_DR_4_branches}. 
In particular, for $n_1 \, a_s < 0.1$ this difference 
is of the order of or larger than $10$, meaning that these additional branches will be more difficult to excite and can thus be safely neglected. 

\subsection{Dynamics in a time-dependent harmonic potential}

In this subsection we recall a result of~\cite{Kagan-Surkov-Shlyapnikov} that justifies 
the description in Sec.~\ref{sub:response} of the time-evolution of the condensate when varying the strength of the harmonic trap. 
Let us consider a BEC in 2 dimensions, with a potential of the form
\begin{align} \label{eq:app:V_time_dep}
V: (t,r) \mapsto \frac{1}{2} \, \omega_\perp(t)^2 \, r^2.
\end{align}
Let us assume we know a stationary, real-valued solution $\Phi_0^{(0)}$ in the case where $\omega_\perp$ takes the constant value $\omega_\perp^{(0)}$. 
One can than look for a solution in the potential of Eq.~\eqref{eq:app:V_time_dep} with a density $\rho_0 \equiv \abs{\Phi_0}^2$ of the form
\begin{align}
\rho_0: (t,r) \left. \mapsto \frac{1}{\sigma(t)^2} \rho_0^{(0)} \left( \frac{x}{\sigma(t)} \right) \right., 
\end{align}
where $\rho_0^{(0)} \equiv \abs{\Phi_0}^2$. 
A straightforward calculation using the continuity equation gives the velocity $u_0 \equiv \partial_r \arg{\Phi_0}$ as
\begin{align*}
u_0: (t,r) \mapsto \frac{\sigma'(t)}{\sigma(t)} \, r. 
\end{align*}
Plugging $\Phi_0 = \sqrt{\rho_0} \, \ee^{\ii \int^r u_0 \, \dd r - \ii \mu \, t}$ in the GPE~\eqref{eq:GPE}, one finds a solution exists if and only if $\sigma$ obeys Eq.~\eqref{eq:sigma_eom}. The latter is thus an exact equation (as far as the GPE correctly describes the physics at play), independent of the Gaussian approximation. 


\section{Thermal effects in TOF experiments 
\label{app:Thermal}}

\subsection{DCE due to a one-time change in $c^{2}$}  

\begin{figure}
\includegraphics[width=0.45\columnwidth]{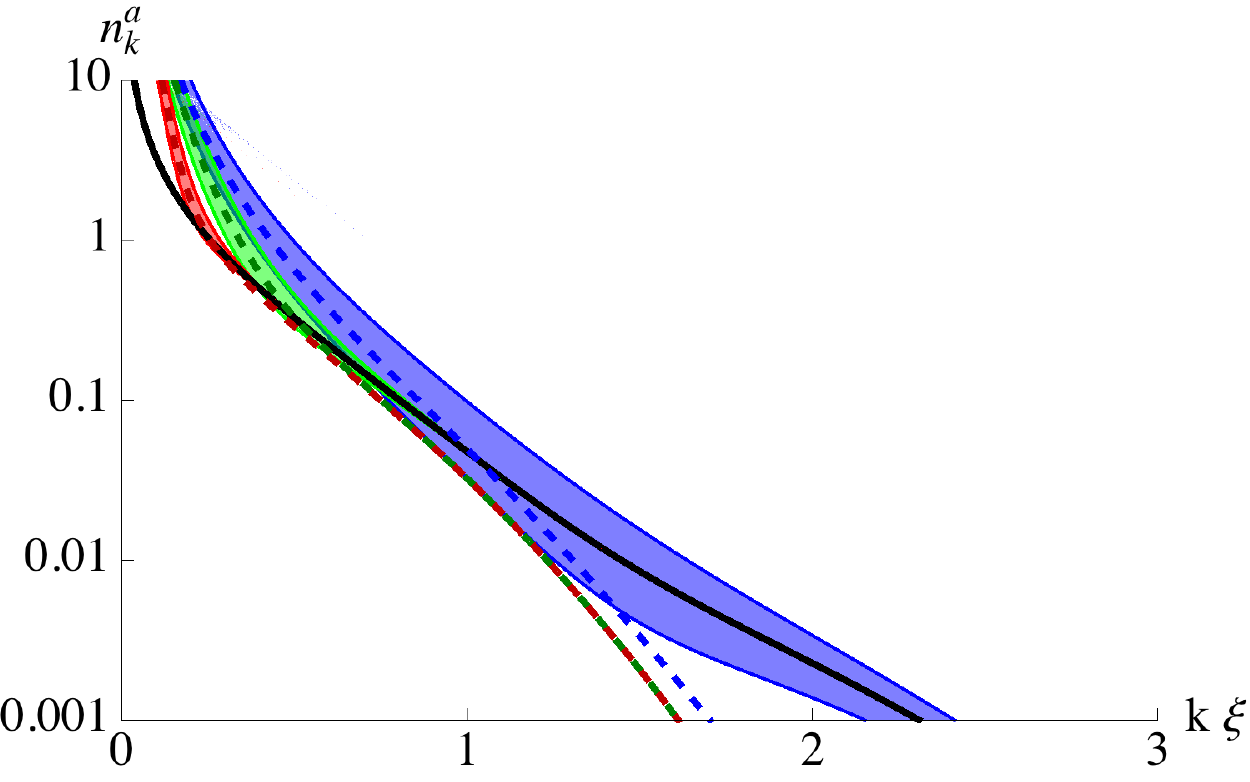} \, \includegraphics[width=0.45\columnwidth]{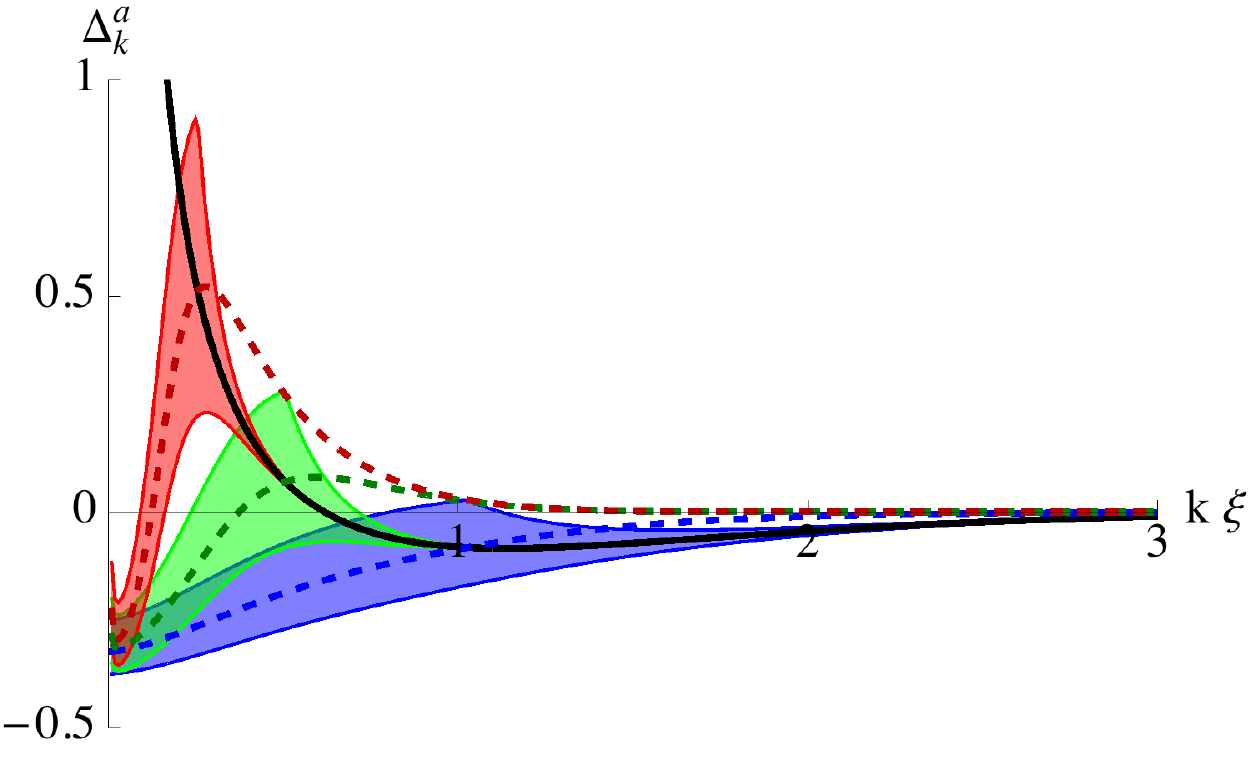}  
\caption{
The final state of the atoms after TOF when the initial state is a thermal bath of phonons.  As in the lower row of Fig. \ref{fig:TOF_InSitu_Opening}, the solid black curves show 
the state before the expansion (but after the initial DCE), 
$\xi$ is the healing length before the expansion, 
the colored dashed curves show the state that would result from the expansion alone, and the colored regions show the envelopes within which $n_{k}^{a}$ and $\Delta_{k}^{a}$ 
will oscillate.  The initial temperature is taken to be $T_{i}/mc_{i}^{2} = 0.5$, while the rates of opening the trap are the same as in Fig. \ref{fig:TOF_InSitu_Opening}: $a/\omega_{\perp} = \infty$ (i.e. sudden switch-off, blue), $0.6$ (green) and $0.2$ (red).
Interestingly, for $k \to 0$, the distribution becomes nonseparable with a value of $\Delta_{k}^{a}$ that is independent of $a/\omega_{\perp}$. 
\label{fig:TOF_InSitu_Opening_temp}}
\end{figure}

In Figure \ref{fig:TOF_InSitu_Opening_temp}, we show non-zero temperature versions of the plots after a slow opening of the trap shown in Fig. \ref{fig:TOF_InSitu_Opening}, but with an initial temperature 
(i.e. before the initial DCE) 
of $T/mc^{2} = 1/2$. Since an initial temperature boosts the final occupation number in exactly the same way for any DCE process, its effect can be viewed as rather trivial since it cannot change the outputs of the initial DCE and of the cloud expansion with respect to each other.  Whichever of these is dominant at zero temperature will therefore be dominant at any temperature, and where they strongly interfere at zero temperature, they will also strongly interfere at non-zero temperatures.  The overall effect is simply to boost $n_{k}^{a}$, 
and thus also to increase 
the nonseparability parameter $\Delta_{k}^{a}$. 
Notice in particular the behavior in the low-$k$ domain, where the atomic state after TOF can be unambiguously nonseparable even though the phononic state before TOF is separable.  Thus we see even more clearly than in Fig. \ref{fig:TOF_InSitu_Opening} how the opening of the trap can pollute the final state, and that control over the rate of expansion is vital if we are to be able to reconstruct the phononic state from TOF measurements.

\subsection{DCE due to a modulation of $c^{2}$}  

\begin{figure}
\includegraphics[width=0.45\columnwidth]{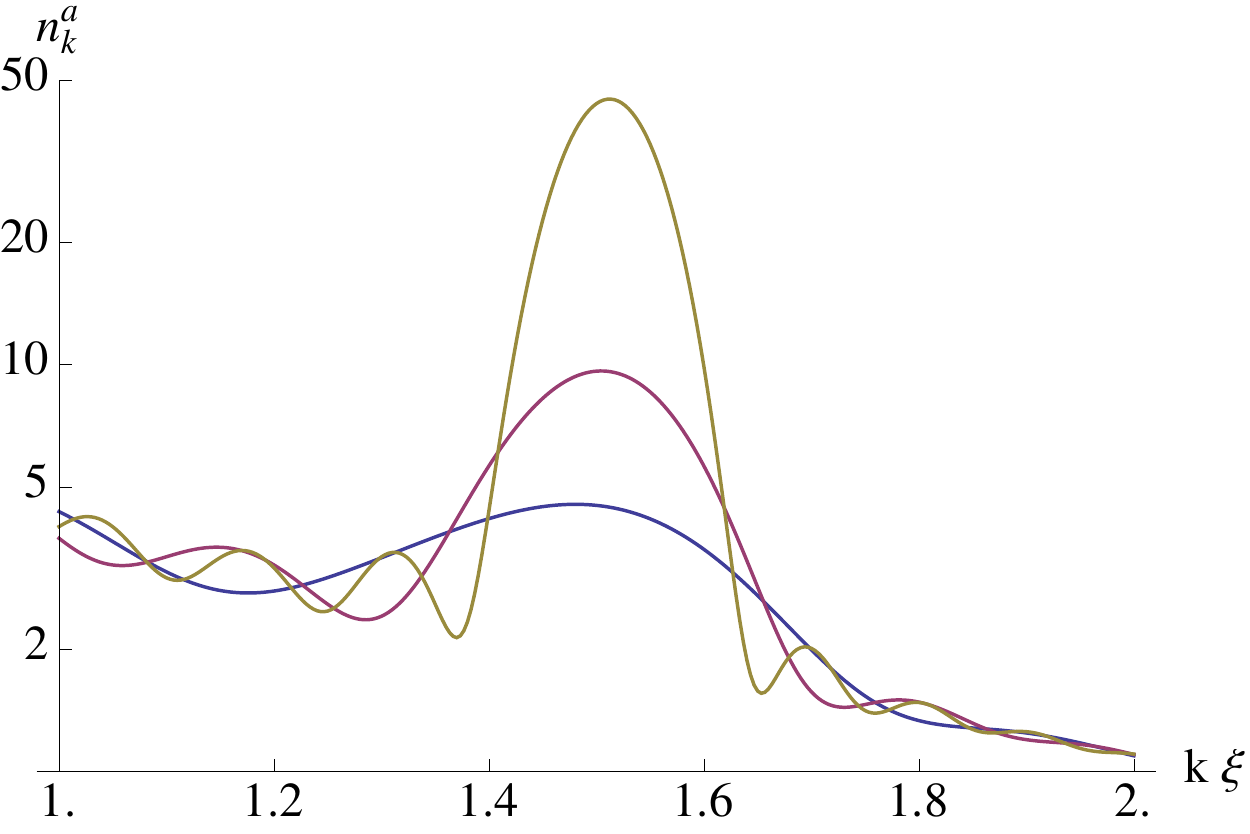} \, \includegraphics[width=0.45\columnwidth]{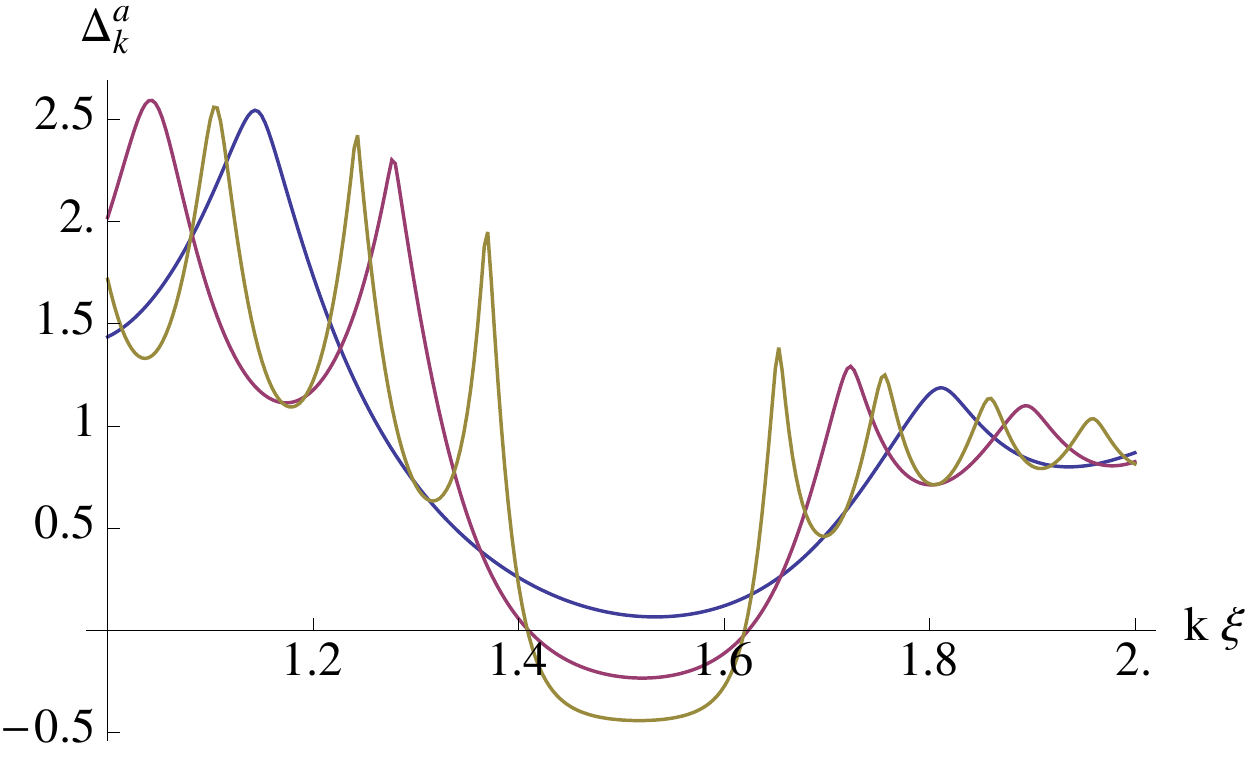}
\caption{Final state of atoms after a modulation of $c^{2}$, at non-zero temperature.  We consider the situation described in Ref.~\cite{Jaskula-et-al}. 
The parameters used are $\omega_{\perp f}/\omega_{\perp i} = \sqrt{2}$, $\omega_{\perp \, i}/mc^{2} = 1.5$, $\delta \left(c^{2}\right) / c^{2} = 0.5$ and $T/\omega_{\perp} = 2.8$.  $\omega_{\perp}$ is immediately switched off after an integer number of oscillations, after which the condensate expands freely.  On the left is shown the final occupation number of atoms, while on the right is shown the nonseparability parameter $\Delta_{k}^{a}$.  
The various curves correspond to a different number of oscillations of the condensate before the trap is opened: $N = 4$ (blue curves), $6$ (red curves) and $10$ (yellow curves).
Higher values of $N$ (namely 60, 30 and 15) were considered in Fig.~\ref{fig:InSitu_Jaskula}. 
\label{fig:Jaskula_temp}}
\end{figure}

When $c^{2}$ is modulated sinusoidally and $\left|\beta_{k}\right|^{2}$ grows exponentially at resonance, an initial temperature can have a very significant impact on the final state.  We have already seen, in Fig. \ref{fig:InSitu_Jaskula}, that introducing a temperature can increase the final occupation number so much that it is pushed into the nonlinear regime much earlier than it would have been if the initial temperature were zero.  Of course, temperature also affects the final degree of nonseparability.  In Figure \ref{fig:Jaskula_temp} are shown the final values of $n_{k}^{a}$ and $\Delta_{k}^{a}$ 
for an initial temperature $T/\omega_{\perp} = 2.8$ (corresponding to that in \cite{Jaskula-et-al}).  The various curves are for different total numbers of oscillations of the condensate, and in order to have reasonable values, these are much less than the actual value of around $60$: we have $N = 4$ (blue), $6$ (red) and $10$ (yellow).  Note that, at resonance, the state goes from separable to nonseparable after between $4$ and $6$ oscillations, whereas it would always be nonseparable if the initial temperature were zero.  (These values of $N$ 
are still small, mainly 
because the amplitude of the oscillations is large: $\delta\left(c^{2}\right)/c^{2} = 0.5$).

\begin{figure}
\includegraphics[width=0.45\columnwidth]{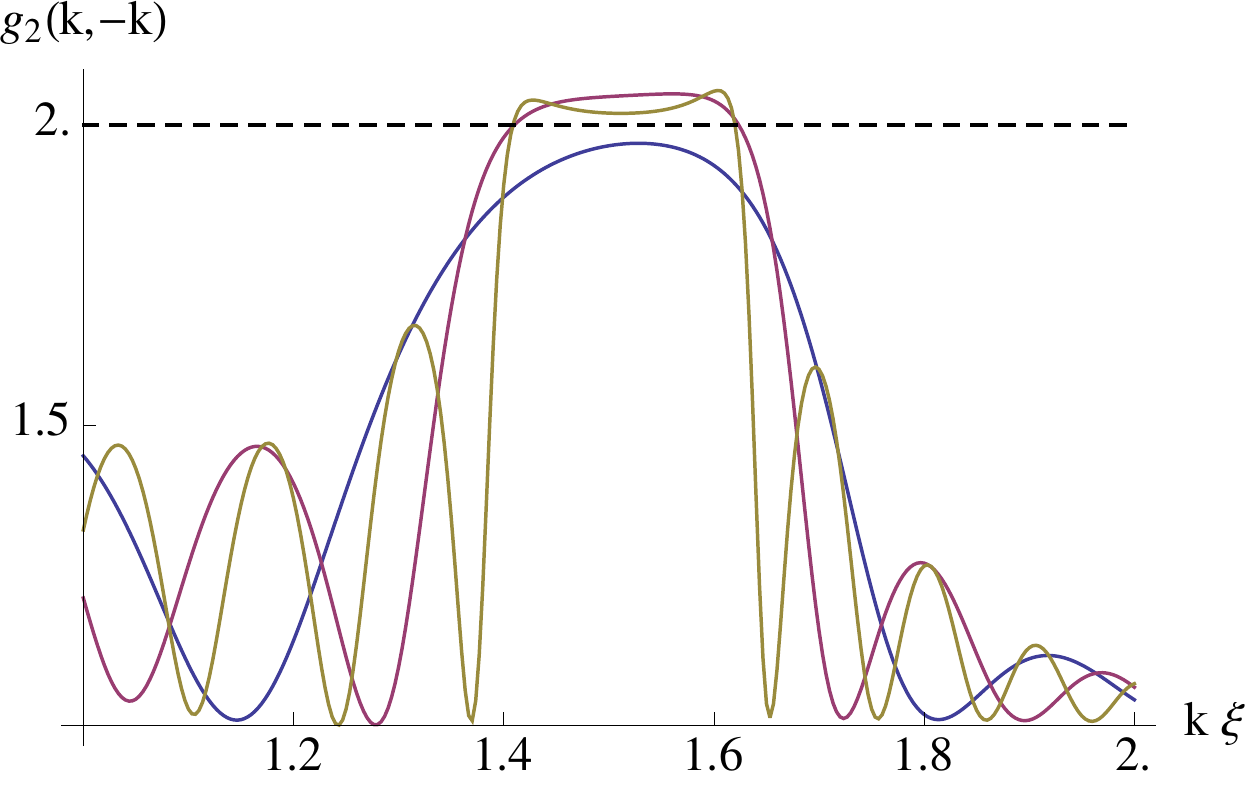} \, \includegraphics[width=0.45\columnwidth]{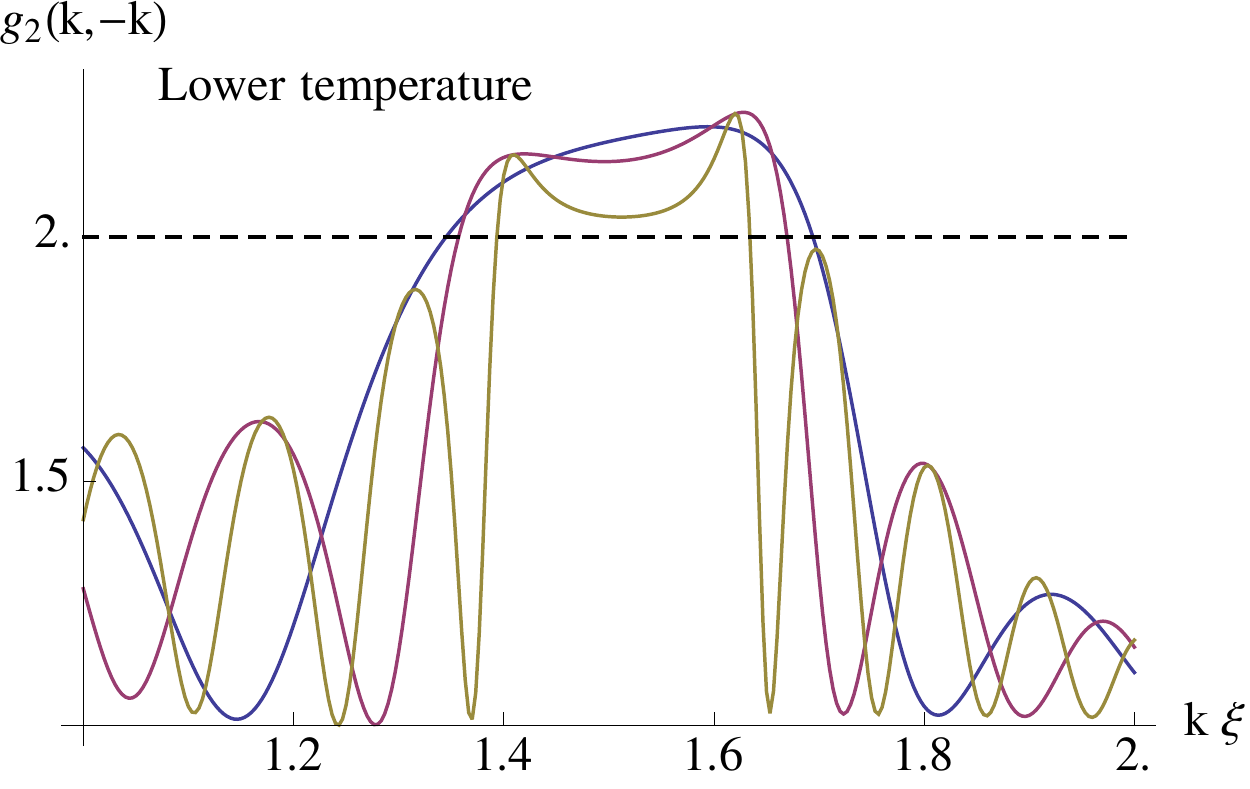}
\caption{Final two-atom correlation function, $g_{2}(k,-k)$.  On the left, the situation is exactly the same as in Fig. \ref{fig:Jaskula_temp}, whereas on the right the temperature has been divided by $2$, i.e. it has $T/\omega_{\perp} = 1.4$.  As in Fig. \ref{fig:Jaskula_temp}, the different curves correspond to different total numbers of oscillations of the condensate: $4$ (blue), $6$ (red) and $10$ (yellow).  Note that $g_{2}(k,-k)$ varies non-monotonically with $N$, particularly at resonance.  Thus, just as with the {\it in situ} density-density correlation function (see Fig. \ref{fig:InSitu_exp} and its caption), nonseparability becomes less visible with larger $n_{k}^{a}$. 
\label{fig:Jaskula_g2_temp}}
\end{figure}

To make a more explicit connection with experiment, in Figure \ref{fig:Jaskula_g2_temp} is plotted the two-atom correlation function:
\begin{equation}
g_{2}(k,-k) = \frac{ \langle \hat{a}_{k}^{\dagger} \hat{a}_{-k}^{\dagger} \hat{a}_{-k} \hat{a}_{k} \rangle }{ \langle \hat{a}_{k}^{\dagger} \hat{a}_{k} \rangle \, \langle \hat{a}_{-k}^{\dagger} \hat{a}_{-k} \rangle } =\frac{C_k^{a}}{n_k^{a} n_{-k}^{a}} \,. 
\label{eq:g2_defn}
\end{equation}
Using Eq.~(\ref{Deltasq}) it is clear that 
$g_{2}(k,-k)$ becomes larger than $2$ when $\Delta_{k}^{a} < 0$.  
There are two points to notice.  Firstly, the value of $g_{2}(k,-k)$, even at resonance, does not increase monotonically with increasing $N$ and, hence, with increasing $n_k^{a}$.  Indeed, as $n_k^{a}$ increases indefinitely, $g_{2}(k,-k)$ approaches $2$ from above.  Therefore, as was noted in Section 
\ref{sub:DCE_modulated}, large $n_k^{a}$ is not necessarily helpful if we want to observe nonseparability, for the nonseparability becomes less visible when the occupation number is large.  Secondly, the temperature can also have a large impact on the visibility of nonseparability.  (This is related to the first point, since a higher initial temperature requires a larger occupation number before the state actually becomes nonseparable, as was noted in \cite{Busch-Parentani-Robertson}.)  On the right of Fig. \ref{fig:Jaskula_g2_temp} is plotted $g_{2}(k,-k)$ with all parameters the same except for the temperature, which has been reduced by a half.  It is clear that $g_{2}(k,-k)$ reaches much further above $2$ than it does in the left plot.

\end{appendices}

\bibliography{biblio}

\begin{thebibliography}{53}%
\makeatletter
\providecommand \@ifxundefined [1]{%
 \@ifx{#1\undefined}
}%
\providecommand \@ifnum [1]{%
 \ifnum #1\expandafter \@firstoftwo
 \else \expandafter \@secondoftwo
 \fi
}%
\providecommand \@ifx [1]{%
 \ifx #1\expandafter \@firstoftwo
 \else \expandafter \@secondoftwo
 \fi
}%
\providecommand \natexlab [1]{#1}%
\providecommand \enquote  [1]{``#1''}%
\providecommand \bibnamefont  [1]{#1}%
\providecommand \bibfnamefont [1]{#1}%
\providecommand \citenamefont [1]{#1}%
\providecommand \href@noop [0]{\@secondoftwo}%
\providecommand \href [0]{\begingroup \@sanitize@url \@href}%
\providecommand \@href[1]{\@@startlink{#1}\@@href}%
\providecommand \@@href[1]{\endgroup#1\@@endlink}%
\providecommand \@sanitize@url [0]{\catcode `\\12\catcode `\$12\catcode
  `\&12\catcode `\#12\catcode `\^12\catcode `\_12\catcode `\%12\relax}%
\providecommand \@@startlink[1]{}%
\providecommand \@@endlink[0]{}%
\providecommand \url  [0]{\begingroup\@sanitize@url \@url }%
\providecommand \@url [1]{\endgroup\@href {#1}{\urlprefix }}%
\providecommand \urlprefix  [0]{URL }%
\providecommand \Eprint [0]{\href }%
\providecommand \doibase [0]{http://dx.doi.org/}%
\providecommand \selectlanguage [0]{\@gobble}%
\providecommand \bibinfo  [0]{\@secondoftwo}%
\providecommand \bibfield  [0]{\@secondoftwo}%
\providecommand \translation [1]{[#1]}%
\providecommand \BibitemOpen [0]{}%
\providecommand \bibitemStop [0]{}%
\providecommand \bibitemNoStop [0]{.\EOS\space}%
\providecommand \EOS [0]{\spacefactor3000\relax}%
\providecommand \BibitemShut  [1]{\csname bibitem#1\endcsname}%
\let\auto@bib@innerbib\@empty
\bibitem [{\citenamefont {Parker}(1968)}]{Parker-1968}%
  \BibitemOpen
  \bibfield  {author} {\bibinfo {author} {\bibfnamefont {L.}~\bibnamefont
  {Parker}},\ }\href {\doibase 10.1103/PhysRevLett.21.562} {\bibfield
  {journal} {\bibinfo  {journal} {Phys. Rev. Lett.}\ }\textbf {\bibinfo
  {volume} {21}},\ \bibinfo {pages} {562} (\bibinfo {year} {1968})}\BibitemShut
  {NoStop}%
\bibitem [{\citenamefont {Birrell}\ and\ \citenamefont
  {Davies}(1982)}]{Birrell-Davies}%
  \BibitemOpen
  \bibfield  {author} {\bibinfo {author} {\bibfnamefont {N.~D.}\ \bibnamefont
  {Birrell}}\ and\ \bibinfo {author} {\bibfnamefont {P.~C.~W.}\ \bibnamefont
  {Davies}},\ }\href@noop {} {\emph {\bibinfo {title} {Quantum fields in curved
  space}}}\ (\bibinfo  {publisher} {Cambridge University Press},\ \bibinfo
  {year} {1982})\BibitemShut {NoStop}%
\bibitem [{\citenamefont {Fulling}(1989)}]{Fulling}%
  \BibitemOpen
  \bibfield  {author} {\bibinfo {author} {\bibfnamefont {S.~A.}\ \bibnamefont
  {Fulling}},\ }\href@noop {} {\emph {\bibinfo {title} {Aspects of Quantum
  Field Theory in Curved Space-Time}}},\ London Mathematical Society Student
  Texts\ (\bibinfo  {publisher} {Cambridge University Press},\ \bibinfo {year}
  {1989})\BibitemShut {NoStop}%
\bibitem [{\citenamefont {Schwinger}(1951)}]{Schwinger-1951}%
  \BibitemOpen
  \bibfield  {author} {\bibinfo {author} {\bibfnamefont {J.}~\bibnamefont
  {Schwinger}},\ }\href {\doibase 10.1103/PhysRev.82.664} {\bibfield  {journal}
  {\bibinfo  {journal} {Phys. Rev.}\ }\textbf {\bibinfo {volume} {82}},\
  \bibinfo {pages} {664} (\bibinfo {year} {1951})}\BibitemShut {NoStop}%
\bibitem [{\citenamefont {Greiner}\ \emph {et~al.}(1985)\citenamefont
  {Greiner}, \citenamefont {M\"{u}ller},\ and\ \citenamefont
  {Rafelski}}]{Greiner}%
  \BibitemOpen
  \bibfield  {author} {\bibinfo {author} {\bibfnamefont {W.}~\bibnamefont
  {Greiner}}, \bibinfo {author} {\bibfnamefont {B.}~\bibnamefont {M\"{u}ller}},
  \ and\ \bibinfo {author} {\bibfnamefont {J.}~\bibnamefont {Rafelski}},\
  }\href@noop {} {\emph {\bibinfo {title} {Quantum Electrodynamics of Strong
  Fields}}},\ Theoretical and Mathematical Physics\ (\bibinfo  {publisher}
  {Springer-Verlag Berlin Heidelberg},\ \bibinfo {year} {1985})\BibitemShut
  {NoStop}%
\bibitem [{\citenamefont {Hawking}(1975)}]{Hawking-1975}%
  \BibitemOpen
  \bibfield  {author} {\bibinfo {author} {\bibfnamefont {S.~W.}\ \bibnamefont
  {Hawking}},\ }\href {\doibase 10.1007/BF02345020} {\bibfield  {journal}
  {\bibinfo  {journal} {Communications in Mathematical Physics}\ }\textbf
  {\bibinfo {volume} {43}},\ \bibinfo {pages} {199} (\bibinfo {year}
  {1975})}\BibitemShut {NoStop}%
\bibitem [{\citenamefont {Brout}\ \emph {et~al.}(1995)\citenamefont {Brout},
  \citenamefont {Massar}, \citenamefont {Parentani},\ and\ \citenamefont
  {Spindel}}]{Primer}%
  \BibitemOpen
  \bibfield  {author} {\bibinfo {author} {\bibfnamefont {R.}~\bibnamefont
  {Brout}}, \bibinfo {author} {\bibfnamefont {S.}~\bibnamefont {Massar}},
  \bibinfo {author} {\bibfnamefont {R.}~\bibnamefont {Parentani}}, \ and\
  \bibinfo {author} {\bibfnamefont {P.}~\bibnamefont {Spindel}},\ }\href
  {\doibase http://dx.doi.org/10.1016/0370-1573(95)00008-5} {\bibfield
  {journal} {\bibinfo  {journal} {Physics Reports}\ }\textbf {\bibinfo {volume}
  {260}},\ \bibinfo {pages} {329 } (\bibinfo {year} {1995})}\BibitemShut
  {NoStop}%
\bibitem [{\citenamefont {Unruh}(1981)}]{Unruh-1981}%
  \BibitemOpen
  \bibfield  {author} {\bibinfo {author} {\bibfnamefont {W.~G.}\ \bibnamefont
  {Unruh}},\ }\href {\doibase 10.1103/PhysRevLett.46.1351} {\bibfield
  {journal} {\bibinfo  {journal} {Phys. Rev. Lett.}\ }\textbf {\bibinfo
  {volume} {46}},\ \bibinfo {pages} {1351} (\bibinfo {year}
  {1981})}\BibitemShut {NoStop}%
\bibitem [{\citenamefont {Barcel{\'o}}\ \emph {et~al.}(2011)\citenamefont
  {Barcel{\'o}}, \citenamefont {Liberati},\ and\ \citenamefont
  {Visser}}]{AnalogueGravity-LivingReview}%
  \BibitemOpen
  \bibfield  {author} {\bibinfo {author} {\bibfnamefont {C.}~\bibnamefont
  {Barcel{\'o}}}, \bibinfo {author} {\bibfnamefont {S.}~\bibnamefont
  {Liberati}}, \ and\ \bibinfo {author} {\bibfnamefont {M.}~\bibnamefont
  {Visser}},\ }\href {\doibase 10.12942/lrr-2011-3} {\bibfield  {journal}
  {\bibinfo  {journal} {Living Reviews in Relativity}\ }\textbf {\bibinfo
  {volume} {14}},\ \bibinfo {pages} {3} (\bibinfo {year} {2011})}\BibitemShut
  {NoStop}%
\bibitem [{\citenamefont {Wilson}\ \emph {et~al.}(2011)\citenamefont {Wilson},
  \citenamefont {Johansson}, \citenamefont {Pourkabirian}, \citenamefont
  {Simoen}, \citenamefont {Johansson}, \citenamefont {Duty}, \citenamefont
  {Nori},\ and\ \citenamefont {Delsing}}]{Wilson-et-al-2011}%
  \BibitemOpen
  \bibfield  {author} {\bibinfo {author} {\bibfnamefont {C.~M.}\ \bibnamefont
  {Wilson}}, \bibinfo {author} {\bibfnamefont {G.}~\bibnamefont {Johansson}},
  \bibinfo {author} {\bibfnamefont {A.}~\bibnamefont {Pourkabirian}}, \bibinfo
  {author} {\bibfnamefont {M.}~\bibnamefont {Simoen}}, \bibinfo {author}
  {\bibfnamefont {J.~R.}\ \bibnamefont {Johansson}}, \bibinfo {author}
  {\bibfnamefont {T.}~\bibnamefont {Duty}}, \bibinfo {author} {\bibfnamefont
  {F.}~\bibnamefont {Nori}}, \ and\ \bibinfo {author} {\bibfnamefont
  {P.}~\bibnamefont {Delsing}},\ }\href
  {http://www.nature.com.proxy.scd.u-psud.fr/nature/journal/v479/n7373/full/nature10561.html}
  {\bibfield  {journal} {\bibinfo  {journal} {Nature}\ }\textbf {\bibinfo
  {volume} {479}},\ \bibinfo {pages} {376} (\bibinfo {year}
  {2011})}\BibitemShut {NoStop}%
\bibitem [{\citenamefont {Jaskula}\ \emph {et~al.}(2012)\citenamefont
  {Jaskula}, \citenamefont {Partridge}, \citenamefont {Bonneau}, \citenamefont
  {Lopes}, \citenamefont {Ruaudel}, \citenamefont {Boiron},\ and\ \citenamefont
  {Westbrook}}]{Jaskula-et-al}%
  \BibitemOpen
  \bibfield  {author} {\bibinfo {author} {\bibfnamefont {J.-C.}\ \bibnamefont
  {Jaskula}}, \bibinfo {author} {\bibfnamefont {G.~B.}\ \bibnamefont
  {Partridge}}, \bibinfo {author} {\bibfnamefont {M.}~\bibnamefont {Bonneau}},
  \bibinfo {author} {\bibfnamefont {R.}~\bibnamefont {Lopes}}, \bibinfo
  {author} {\bibfnamefont {J.}~\bibnamefont {Ruaudel}}, \bibinfo {author}
  {\bibfnamefont {D.}~\bibnamefont {Boiron}}, \ and\ \bibinfo {author}
  {\bibfnamefont {C.~I.}\ \bibnamefont {Westbrook}},\ }\href {\doibase
  10.1103/PhysRevLett.109.220401} {\bibfield  {journal} {\bibinfo  {journal}
  {Phys. Rev. Lett.}\ }\textbf {\bibinfo {volume} {109}},\ \bibinfo {pages}
  {220401} (\bibinfo {year} {2012})}\BibitemShut {NoStop}%
\bibitem [{\citenamefont {L\"{a}hteenm\"{a}ki}\ \emph
  {et~al.}(2013)\citenamefont {L\"{a}hteenm\"{a}ki}, \citenamefont {Paraoanu},
  \citenamefont {Hassel},\ and\ \citenamefont {Hakonen}}]{Lahteenmaki-et-al}%
  \BibitemOpen
  \bibfield  {author} {\bibinfo {author} {\bibfnamefont {P.}~\bibnamefont
  {L\"{a}hteenm\"{a}ki}}, \bibinfo {author} {\bibfnamefont {G.~S.}\
  \bibnamefont {Paraoanu}}, \bibinfo {author} {\bibfnamefont {J.}~\bibnamefont
  {Hassel}}, \ and\ \bibinfo {author} {\bibfnamefont {P.~J.}\ \bibnamefont
  {Hakonen}},\ }\href {\doibase 10.1073/pnas.1212705110} {\bibfield  {journal}
  {\bibinfo  {journal} {Proceedings of the National Academy of Sciences}\
  }\textbf {\bibinfo {volume} {110}},\ \bibinfo {pages} {4234} (\bibinfo {year}
  {2013})}\BibitemShut {NoStop}%
\bibitem [{\citenamefont {Lahav}\ \emph {et~al.}(2010)\citenamefont {Lahav},
  \citenamefont {Itah}, \citenamefont {Blumkin}, \citenamefont {Gordon},
  \citenamefont {Rinott}, \citenamefont {Zayats},\ and\ \citenamefont
  {Steinhauer}}]{Lahav-et-al-2010}%
  \BibitemOpen
  \bibfield  {author} {\bibinfo {author} {\bibfnamefont {O.}~\bibnamefont
  {Lahav}}, \bibinfo {author} {\bibfnamefont {A.}~\bibnamefont {Itah}},
  \bibinfo {author} {\bibfnamefont {A.}~\bibnamefont {Blumkin}}, \bibinfo
  {author} {\bibfnamefont {C.}~\bibnamefont {Gordon}}, \bibinfo {author}
  {\bibfnamefont {S.}~\bibnamefont {Rinott}}, \bibinfo {author} {\bibfnamefont
  {A.}~\bibnamefont {Zayats}}, \ and\ \bibinfo {author} {\bibfnamefont
  {J.}~\bibnamefont {Steinhauer}},\ }\href {\doibase
  10.1103/PhysRevLett.105.240401} {\bibfield  {journal} {\bibinfo  {journal}
  {Phys. Rev. Lett.}\ }\textbf {\bibinfo {volume} {105}},\ \bibinfo {pages}
  {240401} (\bibinfo {year} {2010})}\BibitemShut {NoStop}%
\bibitem [{\citenamefont {Nguyen}\ \emph {et~al.}(2015)\citenamefont {Nguyen},
  \citenamefont {Gerace}, \citenamefont {Carusotto}, \citenamefont {Sanvitto},
  \citenamefont {Galopin}, \citenamefont {Lema\^{\i}tre}, \citenamefont
  {Sagnes}, \citenamefont {Bloch},\ and\ \citenamefont {Amo}}]{Nguyen-et-al}%
  \BibitemOpen
  \bibfield  {author} {\bibinfo {author} {\bibfnamefont {H.~S.}\ \bibnamefont
  {Nguyen}}, \bibinfo {author} {\bibfnamefont {D.}~\bibnamefont {Gerace}},
  \bibinfo {author} {\bibfnamefont {I.}~\bibnamefont {Carusotto}}, \bibinfo
  {author} {\bibfnamefont {D.}~\bibnamefont {Sanvitto}}, \bibinfo {author}
  {\bibfnamefont {E.}~\bibnamefont {Galopin}}, \bibinfo {author} {\bibfnamefont
  {A.}~\bibnamefont {Lema\^{\i}tre}}, \bibinfo {author} {\bibfnamefont
  {I.}~\bibnamefont {Sagnes}}, \bibinfo {author} {\bibfnamefont
  {J.}~\bibnamefont {Bloch}}, \ and\ \bibinfo {author} {\bibfnamefont
  {A.}~\bibnamefont {Amo}},\ }\href {\doibase 10.1103/PhysRevLett.114.036402}
  {\bibfield  {journal} {\bibinfo  {journal} {Phys. Rev. Lett.}\ }\textbf
  {\bibinfo {volume} {114}},\ \bibinfo {pages} {036402} (\bibinfo {year}
  {2015})}\BibitemShut {NoStop}%
\bibitem [{\citenamefont {Steinhauer}(2016)}]{Steinhauer-2016}%
  \BibitemOpen
  \bibfield  {author} {\bibinfo {author} {\bibfnamefont {J.}~\bibnamefont
  {Steinhauer}},\ }\href
  {https://www.nature.com/nphys/journal/v12/n10/abs/nphys3863.html} {\bibfield
  {journal} {\bibinfo  {journal} {Nat. Phys.}\ }\textbf {\bibinfo {volume}
  {12}},\ \bibinfo {pages} {959} (\bibinfo {year} {2016})}\BibitemShut
  {NoStop}%
\bibitem [{\citenamefont {Fedichev}\ and\ \citenamefont
  {Fischer}(2004)}]{Fedichev-Fischer}%
  \BibitemOpen
  \bibfield  {author} {\bibinfo {author} {\bibfnamefont {P.~O.}\ \bibnamefont
  {Fedichev}}\ and\ \bibinfo {author} {\bibfnamefont {U.~R.}\ \bibnamefont
  {Fischer}},\ }\href {\doibase 10.1103/PhysRevA.69.033602} {\bibfield
  {journal} {\bibinfo  {journal} {Phys. Rev. A}\ }\textbf {\bibinfo {volume}
  {69}},\ \bibinfo {pages} {033602} (\bibinfo {year} {2004})}\BibitemShut
  {NoStop}%
\bibitem [{\citenamefont {Carusotto}\ \emph {et~al.}(2010)\citenamefont
  {Carusotto}, \citenamefont {Balbinot}, \citenamefont {Fabbri},\ and\
  \citenamefont {Recati}}]{Carusotto-DCE}%
  \BibitemOpen
  \bibfield  {author} {\bibinfo {author} {\bibfnamefont {I.}~\bibnamefont
  {Carusotto}}, \bibinfo {author} {\bibfnamefont {R.}~\bibnamefont {Balbinot}},
  \bibinfo {author} {\bibfnamefont {A.}~\bibnamefont {Fabbri}}, \ and\ \bibinfo
  {author} {\bibfnamefont {A.}~\bibnamefont {Recati}},\ }\href {\doibase
  10.1140/epjd/e2009-00314-3} {\bibfield  {journal} {\bibinfo  {journal} {The
  European Physical Journal D}\ }\textbf {\bibinfo {volume} {56}},\ \bibinfo
  {pages} {391} (\bibinfo {year} {2010})}\BibitemShut {NoStop}%
\bibitem [{\citenamefont {Werner}(1989)}]{Werner}%
  \BibitemOpen
  \bibfield  {author} {\bibinfo {author} {\bibfnamefont {R.~F.}\ \bibnamefont
  {Werner}},\ }\href {\doibase 10.1103/PhysRevA.40.4277} {\bibfield  {journal}
  {\bibinfo  {journal} {Phys. Rev. A}\ }\textbf {\bibinfo {volume} {40}},\
  \bibinfo {pages} {4277} (\bibinfo {year} {1989})}\BibitemShut {NoStop}%
\bibitem [{\citenamefont {Simon}(2000)}]{Simon}%
  \BibitemOpen
  \bibfield  {author} {\bibinfo {author} {\bibfnamefont {R.}~\bibnamefont
  {Simon}},\ }\href {\doibase 10.1103/PhysRevLett.84.2726} {\bibfield
  {journal} {\bibinfo  {journal} {Phys. Rev. Lett.}\ }\textbf {\bibinfo
  {volume} {84}},\ \bibinfo {pages} {2726} (\bibinfo {year}
  {2000})}\BibitemShut {NoStop}%
\bibitem [{\citenamefont {Horodecki}\ \emph {et~al.}(2009)\citenamefont
  {Horodecki}, \citenamefont {Horodecki}, \citenamefont {Horodecki},\ and\
  \citenamefont {Horodecki}}]{Horodecki-Review}%
  \BibitemOpen
  \bibfield  {author} {\bibinfo {author} {\bibfnamefont {R.}~\bibnamefont
  {Horodecki}}, \bibinfo {author} {\bibfnamefont {P.}~\bibnamefont
  {Horodecki}}, \bibinfo {author} {\bibfnamefont {M.}~\bibnamefont
  {Horodecki}}, \ and\ \bibinfo {author} {\bibfnamefont {K.}~\bibnamefont
  {Horodecki}},\ }\href {\doibase 10.1103/RevModPhys.81.865} {\bibfield
  {journal} {\bibinfo  {journal} {Rev. Mod. Phys.}\ }\textbf {\bibinfo {volume}
  {81}},\ \bibinfo {pages} {865} (\bibinfo {year} {2009})}\BibitemShut
  {NoStop}%
\bibitem [{\citenamefont {Busch}\ and\ \citenamefont
  {Parentani}(2013)}]{Busch-Parentani-2013}%
  \BibitemOpen
  \bibfield  {author} {\bibinfo {author} {\bibfnamefont {X.}~\bibnamefont
  {Busch}}\ and\ \bibinfo {author} {\bibfnamefont {R.}~\bibnamefont
  {Parentani}},\ }\href {\doibase 10.1103/PhysRevD.88.045023} {\bibfield
  {journal} {\bibinfo  {journal} {Phys. Rev. D}\ }\textbf {\bibinfo {volume}
  {88}},\ \bibinfo {pages} {045023} (\bibinfo {year} {2013})}\BibitemShut
  {NoStop}%
\bibitem [{\citenamefont {Busch}\ \emph
  {et~al.}(2014{\natexlab{a}})\citenamefont {Busch}, \citenamefont
  {Parentani},\ and\ \citenamefont {Robertson}}]{Busch-Parentani-Robertson}%
  \BibitemOpen
  \bibfield  {author} {\bibinfo {author} {\bibfnamefont {X.}~\bibnamefont
  {Busch}}, \bibinfo {author} {\bibfnamefont {R.}~\bibnamefont {Parentani}}, \
  and\ \bibinfo {author} {\bibfnamefont {S.}~\bibnamefont {Robertson}},\ }\href
  {\doibase 10.1103/PhysRevA.89.063606} {\bibfield  {journal} {\bibinfo
  {journal} {Phys. Rev. A}\ }\textbf {\bibinfo {volume} {89}},\ \bibinfo
  {pages} {063606} (\bibinfo {year} {2014}{\natexlab{a}})}\BibitemShut
  {NoStop}%
\bibitem [{\citenamefont {Busch}\ \emph
  {et~al.}(2014{\natexlab{b}})\citenamefont {Busch}, \citenamefont
  {Carusotto},\ and\ \citenamefont {Parentani}}]{Busch-Carusotto-Parentani}%
  \BibitemOpen
  \bibfield  {author} {\bibinfo {author} {\bibfnamefont {X.}~\bibnamefont
  {Busch}}, \bibinfo {author} {\bibfnamefont {I.}~\bibnamefont {Carusotto}}, \
  and\ \bibinfo {author} {\bibfnamefont {R.}~\bibnamefont {Parentani}},\ }\href
  {\doibase 10.1103/PhysRevA.89.043819} {\bibfield  {journal} {\bibinfo
  {journal} {Phys. Rev. A}\ }\textbf {\bibinfo {volume} {89}},\ \bibinfo
  {pages} {043819} (\bibinfo {year} {2014}{\natexlab{b}})}\BibitemShut
  {NoStop}%
\bibitem [{\citenamefont {\'{S}wis{\l}ocki}\ and\ \citenamefont
  {Deuar}(2016)}]{Swislocki-Deuar}%
  \BibitemOpen
  \bibfield  {author} {\bibinfo {author} {\bibfnamefont {T.}~\bibnamefont
  {\'{S}wis{\l}ocki}}\ and\ \bibinfo {author} {\bibfnamefont {P.}~\bibnamefont
  {Deuar}},\ }\href {http://stacks.iop.org/0953-4075/49/i=14/a=145303}
  {\bibfield  {journal} {\bibinfo  {journal} {J. Phys. B}\ }\textbf {\bibinfo
  {volume} {49}},\ \bibinfo {pages} {145303} (\bibinfo {year}
  {2016})}\BibitemShut {NoStop}%
\bibitem [{\citenamefont {Zi\'{n}}\ and\ \citenamefont
  {Pylak}(2017)}]{Zin-Pylak}%
  \BibitemOpen
  \bibfield  {author} {\bibinfo {author} {\bibfnamefont {P.}~\bibnamefont
  {Zi\'{n}}}\ and\ \bibinfo {author} {\bibfnamefont {M.}~\bibnamefont
  {Pylak}},\ }\href {http://stacks.iop.org/0953-4075/50/i=8/a=085301}
  {\bibfield  {journal} {\bibinfo  {journal} {J. Phys. B}\ }\textbf {\bibinfo
  {volume} {50}},\ \bibinfo {pages} {085301} (\bibinfo {year}
  {2017})}\BibitemShut {NoStop}%
\bibitem [{\citenamefont {Campo}\ and\ \citenamefont
  {Parentani}(2005)}]{Campo-Parentani-2005}%
  \BibitemOpen
  \bibfield  {author} {\bibinfo {author} {\bibfnamefont {D.}~\bibnamefont
  {Campo}}\ and\ \bibinfo {author} {\bibfnamefont {R.}~\bibnamefont
  {Parentani}},\ }\href {\doibase 10.1103/PhysRevD.72.045015} {\bibfield
  {journal} {\bibinfo  {journal} {Phys. Rev. D}\ }\textbf {\bibinfo {volume}
  {72}},\ \bibinfo {pages} {045015} (\bibinfo {year} {2005})}\BibitemShut
  {NoStop}%
\bibitem [{\citenamefont {Campo}\ and\ \citenamefont
  {Parentani}(2008{\natexlab{a}})}]{Campo-Parentani-2008-I}%
  \BibitemOpen
  \bibfield  {author} {\bibinfo {author} {\bibfnamefont {D.}~\bibnamefont
  {Campo}}\ and\ \bibinfo {author} {\bibfnamefont {R.}~\bibnamefont
  {Parentani}},\ }\href {\doibase 10.1103/PhysRevD.78.065044} {\bibfield
  {journal} {\bibinfo  {journal} {Phys. Rev. D}\ }\textbf {\bibinfo {volume}
  {78}},\ \bibinfo {pages} {065044} (\bibinfo {year}
  {2008}{\natexlab{a}})}\BibitemShut {NoStop}%
\bibitem [{\citenamefont {Campo}\ and\ \citenamefont
  {Parentani}(2008{\natexlab{b}})}]{Campo-Parentani-2008-II}%
  \BibitemOpen
  \bibfield  {author} {\bibinfo {author} {\bibfnamefont {D.}~\bibnamefont
  {Campo}}\ and\ \bibinfo {author} {\bibfnamefont {R.}~\bibnamefont
  {Parentani}},\ }\href {\doibase 10.1103/PhysRevD.78.065045} {\bibfield
  {journal} {\bibinfo  {journal} {Phys. Rev. D}\ }\textbf {\bibinfo {volume}
  {78}},\ \bibinfo {pages} {065045} (\bibinfo {year}
  {2008}{\natexlab{b}})}\BibitemShut {NoStop}%
\bibitem [{\citenamefont {Adamek}\ \emph {et~al.}(2013)\citenamefont {Adamek},
  \citenamefont {Busch},\ and\ \citenamefont
  {Parentani}}]{Adamek-Busch-Parentani}%
  \BibitemOpen
  \bibfield  {author} {\bibinfo {author} {\bibfnamefont {J.}~\bibnamefont
  {Adamek}}, \bibinfo {author} {\bibfnamefont {X.}~\bibnamefont {Busch}}, \
  and\ \bibinfo {author} {\bibfnamefont {R.}~\bibnamefont {Parentani}},\ }\href
  {\doibase 10.1103/PhysRevD.87.124039} {\bibfield  {journal} {\bibinfo
  {journal} {Phys. Rev. D}\ }\textbf {\bibinfo {volume} {87}},\ \bibinfo
  {pages} {124039} (\bibinfo {year} {2013})}\BibitemShut {NoStop}%
\bibitem [{\citenamefont {Pitaevskii}\ and\ \citenamefont
  {Stringari}(2003)}]{Pitaevskii-Stringari-BEC}%
  \BibitemOpen
  \bibfield  {author} {\bibinfo {author} {\bibfnamefont {L.~P.}\ \bibnamefont
  {Pitaevskii}}\ and\ \bibinfo {author} {\bibfnamefont {S.}~\bibnamefont
  {Stringari}},\ }\href@noop {} {\emph {\bibinfo {title} {Bose-Einstein
  Condensation}}}\ (\bibinfo  {publisher} {Oxford University Press},\ \bibinfo
  {year} {2003})\BibitemShut {NoStop}%
\bibitem [{\citenamefont {Robertson}\ \emph {et~al.}(2017)\citenamefont
  {Robertson}, \citenamefont {Michel},\ and\ \citenamefont
  {Parentani}}]{Robertson-Michel-Parentani-steering}%
  \BibitemOpen
  \bibfield  {author} {\bibinfo {author} {\bibfnamefont {S.}~\bibnamefont
  {Robertson}}, \bibinfo {author} {\bibfnamefont {F.}~\bibnamefont {Michel}}, \
  and\ \bibinfo {author} {\bibfnamefont {R.}~\bibnamefont {Parentani}},\ }\href
  {https://arxiv.org/abs/1705.06648} {\bibfield  {journal} {\bibinfo  {journal}
  {arXiv:1705.06648}\ } (\bibinfo {year} {2017})}\BibitemShut {NoStop}%
\bibitem [{\citenamefont {Menotti}\ and\ \citenamefont
  {Stringari}(2002)}]{Menotti-Stringari}%
  \BibitemOpen
  \bibfield  {author} {\bibinfo {author} {\bibfnamefont {C.}~\bibnamefont
  {Menotti}}\ and\ \bibinfo {author} {\bibfnamefont {S.}~\bibnamefont
  {Stringari}},\ }\href {\doibase 10.1103/PhysRevA.66.043610} {\bibfield
  {journal} {\bibinfo  {journal} {Phys. Rev. A}\ }\textbf {\bibinfo {volume}
  {66}},\ \bibinfo {pages} {043610} (\bibinfo {year} {2002})}\BibitemShut
  {NoStop}%
\bibitem [{\citenamefont {Tozzo}\ and\ \citenamefont
  {Dalfovo}(2004)}]{Tozzo-Dalfovo}%
  \BibitemOpen
  \bibfield  {author} {\bibinfo {author} {\bibfnamefont {C.}~\bibnamefont
  {Tozzo}}\ and\ \bibinfo {author} {\bibfnamefont {F.}~\bibnamefont
  {Dalfovo}},\ }\href {\doibase 10.1103/PhysRevA.69.053606} {\bibfield
  {journal} {\bibinfo  {journal} {Phys. Rev. A}\ }\textbf {\bibinfo {volume}
  {69}},\ \bibinfo {pages} {053606} (\bibinfo {year} {2004})}\BibitemShut
  {NoStop}%
\bibitem [{\citenamefont {Gerbier}(2004)}]{Gerbier}%
  \BibitemOpen
  \bibfield  {author} {\bibinfo {author} {\bibfnamefont {F.}~\bibnamefont
  {Gerbier}},\ }\href
  {http://iopscience.iop.org/article/10.1209/epl/i2004-10035-7?pageTitle=IOPscience}
  {\bibfield  {journal} {\bibinfo  {journal} {Europhys. Lett.}\ }\textbf
  {\bibinfo {volume} {66}},\ \bibinfo {pages} {771} (\bibinfo {year}
  {2004})}\BibitemShut {NoStop}%
\bibitem [{\citenamefont {Kagan}\ \emph {et~al.}(1996)\citenamefont {Kagan},
  \citenamefont {Surkov},\ and\ \citenamefont
  {Shlyapnikov}}]{Kagan-Surkov-Shlyapnikov}%
  \BibitemOpen
  \bibfield  {author} {\bibinfo {author} {\bibfnamefont {Y.}~\bibnamefont
  {Kagan}}, \bibinfo {author} {\bibfnamefont {E.~L.}\ \bibnamefont {Surkov}}, \
  and\ \bibinfo {author} {\bibfnamefont {G.~V.}\ \bibnamefont {Shlyapnikov}},\
  }\href {\doibase 10.1103/PhysRevA.54.R1753} {\bibfield  {journal} {\bibinfo
  {journal} {Phys. Rev. A}\ }\textbf {\bibinfo {volume} {54}},\ \bibinfo
  {pages} {R1753} (\bibinfo {year} {1996})}\BibitemShut {NoStop}%
\bibitem [{\citenamefont {Dalfovo}\ \emph {et~al.}(1999)\citenamefont
  {Dalfovo}, \citenamefont {Giorgini}, \citenamefont {Pitaevskii},\ and\
  \citenamefont {Stringari}}]{Dalfovo-et-al-1999}%
  \BibitemOpen
  \bibfield  {author} {\bibinfo {author} {\bibfnamefont {F.}~\bibnamefont
  {Dalfovo}}, \bibinfo {author} {\bibfnamefont {S.}~\bibnamefont {Giorgini}},
  \bibinfo {author} {\bibfnamefont {L.~P.}\ \bibnamefont {Pitaevskii}}, \ and\
  \bibinfo {author} {\bibfnamefont {S.}~\bibnamefont {Stringari}},\ }\href
  {https://journals.aps.org/rmp/abstract/10.1103/RevModPhys.71.463} {\bibfield
  {journal} {\bibinfo  {journal} {Rev. Mod. Phys.}\ }\textbf {\bibinfo {volume}
  {71}},\ \bibinfo {pages} {463} (\bibinfo {year} {1999})}\BibitemShut
  {NoStop}%
\bibitem [{\citenamefont {Petrov}\ \emph {et~al.}(2000)\citenamefont {Petrov},
  \citenamefont {Shlyapnikov},\ and\ \citenamefont
  {Walraven}}]{Petrov-Shlyapnikov-Walraven}%
  \BibitemOpen
  \bibfield  {author} {\bibinfo {author} {\bibfnamefont {D.~S.}\ \bibnamefont
  {Petrov}}, \bibinfo {author} {\bibfnamefont {G.~V.}\ \bibnamefont
  {Shlyapnikov}}, \ and\ \bibinfo {author} {\bibfnamefont {J.~T.~M.}\
  \bibnamefont {Walraven}},\ }\href {\doibase 10.1103/PhysRevLett.85.3745}
  {\bibfield  {journal} {\bibinfo  {journal} {Phys. Rev. Lett.}\ }\textbf
  {\bibinfo {volume} {85}},\ \bibinfo {pages} {3745} (\bibinfo {year}
  {2000})}\BibitemShut {NoStop}%
\bibitem [{\citenamefont {Campo}\ and\ \citenamefont
  {Parentani}(2006)}]{Campo-Parentani-2006}%
  \BibitemOpen
  \bibfield  {author} {\bibinfo {author} {\bibfnamefont {D.}~\bibnamefont
  {Campo}}\ and\ \bibinfo {author} {\bibfnamefont {R.}~\bibnamefont
  {Parentani}},\ }\href {\doibase 10.1103/PhysRevD.74.025001} {\bibfield
  {journal} {\bibinfo  {journal} {Phys. Rev. D}\ }\textbf {\bibinfo {volume}
  {74}},\ \bibinfo {pages} {025001} (\bibinfo {year} {2006})}\BibitemShut
  {NoStop}%
\bibitem [{\citenamefont {de~Nova}\ \emph {et~al.}(2015)\citenamefont
  {de~Nova}, \citenamefont {Sols},\ and\ \citenamefont
  {Zapata}}]{deNova-Sols-Zapata}%
  \BibitemOpen
  \bibfield  {author} {\bibinfo {author} {\bibfnamefont {J.~R.~M.}\
  \bibnamefont {de~Nova}}, \bibinfo {author} {\bibfnamefont {F.}~\bibnamefont
  {Sols}}, \ and\ \bibinfo {author} {\bibfnamefont {I.}~\bibnamefont
  {Zapata}},\ }\href {http://stacks.iop.org/1367-2630/17/i=10/a=105003}
  {\bibfield  {journal} {\bibinfo  {journal} {New J. Phys.}\ }\textbf {\bibinfo
  {volume} {17}},\ \bibinfo {pages} {105003} (\bibinfo {year}
  {2015})}\BibitemShut {NoStop}%
\bibitem [{\citenamefont {Carusotto}\ \emph {et~al.}(2008)\citenamefont
  {Carusotto}, \citenamefont {Fagnocchi}, \citenamefont {Recati}, \citenamefont
  {Balbinot},\ and\ \citenamefont {Fabbri}}]{Carusotto-BH}%
  \BibitemOpen
  \bibfield  {author} {\bibinfo {author} {\bibfnamefont {I.}~\bibnamefont
  {Carusotto}}, \bibinfo {author} {\bibfnamefont {S.}~\bibnamefont
  {Fagnocchi}}, \bibinfo {author} {\bibfnamefont {A.}~\bibnamefont {Recati}},
  \bibinfo {author} {\bibfnamefont {R.}~\bibnamefont {Balbinot}}, \ and\
  \bibinfo {author} {\bibfnamefont {A.}~\bibnamefont {Fabbri}},\ }\href
  {http://stacks.iop.org/1367-2630/10/i=10/a=103001} {\bibfield  {journal}
  {\bibinfo  {journal} {New J. Phys.}\ }\textbf {\bibinfo {volume} {10}},\
  \bibinfo {pages} {103001} (\bibinfo {year} {2008})}\BibitemShut {NoStop}%
\bibitem [{\citenamefont {de~Nova}\ \emph {et~al.}(2014)\citenamefont
  {de~Nova}, \citenamefont {Sols},\ and\ \citenamefont
  {Zapata}}]{deNova-Sols-Zapata-CS}%
  \BibitemOpen
  \bibfield  {author} {\bibinfo {author} {\bibfnamefont {J.~R.~M.}\
  \bibnamefont {de~Nova}}, \bibinfo {author} {\bibfnamefont {F.}~\bibnamefont
  {Sols}}, \ and\ \bibinfo {author} {\bibfnamefont {I.}~\bibnamefont
  {Zapata}},\ }\href {\doibase 10.1103/PhysRevA.89.043808} {\bibfield
  {journal} {\bibinfo  {journal} {Phys. Rev. A}\ }\textbf {\bibinfo {volume}
  {89}},\ \bibinfo {pages} {043808} (\bibinfo {year} {2014})}\BibitemShut
  {NoStop}%
\bibitem [{\citenamefont {Boiron}\ \emph {et~al.}(2015)\citenamefont {Boiron},
  \citenamefont {Fabbri}, \citenamefont {Larr\'e}, \citenamefont {Pavloff},
  \citenamefont {Westbrook},\ and\ \citenamefont {Zi\ifmmode~\acute{n}\else
  \'{n}\fi{}}}]{Boiron-et-al}%
  \BibitemOpen
  \bibfield  {author} {\bibinfo {author} {\bibfnamefont {D.}~\bibnamefont
  {Boiron}}, \bibinfo {author} {\bibfnamefont {A.}~\bibnamefont {Fabbri}},
  \bibinfo {author} {\bibfnamefont {P.-E.}\ \bibnamefont {Larr\'e}}, \bibinfo
  {author} {\bibfnamefont {N.}~\bibnamefont {Pavloff}}, \bibinfo {author}
  {\bibfnamefont {C.~I.}\ \bibnamefont {Westbrook}}, \ and\ \bibinfo {author}
  {\bibfnamefont {P.}~\bibnamefont {Zi\ifmmode~\acute{n}\else \'{n}\fi{}}},\
  }\href {\doibase 10.1103/PhysRevLett.115.025301} {\bibfield  {journal}
  {\bibinfo  {journal} {Phys. Rev. Lett.}\ }\textbf {\bibinfo {volume} {115}},\
  \bibinfo {pages} {025301} (\bibinfo {year} {2015})}\BibitemShut {NoStop}%
\bibitem [{\citenamefont {Steinhauer}(2015)}]{Steinhauer-2015}%
  \BibitemOpen
  \bibfield  {author} {\bibinfo {author} {\bibfnamefont {J.}~\bibnamefont
  {Steinhauer}},\ }\href {\doibase 10.1103/PhysRevD.92.024043} {\bibfield
  {journal} {\bibinfo  {journal} {Phys. Rev. D}\ }\textbf {\bibinfo {volume}
  {92}},\ \bibinfo {pages} {024043} (\bibinfo {year} {2015})}\BibitemShut
  {NoStop}%
\bibitem [{\citenamefont {Schley}\ \emph {et~al.}(2013)\citenamefont {Schley},
  \citenamefont {Berkovitz}, \citenamefont {Rinott}, \citenamefont {Shammass},
  \citenamefont {Blumkin},\ and\ \citenamefont {Steinhauer}}]{Schley-et-al}%
  \BibitemOpen
  \bibfield  {author} {\bibinfo {author} {\bibfnamefont {R.}~\bibnamefont
  {Schley}}, \bibinfo {author} {\bibfnamefont {A.}~\bibnamefont {Berkovitz}},
  \bibinfo {author} {\bibfnamefont {S.}~\bibnamefont {Rinott}}, \bibinfo
  {author} {\bibfnamefont {I.}~\bibnamefont {Shammass}}, \bibinfo {author}
  {\bibfnamefont {A.}~\bibnamefont {Blumkin}}, \ and\ \bibinfo {author}
  {\bibfnamefont {J.}~\bibnamefont {Steinhauer}},\ }\href {\doibase
  10.1103/PhysRevLett.111.055301} {\bibfield  {journal} {\bibinfo  {journal}
  {Phys. Rev. Lett.}\ }\textbf {\bibinfo {volume} {111}},\ \bibinfo {pages}
  {055301} (\bibinfo {year} {2013})}\BibitemShut {NoStop}%
\bibitem [{\citenamefont {Macher}\ and\ \citenamefont
  {Parentani}(2009)}]{Macher-Parentani-BEC}%
  \BibitemOpen
  \bibfield  {author} {\bibinfo {author} {\bibfnamefont {J.}~\bibnamefont
  {Macher}}\ and\ \bibinfo {author} {\bibfnamefont {R.}~\bibnamefont
  {Parentani}},\ }\href {\doibase 10.1103/PhysRevA.80.043601} {\bibfield
  {journal} {\bibinfo  {journal} {Phys. Rev. A}\ }\textbf {\bibinfo {volume}
  {80}},\ \bibinfo {pages} {043601} (\bibinfo {year} {2009})}\BibitemShut
  {NoStop}%
\bibitem [{\citenamefont {Finke}\ \emph {et~al.}(2016)\citenamefont {Finke},
  \citenamefont {Jain},\ and\ \citenamefont
  {Weinfurtner}}]{Finke-Jain-Weinfurtner}%
  \BibitemOpen
  \bibfield  {author} {\bibinfo {author} {\bibfnamefont {A.}~\bibnamefont
  {Finke}}, \bibinfo {author} {\bibfnamefont {P.}~\bibnamefont {Jain}}, \ and\
  \bibinfo {author} {\bibfnamefont {S.}~\bibnamefont {Weinfurtner}},\ }\href
  {http://stacks.iop.org/1367-2630/18/i=11/a=113017} {\bibfield  {journal}
  {\bibinfo  {journal} {New J. Phys.}\ }\textbf {\bibinfo {volume} {18}},\
  \bibinfo {pages} {113017} (\bibinfo {year} {2016})}\BibitemShut {NoStop}%
\bibitem [{\citenamefont {Lopes}\ \emph {et~al.}(2015)\citenamefont {Lopes},
  \citenamefont {Imanaliev}, \citenamefont {Aspect}, \citenamefont {Cheneau},
  \citenamefont {Boiron},\ and\ \citenamefont {Westbrook}}]{Lopes-et-al}%
  \BibitemOpen
  \bibfield  {author} {\bibinfo {author} {\bibfnamefont {R.}~\bibnamefont
  {Lopes}}, \bibinfo {author} {\bibfnamefont {A.}~\bibnamefont {Imanaliev}},
  \bibinfo {author} {\bibfnamefont {A.}~\bibnamefont {Aspect}}, \bibinfo
  {author} {\bibfnamefont {M.}~\bibnamefont {Cheneau}}, \bibinfo {author}
  {\bibfnamefont {D.}~\bibnamefont {Boiron}}, \ and\ \bibinfo {author}
  {\bibfnamefont {C.}~\bibnamefont {Westbrook}},\ }\href
  {http://www.nature.com/nature/journal/v520/n7545/full/nature14331.html}
  {\bibfield  {journal} {\bibinfo  {journal} {Nature}\ }\textbf {\bibinfo
  {volume} {520}},\ \bibinfo {pages} {66} (\bibinfo {year} {2015})}\BibitemShut
  {NoStop}%
\bibitem [{\citenamefont {Mukhanov}\ \emph {et~al.}(1992)\citenamefont
  {Mukhanov}, \citenamefont {Feldman},\ and\ \citenamefont
  {Brandenberger}}]{Mukhanov-PhysRep}%
  \BibitemOpen
  \bibfield  {author} {\bibinfo {author} {\bibfnamefont {V.}~\bibnamefont
  {Mukhanov}}, \bibinfo {author} {\bibfnamefont {H.}~\bibnamefont {Feldman}}, \
  and\ \bibinfo {author} {\bibfnamefont {R.}~\bibnamefont {Brandenberger}},\
  }\href {http://www.sciencedirect.com/science/article/pii/037015739290044Z}
  {\bibfield  {journal} {\bibinfo  {journal} {Physics Reports}\ }\textbf
  {\bibinfo {volume} {215}},\ \bibinfo {pages} {203 } (\bibinfo {year}
  {1992})}\BibitemShut {NoStop}%
\bibitem [{\citenamefont {Campo}\ and\ \citenamefont
  {Parentani}(2004)}]{Campo-Parentani-2004}%
  \BibitemOpen
  \bibfield  {author} {\bibinfo {author} {\bibfnamefont {D.}~\bibnamefont
  {Campo}}\ and\ \bibinfo {author} {\bibfnamefont {R.}~\bibnamefont
  {Parentani}},\ }\href
  {https://journals.aps.org/prd/abstract/10.1103/PhysRevD.70.105020} {\bibfield
   {journal} {\bibinfo  {journal} {Phys. Rev. D}\ }\textbf {\bibinfo {volume}
  {70}},\ \bibinfo {pages} {105020} (\bibinfo {year} {2004})}\BibitemShut
  {NoStop}%
\bibitem [{\citenamefont {Hung}\ \emph {et~al.}(2013)\citenamefont {Hung},
  \citenamefont {Gurarie},\ and\ \citenamefont {Chin}}]{Hung-Gurarie-Chin}%
  \BibitemOpen
  \bibfield  {author} {\bibinfo {author} {\bibfnamefont {C.-L.}\ \bibnamefont
  {Hung}}, \bibinfo {author} {\bibfnamefont {V.}~\bibnamefont {Gurarie}}, \
  and\ \bibinfo {author} {\bibfnamefont {C.}~\bibnamefont {Chin}},\ }\href
  {\doibase 10.1126/science.1237557} {\bibfield  {journal} {\bibinfo  {journal}
  {Science}\ }\textbf {\bibinfo {volume} {341}},\ \bibinfo {pages} {1213}
  (\bibinfo {year} {2013})}\BibitemShut {NoStop}%
\bibitem [{\citenamefont {Kofman}\ \emph {et~al.}(1997)\citenamefont {Kofman},
  \citenamefont {Linde},\ and\ \citenamefont
  {Starobinsky}}]{Kofman-Linde-Starobinsky}%
  \BibitemOpen
  \bibfield  {author} {\bibinfo {author} {\bibfnamefont {L.}~\bibnamefont
  {Kofman}}, \bibinfo {author} {\bibfnamefont {A.}~\bibnamefont {Linde}}, \
  and\ \bibinfo {author} {\bibfnamefont {A.~A.}\ \bibnamefont {Starobinsky}},\
  }\href {\doibase 10.1103/PhysRevD.56.3258} {\bibfield  {journal} {\bibinfo
  {journal} {Phys. Rev. D}\ }\textbf {\bibinfo {volume} {56}},\ \bibinfo
  {pages} {3258} (\bibinfo {year} {1997})}\BibitemShut {NoStop}%
\bibitem [{\citenamefont {Vogels}\ \emph {et~al.}(2002)\citenamefont {Vogels},
  \citenamefont {Xu}, \citenamefont {Raman}, \citenamefont {Abo-Shaeer},\ and\
  \citenamefont {Ketterle}}]{Vogels-et-al}%
  \BibitemOpen
  \bibfield  {author} {\bibinfo {author} {\bibfnamefont {J.~M.}\ \bibnamefont
  {Vogels}}, \bibinfo {author} {\bibfnamefont {K.}~\bibnamefont {Xu}}, \bibinfo
  {author} {\bibfnamefont {C.}~\bibnamefont {Raman}}, \bibinfo {author}
  {\bibfnamefont {J.~R.}\ \bibnamefont {Abo-Shaeer}}, \ and\ \bibinfo {author}
  {\bibfnamefont {W.}~\bibnamefont {Ketterle}},\ }\href {\doibase
  10.1103/PhysRevLett.88.060402} {\bibfield  {journal} {\bibinfo  {journal}
  {Phys. Rev. Lett.}\ }\textbf {\bibinfo {volume} {88}},\ \bibinfo {pages}
  {060402} (\bibinfo {year} {2002})}\BibitemShut {NoStop}%
\bibitem [{\citenamefont {Press}\ \emph {et~al.}(2007)\citenamefont {Press},
  \citenamefont {Teukolsky}, \citenamefont {Vetterling},\ and\ \citenamefont
  {Flannery}}]{Numerical_Recipes}%
  \BibitemOpen
  \bibfield  {author} {\bibinfo {author} {\bibfnamefont {W.~H.}\ \bibnamefont
  {Press}}, \bibinfo {author} {\bibfnamefont {S.~A.}\ \bibnamefont
  {Teukolsky}}, \bibinfo {author} {\bibfnamefont {W.~T.}\ \bibnamefont
  {Vetterling}}, \ and\ \bibinfo {author} {\bibfnamefont {B.~P.}\ \bibnamefont
  {Flannery}},\ }\href@noop {} {\emph {\bibinfo {title} {Numerical Recipes: The
  Art of Scientific Computing}}}\ (\bibinfo  {publisher} {Cambridge University
  Press},\ \bibinfo {year} {2007})\BibitemShut {NoStop}%
\end{thebibliography}%

\end{document}